\documentclass[12pt]{article}
\usepackage{amsmath}
\usepackage{epsfig}
\usepackage{amssymb}
\usepackage{hyperref}
\usepackage{bbold}
\usepackage{braket}
\input{epsf}
\def\be{\begin{equation}}
\def\ee{\end{equation}}
\def\beq{\begin{equation}}
\def\eeq{\end{equation}}
\def\beqa{\begin{eqnarray}}
\def\eeqa{\end{eqnarray}}
\def\ba{\begin{eqnarray}}
\def\ea{\end{eqnarray}}
\def\bea{\begin{eqnarray}}
\def\eea{\end{eqnarray}}

\newcommand\as{\alpha_s}
\newcommand\f[2]{\frac{#1}{#2}}
\def\la{\lambda}

\def\beq{\begin{equation}}
\def\eeq{\end{equation}}
\def\beeq{\begin{eqnarray}}
\def\eeeq{\end{eqnarray}}
\def\to{\rightarrow}
\def\nn{\nonumber}

\def\b0{b_0}
\def\bone{b_1}
\def\btwo{b_2}

\def\b0{b_0}
\def\bone{b_1}
\def\btwo{b_2}

\def\la{\lambda}

\begin{document}

\begin{titlepage}
\renewcommand{\thefootnote}{\fnsymbol{footnote}}
\begin{flushright}
YITP-SB-18-35 \\
     \end{flushright}
\par \vspace{10mm}
\begin{center}
{\large \bf
Threshold Resummation at NNLL for Single-particle Production \\[4mm] 
in Hadronic Collisions}\\

\vspace{8mm}

\today
\end{center}

\par \vspace{2mm}
\begin{center}
{\bf Patriz Hinderer${}^{\,a}$,}
\hskip .2cm
{\bf Felix Ringer${}^{\,b}$,}
\hskip .2cm
{\bf George Sterman${}^{\,c}$,}
\hskip .2cm
{\bf Werner Vogelsang${}^{\,a}$  }\\[5mm]
\vspace{5mm}
${}^{a}\,$ Institute for Theoretical Physics, T\"ubingen University, 
Auf der Morgenstelle 14, \\ 72076 T\"ubingen, Germany\\[2mm]
${}^b$ Nuclear Science Division, Lawrence Berkeley National Laboratory, Berkeley, CA 94720, U.S.A. \\[2mm]
${}^{c}\,$ C.N.\ Yang Institute for Theoretical Physics and Department of Physics and Astronomy\\
Stony Brook University, Stony Brook, 
New York 11794 -- 3840, U.S.A.\\
\end{center}


\vspace{9mm}
\begin{center} {\large \bf Abstract} \end{center}
We develop threshold resummation for single-particle inclusive cross sections in hadron-hadron collisions to the level of next-to-next-to-leading logarithm, up to 
full matching with two-loop hard functions.
We define and calculate all one-loop soft functions for all partonic channels.   This enables us to separate the hard and soft functions at one loop.
Along with these results, the one-loop finite parts of jet functions are used to check that the full soft, collinear and virtual corrections are reproduced to one loop for all partonic reactions.
We exhibit these NLO results explicitly.     NLO expansions of the resummed cross section match the exact NLO results extremely well
numerically, and two loop expansions result in substantial corrections over many kinematic configurations.  
Explicit results are given in Mellin moment space, and a number of options for generating resummed cross sections are discussed.

\end{titlepage}  

\setcounter{footnote}{2}
\renewcommand{\thefootnote}{\fnsymbol{footnote}}


\section{Introduction}

Single-particle inclusive (1PI) cross sections are among the fundamental processes in QCD, with factorization and evolution properties that are  basic results of quantum field theory \cite{Mueller:1978xu,Ellis:1991qj}.   It is the purpose of this paper to study the resummation of 1PI cross sections for hadrons in hadron-hadron scattering at the level of next-to-next-to-leading logarithmic (NNLL) resummation at partonic threshold \cite{Sterman:1986aj,Catani:1989ne}.  We will derive explicit NNLL-resummed partonic hard-scattering functions for all partonic $2\to 2$ reactions in terms of moments with respect to the variable $\hat s_4$, which characterizes the kinematic distance to partonic threshold for the production of a final-state particle with observed transverse momentum and rapidity. 

In hadronic scattering at NLO \cite{Aversa:1988vb,Jager:2002xm}, large corrections to single-inclusive 
cross sections are primarily associated with the kinematics of partonic threshold \cite{Sterman:1986aj,Catani:1989ne,Laenen:1998qw,Catani:1998tm,Bonciani:2003nt}. 
The resummation of such threshold corrections for 1PI cross sections was systematized  in Refs.\ \cite{Laenen:1998qw,Catani:1998tm,Kidonakis:1999hq,Sterman:2000pt,deFlorian:2005wf} for the case of prompt photons, 
where the resummation was carried out at next-to-leading logarithmic (NLL) level. 
In particular, the NLO expansion of the NLL resummed cross section can readily be compared to the full 
NLO result \cite{Gordon:1993qc,Aurenche:1983ws} for this process. Resummation for 
single-inclusive hadron production was investigated at NLL for hadronic scattering in Ref.~\cite{deFlorian:2005yj}
and for photoproduction in Ref.~\cite{deFlorian:2013taa}. Especially large corrections from resummation were 
found in~\cite{deFlorian:2005yj} for the case of the rapidity-integrated cross section. 

Important extensions of resummation techniques for 1PI cross sections in hadronic scattering to 
next-to-next-to-leading logarithm (NNLL) were made for hadron production in \cite{Catani:2013vaa} 
in conventional ``direct'' QCD, and for prompt photon cross sections in \cite{Becher:2009th} and top production
in \cite{Ferroglia:2013awa}
using soft-collinear effective theory (SCET). The effective theory treatments provide similar results in a Mellin transform space for the partonic cross section, as described below, but differ in their implementation of the transform back to the factorized cross section.   Here we will follow a direct QCD approach, although we will make contact with the effective theory results of Ref.~\cite{Becher:2009th}.   
We will rederive the necessary results found in Ref.~\cite{Catani:2013vaa} and the analysis of wide-angle soft radiation in \cite{Ferroglia:2013awa},
develop further the all-orders factorization properties of the relevant cross sections in terms of direct
QCD matrix elements, and provide a formal moment inversion prescription.   In addition, we will provide explicit 
expressions for the relevant hard-scattering and soft functions necessary to describe 1PI hadron production
cross sections at NNLL level in hadronic scattering, and exhibit finite NLO  contributions that enter the resummed NNLL cross section in moment space with the running coupling evolved to a soft scale.

The path taken in this paper follows the general lines of threshold resummation for dihadron pairs in Ref.\ \cite{Almeida:2009jt} at NLL and \cite{Hinderer:2014qta} at NNLL.  In the development of the formalism, however, we find several significant differences from the dihadron case.   In particular, we conclude that a direct application of the inverse transform to momentum space following the method of \cite{Catani:1996yz} is not as appropriate for 1PI cross sections as for the dihadron case.   This is manifested by potentially large values of the 1PI cross section in an unphysical region, even though these contributions remain formally power-suppressed in moment space \cite{Catani:1996yz,Abbate:2007qv}.  
We leave for future work the development of an extension of the method of Ref.\ \cite{Catani:1996yz} to this case.  We believe that the studies in this paper will be relevant to classic fixed target data  on 1PI cross sections for hadron-hadron scattering, as well as to higher-energy collider data.   Extensions of the formalism to photoproduction at NNLL, both resolved and direct, are straightforward.   

For very high-energy colliders, jet inclusive cross sections are of special interest, and are clearly related to the results presented here. Fixed-order jet cross sections
have been brought to the level of two loops, \cite{Currie:2016bfm,Abelof:2016pby}, and the impact of threshold resummation has been discussed at least to NLL accuracy in direct QCD \cite{KOS,Kidonakis:2000gi,deFlorian:2007fv,Kumar:2013hia,deFlorian:2013qia,Biekotter:2015nra,Nagy:2017dxh} 
and effective theory formalisms \cite{Becher:2015hka,Becher:2016omr,Liu:2017pbb,Dai:2017dpc,Balsiger:2018ezi,Kang:2018jwa}. 
Inclusive jet cross sections share the underlying kinematics of 
single-particle inclusive cross sections, and we anticipate that the formalism for single particles described in this paper will have useful applications to single jets.

We begin with a review of the kinematics and the factorization properties of single-particle inclusive cross sections in Sec.\ \ref{sec:fact}.   The ``refactorization" relations upon which threshold resummation for 1PI cross sections is based are reviewed in Sec.\ \ref{sec:refactorization}, where we emphasize similarities and differences compared to dihadron and related cross sections for which threshold resummations have been carried out.   In  Sec.\ \ref{sec:resum}, we rederive results for the relevant jet functions 
in moment space.   
 The definition of the direct QCD soft function turns out to require a slightly modified treatment of soft gluon phase space, leading to a soft function that differs from that for dihadron cross 
sections \cite{Hinderer:2014qta}, for example.  This construction is the subject of Sec.\ \ref{sec:define-S1pi}.    The resulting calculations for the one-loop soft function are given for $qq'$ scattering in Sec.\ \ref{sec:calc-S1pi} and summarized for all other partonic 
processes in an Appendix.
The determination of the finite matrices that describe the soft function for all $2\to 2$ partonic processes is a main result of this paper.  
We collect the full resummed cross section in moment space in Sec.\ \ref{sec:hard-resum}, and
in Sec.\ \ref{sec:compare} we confirm that all singular behavior at threshold is reproduced at NLO by the one-loop expansions of our hard, soft and jet functions. 
The results of Sec.\ \ref{sec:hard-resum} are presented in an alternate form in Sec.\ \ref{sec:N-indep}, with fixed jet and soft evolution scales,
confirming that they are consistent with those of \cite{Becher:2009th} when specialized to prompt photon production.
We review in Sec.\ \ref{secmell} several approaches to the application of our results to quantitative cross sections. Section~\ref{secnumres}
presents exploratory numerical tests of the fixed-order (NNLO) expansion of the resummed cross section obtained
in our formalism.  
In an additional Appendix, we provide for completeness the explicit NLO singular contributions at threshold, which are reproduced for all partonic processes by the resummation studied here.

\section{Factorization and Moments \label{sec:fact}}

\subsection{Factorization and kinematics}

The classic, collinear-factorized form of the single-particle inclusive cross section $pp\to h X$ is
\beeq
p_T^3\frac{d^2\sigma^{pp\to hX}}{dp_T    d{\eta} }
&=&\sum_{abc}
\int_0^1 dx_a dx_bdz_c    \, z_c^2\;  
f_a(x_a,\mu_F)f_b(x_b,\mu_F)\, D_c^h(z_c,\mu_F)
 \nn \\[2mm]
&\times & \omega^{ab\to c} \left(\hat{\eta}, \hat{x}_T  \cosh \hat\eta,  \frac{\mu_F^2}{\hat s}\right)\;,
\label{taufac} 
\eeeq
with $p_T$ the transverse momentum of the observed hadron $h$, and $\eta$ its rapidity. 
The partonic inclusive hard-scattering function $\omega^{ab\to c}$, which is accessible to perturbative QCD, 
describes the production of parton $c$ with transverse momentum $\hat p_T = p_T/z_c$ and rapidity 
$\hat \eta$ in the partonic center-of-mass frame, the latter related to $\eta$ by
\bea
 \hat \eta\ =\ \eta\ -\ \frac{1}{2}\, \ln \frac{x_a}{x_b}\, .
 \label{eq:hat-eta-def}
\eea
Parton $c$ subsequently fragments into hadron $h$. We define, for hadronic and partonic kinematics,
\bea
x_T\ &=& \frac{2p_T}{\sqrt{s}}\,, \nonumber \\[2mm]
\hat x_T\ &=&\ \frac{2\hat p_T}{\sqrt{\hat s}}\ \equiv\ \frac{2p_T}{z_c\sqrt{x_ax_bs}}  \, ,
\label{eq:xt-defs}
\eea
with $s$ and $\hat{s}=x_ax_b s$ the hadronic and partonic center-of-mass energies squared, respectively.
In Eq.\ (\ref{taufac}),  $\mu_F$ is the factorization scale
that ties together the partonic inclusive hard-scattering functions and parton distributions $f_{a,b}$ 
and fragmentation functions  $D_c^h$.   
We omit dependence on the scale at which the perturbative coupling is evaluated, normally denoted by $\mu_R$.   In principle, $\omega$ is independent of this choice, although of course to any
fixed order, dependence on $\mu_R$ appears at the next order.   

The conventional partonic variable for 1PI resummation is defined in terms of the 
momenta $p_a,p_b,p_c$ in the partonic reaction $a+b\to c+{\textrm{anything}}$ as
\bea
\hat s_4\ &=&\  (p_a+p_b-p_c)^2 \nn\\[2mm]
&=&\ \hat s \, \left( 1 - \hat x_T \cosh \hat\eta \right)\, ,
\label{eq:s4-def}
\eea  
which, as indicated, is the square of the invariant mass of all radiation additional to $p_c$ 
in the partonic final state.  In the partonic threshold limit, this quantity vanishes, 
and the hard-scattering function becomes singular.

\subsection{Moment analysis of the inclusive hard-scattering function}

We write the cross section in Eq.\ (\ref{taufac}) in short as
\be
p_T^3\frac{d^2\sigma^{pp\to hX}}{dp_Td\eta}\,=\, \sum_{abc}\int dx_a \,f_a(x_a) \int dx_b \,f_b(x_b) 
\int_z^1 {dz_c}{z^2_c} \,D_c^h(z_c)\,\omega^{ab\to c}(\hat\eta,\hat{x}_T\cosh\hat\eta)\,,
\label{eq:taufac-rewrite}
\ee
where we have for simplicity omitted all dependence on $\mu_F$. Using Eq.~(\ref{eq:s4-def}), 
the second argument in the inclusive hard-scattering function $\omega^{ab\to c}$ may also be written as
\be
\hat{x}_T\cosh\hat\eta\,=\,1-\frac{\hat{s}_4}{\hat{s}}\,.
\label{eq:xt_s4}
\ee
In~(\ref{eq:taufac-rewrite}), the lower limit of the integration over the fragmentation variable $z_c$ is given by
\be
z\,\equiv\,\frac{x_T}{\sqrt{x_ax_b}}\cosh\hat\eta\ =\ z_c\left(  1-\frac{\hat{s}_4}{\hat{s}}  \right) \, ,
\label{eq:z-def}
\ee
so that $z_c>z$ ensures that $\hat s_4>0$.   In these terms, we may also write the cross section (\ref{eq:taufac-rewrite}) as
\be\label{crossec}
p_T^3\frac{d^2\sigma^{pp\to hX}}{dp_Td\eta}\,=\, \sum_{abc}\int dx_a \,f_a(x_a) \int dx_b \,f_b(x_b) 
\int_z^1 {dz_c}{z^2_c} \,D_c^h(z_c)\,\omega^{ab\to c}\left(\hat\eta,\frac{z}{z_c}\right)\,,
\ee
so that the last integral takes the form of a genuine convolution of the fragmentation function with the 
hard-scattering function, at fixed $\hat\eta$, which we denote as
\bea
\Omega^{ab\to c}\left(\hat\eta,z\right)\ \equiv \ \int_z^1 {dz_c}\, {z^2_c} \,D_c^h(z_c)\,\omega^{ab\to c}\left(\hat\eta,\frac{z}{z_c}\right)\, .
\label{eq:Omega-def}
\eea
Moments of $\Omega^{ab\to c}$ then factor into simple products:
\bea
\int_0^1 dz\,z^{N-1}\, \Omega^{ab\to c}\left(\hat\eta, z \right)\,=\,
\tilde{D}_c^h(N+3)\,\tilde{\omega}^{ab\to c}(\hat\eta,N)\, ,
\label{eq:Omega-D-moment}
\eea
with $\tilde{D}_c^h(N)$ the moment of a fragmentation function for hadron $h$,
and $\tilde\omega_{ab\to c}$ the moment of the corresponding inclusive hard-scattering function,
\bea
\tilde{\omega}^{ab\to c}(\hat\eta,N)\,&\equiv&\,\int_0^1 d y \, y^{N-1}\,\omega^{ab\to c}(\hat\eta, y)
\nn \\[2mm]
&=&\,
\int_0^{\hat s} \,\frac{d\hat{s}_4}{\hat{s}}\,\left(1-\frac{\hat{s}_4}{\hat{s}}\right)^{N-1}\,\omega^{ab\to c}\left(\hat\eta, 1-\frac{\hat{s}_4}{\hat{s}}\right)\,.
\label{eq:s4-moments}
\eea
As we review below, these are precisely the moments that the resummation formalism gives us
near partonic threshold, where $\hat s_4\to 0$ or alternately $N$ is large in Eq.\ (\ref{eq:s4-moments}).
After resummation, we will perform the Mellin inverse of (\ref{eq:Omega-D-moment}),
\be
\Omega^{ab\to c}(\hat\eta,z)\,\equiv\,
\frac{1}{2\pi i}\int_{\cal C}dN \,z^{-N}\,\tilde{D}_c^h(N+3)\,\tilde{\omega}^{ab\to c}(\hat\eta,N)\,,
\label{eq:Omega-inverse}
\ee
with ${\cal C}$ a contour to the right of all singularities of $\tilde \omega$.
In principle, the moments of the fragmentation functions fall off sufficiently fast so that the integration
can be carried out numerically without problem in Eq.\ (\ref{eq:Omega-inverse}).
So long as the short-distance function has support only for $z<1$,
as is the case for any finite-order expansion of the resummed expression, the procedure is unique.  Assuming that we may use the standard Mellin contour for~(\ref{eq:Omega-inverse}), which, when tilted to the left, acquires an ever larger 
negative real part along the contour, the Mellin integral will converge very rapidly. The result of this procedure can then be convoluted with the parton distributions in Eq.\ (\ref{eq:taufac-rewrite}), to give
\be
p_T^3\frac{d^2\sigma^{pp\to hX}}{dp_Td\eta}\,=\,\sum_{abc} \int_{\frac{x_T{\mathrm{e}}^{\eta}}{2-x_T {\mathrm{e}}^{-\eta}}}^1 dx_a \,f_a(x_a) \int_{\frac{x_ax_T{\mathrm{e}}^{-\eta}}{2x_a-x_T {\mathrm{e}}^{\eta}}}^1 dx_b \,f_b(x_b) \,
\Omega^{ab\to c}\left(\hat\eta,z=\frac{x_T}{\sqrt{x_ax_b}}\cosh\hat\eta\right)\, ,
\label{eq:Omega-eq-1}
\ee
where we have exhibited the ranges of partonic fractional momenta for the incoming hadrons in terms of the quantities $x_T$ and $\eta$ that define the cross section.   

We note that the  procedure described here for moment inversion is different from the one used for dihadron production 
considered in Refs.\ \cite{Almeida:2009jt} and \cite{Hinderer:2014qta}.  For those processes, it was convenient to take a double transform: a Mellin transform in the ratio of the pair invariant mass to the center of mass energy and a Fourier transform in the average rapidity of the observed particles.   After resummation, the double inverse transform may be carried out numerically as in Ref.\ \cite{Sterman:2000pt}.   The reasons for following a different path here will become clear from the refactorization discussion in the following section, and in the explicit form of the moment-space resummation in Sec.\ \ref{sec:zspace}.

\section{Refactorization at Partonic Threshold }
\label{sec:refactorization}

\subsection{Refactorization in momentum space}
\label{sec:refact-mtm-sp}

Partonic threshold is reached when $\hat x_T \cosh \hat\eta=1$ and the $2\to 2$ partonic subprocess is elastic, leaving no energy available for 
radiation.    This restriction of phase space leads, as usual, to plus distributions in the partonic variable, $1-y \equiv \hat s_4/\hat s$, the most 
singular of which are double logarithmic, $\alpha_s^n \big[\ln^{2n-1}(1-y)/(1-y)\big]_+$.    
Threshold resummation organizes these leading, and in principle all nonleading, distributions of $1-y$ at fixed rapidity for the observed hadron.   

For such a single-particle inclusive cross section we have a further factorization of the inclusive hard-scattering function $\omega_{ab\to c}$ near partonic threshold, where $\hat s_4$ vanishes.    This is the lightlike limit for all radiation that accompanies the fragmenting parton $c$ in the final state of the partonic subprocess.    As shown in Refs.\ \cite{Laenen:1998qw,Catani:1998tm}, near partonic threshold each particle in the final state may be associated with one of five factoring functions.   The first two, labelled here $J_{\mathrm{in}}^{(a)}$ 
and $J_{\mathrm{in}}^{(b)}$, are associated with the two incoming partons.   The third, $S_{ab\to cr}$,
is a matrix in the space of color tensors, and generates coherent, wide-angle radiation.  The fourth, $J_{\mathrm{rec}}^{(r)}$, 
describes a jet of particles recoiling against the direction of $p_c$ in the partonic center of mass frame, and the fifth, 
$J_{\mathrm{fr}}^{(c)}$, describes radiation collinear to the fragmenting jet itself.    The key to threshold resummation in this context, and elsewhere, is that in computations of the inclusive hard-scattering function, $\omega^{ab \to c}$, every quantum of final-state radiation can be associated with one of these functions.  This feature is directly related to the definition of $\hat s_4$ in Eq.\ (\ref{eq:s4-def}) by
\bea
p_a+p_b-p_c\ =\ k_a+k_b+k_S+k_r+k_c\, ,
\label{eq:fs-rad}
\eea
where $k_a$ and $k_b$ are the momenta of particles associated with the incoming jets, $k_S$ with the function that generates wide-angle radiation, $k_r$ with the jet recoiling against the observed particle, and finally $k_c$ with radiation collinear to parton $p_c$.    The natural metric for the approach to partonic threshold is $\hat s_4$, defined in Eq.~(\ref{eq:s4-def}) as the invariant mass associated with Eq.\ (\ref{eq:fs-rad}).   

The next key observation is that in the limit of partonic threshold, the only momentum with fixed, nonvanishing components is $k_r$, associated with the recoil jet.   Denoting $\lambda\equiv \hat s_4/\hat s$, to leading power in $\lambda$, $\hat s_4=(p_a+p_b-p_c)^2$
may be written as a sum,
\bea
\hat s_4\ =\ k_r^2\ +\ \sum_{i=a,b,c,S} 2k_r\cdot k_i\ + {\cal O}(\lambda^2)\, .
\label{eq:s4-expand}
\eea
This relation produces a kinematic convolution near partonic threshold between the jet and soft functions into which the cross section factorizes.  Of course, there are ambiguities in the definitions of the functions that generate this radiation, especially for wide-angle soft radiation, in the contribution of each function to a portion of phase space.   It is precisely this issue that is resolved by the arguments for factorization in Refs.\  \cite{Laenen:1998qw,Catani:1998tm}, for example, and is built into the effective theory treatment of prompt photons in Ref.\ \cite{Becher:2009th}.

We will use the formalism developed in Ref.\ \cite{Laenen:1998qw}, reexpressed in covariant 
gauges for threshold resummation in Refs.\  \cite{Almeida:2009jt,Hinderer:2014qta}. 
In this formalism, the incoming jet functions $J_{\mathrm{in}}^{(a)}$ 
and $J_{\mathrm{in}}^{(b)}$ develop logarithms not in $k_r\cdot k_{a,b}$ directly, but in variables $w_a$ and $w_b$, determined by the corresponding terms in (\ref{eq:s4-expand}) through the kinematic relations
\bea
\frac{2k_r\cdot k_a}{\hat s}\ &=&\ w_a\left(\frac{-\hat u}{\hat s}\right)\, ,
\nn\\[2mm]
\frac{2k_r\cdot k_b}{\hat s}\ &=&\ w_b\left(\frac{-\hat t}{\hat s}\right)\, ,
\label{eq:wab-def}
\eea
where $\hat u=(p_b-p_c)^2$ and $\hat t=(p_a-p_c)^2$. 
The fragmenting jet is a function of the fraction of momentum available for collinear radiation, 
and is denoted by $J_{\mathrm{fr}}^{(c)}(w_c)$, with
\bea
w_c\ \equiv\ \frac{2k_r\cdot k_c}{\hat s}\, .
\label{eq:wC-def}
\eea
The recoil jet is a function of its invariant mass directly. We denote it as $J_{\mathrm{rec}}^{(r)}(w_r)$, where
\bea
w_r \equiv\ \frac{k_r^2}{\hat s}\, .
\label{eq:wR-def}
\eea
Finally, the soft function depends on the ratio
\bea
w_S\ \equiv\ \frac{2\beta_r \cdot k_S}{\sqrt{\hat s}}\, ,
\label{eq:wS-def}
\eea
with $\beta_r$ a light-like velocity vector in the direction of the recoil jet.   Although the
variable $w_S$ vanishes when $k_S$ is collinear to $\beta_r$, the soft function is constructed so that there are no enhancements in this limit.
We can take the all-log resummations for each of  these functions from the literature, and we will describe them below.

In the notation we have just reviewed, we can write our re-factorized short distance function as \cite{Laenen:1998qw,KS}
\bea
\omega^{ab\to c} \left( \hat{\eta}, 1-\frac{\hat{s}_4}{\hat{s}} , \frac{\mu_F^2}{\hat{s}}\right)
&=&
\int dw_a\; dw_b\; dw_S\; dw_r\; dw_c\nn\\[2mm]
& \times& \delta \left(\frac{\hat s_4}{\hat s} - w_a\left(\frac{-\hat u}{\hat s}\right) - w_b\left(\frac{-\hat t}{\hat s}\,\right) 
- \sum_{i=S,r,c}w_i \right) \ \nn\\[2mm]
&\times&
J_{\rm in}^{(a)} \left (w_a,\frac{w_a \sqrt{\hat{ s}}}{\mu_{rf}}, \frac{\mu_F}{\mu_{rf}},\as(\mu_{rf}) \right)\;  
J_{\rm in}^{(b)} \left (w_b,\frac{w_b \sqrt{\hat{ s}}}{\mu_{rf}}, \frac{\mu_F}{\mu_{rf}},\as(\mu_{rf}) \right )\nn\\[2mm]
&\times& J_{\rm fr}^{(c)} \left ( w_c,\frac{w_c \sqrt{\hat{ s}}}{\mu_{rf}}, \frac{\mu_F}{\mu_{rf}},\as(\mu_{rf}) \right )\; J_{\rm rec}^{(r)}\left (w_r, \frac{w_r \hat{s}}{\mu_{rf}^2},\as(\mu_{rf}) \right) \nn\\[2mm]
& \times & {\mathrm{Tr}} \left\{ H \left ( \as(\mu_{rf}),\frac{\sqrt{\hat{s}}}{\mu_{rf}} ,\frac{\mu_F}{\mu_{rf}},  \hat\eta  \right)\, 
S\left(w_S,\frac{w_S \sqrt{\hat {s}}}{\mu_{rf}},\as(\mu_{rf}), \hat\eta \right)   \right\}_{ab \to cr} \nn \\[2mm]
 &+& {\cal O}\left( \left (\frac{\hat{s}_4}{\hat{s}} \right)^0\right)\, ,
\label{threshconv}
\eea
where $H$  is a matrix in the space of color tensors, appearing in a trace of possible color factors with the soft matrix $S$ \cite{KS}.   The matrix $H$ serves as a purely short-distance, non-radiative coefficient function for the soft matrix and jet functions.   We will generally refer to it as the ``hard function'' below, although it should not be confused with the full, inclusive hard-scattering function $\omega^{ab\to c}$, 
of whose refactorization $H$ is a part. Both $\omega^{ab\to c}$ and $H$
can be computed order by order in perturbation theory. The inclusive hard-scattering function $\omega^{ab\to c}$ is known explicitly to 
NLO \cite{Aversa:1988vb,Jager:2002xm}, while the hard function $H$ has been computed to 
NLO in~\cite{Kunszt:1993sd,Kelley:2010fn,Hinderer:2014qta}, and even to NNLO in~\cite{Broggio:2014hoa}.  

As indicated, corrections to the refactorized form given in Eq.~(\ref{threshconv}) are not power-divergent at partonic threshold, $\hat s_4=0$, although they may behave logarithmically in $\hat s_4/\hat s$.  The freedom associated with the factorization of $\omega^{ab\to c}$ in Eq.~(\ref{threshconv}) in 
  terms of four jet functions, a hard matrix, and a soft matrix
is encoded into a ``refactorization scale", labelled here by $\mu_{rf}$.    The hard scattering function, however, is independent of $\mu_{rf}$,
 \bea
\mu_{rf}\,  \frac{d}{d \mu_{rf}} \, \omega^{ab\to c} \left( \hat{\eta}, 1-\frac{\hat{s}_4}{\hat{s}} , \frac{\mu_F^2}{\hat{s}}\right)\ =\ 0\, .
 \label{eq:murf-indep}
 \eea
 For matching comparisons to fixed-order calculations, and other purposes associated with numerical evaluation of the cross section, we will introduce a standard ``renormalization scale", which in principle can be chosen independently.   For example, to match to an NLO calculation at a renormalization scale like $p_T$, we expand each coupling $\as(\mu_{rf})$ in terms of $\as(p_T)$.   Notice that every function on the right-hand side of Eq.\ (\ref{threshconv}) is formally independent of the choice of $\mu_R$ in these terms, because $\as(\mu_{rf})$ must take the same value when reexpanded in terms of the coupling at another scale.   
 
An important feature of Eq.\ (\ref{threshconv}) is the dependence on the center-of-mass rapidity, $\hat \eta$, in the hard and soft functions.   Through Eq.\ (\ref{eq:hat-eta-def}), this means that these functions depend directly on the observed rapidity of the particle.    For dihadron production \cite{Hinderer:2014qta}, in contrast, hard and soft functions depend only on boost-invariant differences in the rapidities of the hadrons.    As a result, for those processes it is straightforward to perform a Fourier transform in the average rapidity,
\bea
\bar \eta \ =\ \frac{1}{2}\left (\eta_1+\eta_2\right)\, ,
\eea
with $\eta_i$ the physical rapidity of particle $i$,
as mentioned at the end of Sec.\ \ref{sec:fact}.   In principle, such a transform can be carried out here as well, following the example of Ref.\ \cite{Sterman:2000pt} for prompt photons, but the analytic forms would be more challenging for the exponentiated soft functions, which are matrices in this case.   Similarly, $\hat \eta$ appears implicitly in the delta function that defines the convolution in Eq.\ (\ref{threshconv}), so that $\eta$-dependence enters indirectly in the Mellin moments.   These differences led us to forego the use of a Fourier transform in rapidity as in~\cite{Sterman:2000pt}, 
instead using the calculation of the function $\Omega$ in Eq.\ (\ref{eq:Omega-eq-1}).
 
With the exception of the hard function, each of the factors in Eq.\ (\ref{threshconv}) can be computed from universal matrix elements in full QCD, which will be identified below, and they all obey  evolution equations that control their behavior in the kinematic limits that provide logarithms of  $\hat s_4/\hat s$.   The hard function $H$ is computed directly from the  virtual corrections
to the $2\to 2$ scattering processes $ab\to cr$, decomposed in terms of color tensors \cite{Hinderer:2014qta,Kunszt:1993sd,Kelley:2010fn,Broggio:2014hoa}. The refactorization described here has precise analogs in the formalism of soft-collinear effective theory applied at this level for prompt photon production in Ref.\ \cite{Becher:2009th}.

In our analysis we use the feature that convolutions like the right-hand side of Eq.\ (\ref{threshconv}) are factorized into products in 
Mellin- or equivalently Laplace moment space \cite{Laenen:1998qw,Catani:1998tm,Catani:2013vaa} through
\bea
\label{eq:F(N)}
\tilde F(N)\ \ =\ \int_0^1 dw\, (1-w)^{N-1} F(w)\ =\ \int_0^\infty dw\ {\mathrm{e}}^{-N w}\ F(w)\ +\ {\cal O}\left( \frac{1}{N}\right)\, ,
\eea
for the functions in Eq.\ (\ref{threshconv}), each of which has distributions of the form $ [ \ln^n w/w]_+$.  
Corrections to the second equality are suppressed by inverse powers of the moment variable $N$.    Note that the kinematic argument of the soft function is $w_S \sqrt{\hat{s}}/\mu_{rf}$.   
This dependence is also found, for example, in inclusive electroweak, dijet  \cite{KOS}  and dihadron \cite{Almeida:2009jt,Hinderer:2014qta} cross sections.   
In this notation, the $\hat s_4$ moments, Eq.\ (\ref{eq:s4-moments}), of the factorized hard scattering function (\ref{threshconv}) are now products in terms of hard-scattering and soft functions, and initial state and final state jet functions, identified above,
\bea
\tilde{\omega}^{ab\to c}(\hat\eta,N)\
&=& \tilde J_{\rm in}^{(a)} \left (\frac{ \hat{s}}{\bar{N}_a^2 \mu_{rf}^2}, \frac{\mu_F}{\mu_{rf}},\as(\mu_{rf}) \right)\;  
\tilde J_{\rm in}^{(b)} \left (\frac{\hat{s}}{\bar{N}_b^2 \mu_{rf}^2}, \frac{\mu_F}{\mu_{rf}},\as(\mu_{rf}) \right )\nn\\[2mm]
&\times& \tilde J_{\rm fr}^{(c)} \left ( \frac{ \hat{s}}{\bar{N}^2 \mu_{rf}^2}, \frac{\mu_F}{\mu_{rf}},\as(\mu_{rf}) \right )\; 
\tilde J_{\rm rec}^{(r)}\left (\frac{ \hat{s}}{\bar{N} \mu_{rf}^2},\frac{\hat s}{\bar{N}^2\mu_{rf}^2},\as(\mu_{rf}) \right) \nn\\[2mm]
& \times & {\mathrm{Tr}} \left\{ H \left ( \as(\mu_{rf}),\frac{\sqrt{\hat{s}}}{\mu_{rf}} ,\frac{\mu_F}{\mu_{rf}},  \hat\eta  \right)\, 
\tilde S\left(\frac{ \hat {s}}{\bar{N}^2 \mu_{rf}^2},\as(\mu_{rf}), \hat\eta \right)   \right\}_{ab \to cr} \nn \\[2mm]
\nn\\[2mm]
&+& {\cal O}\left( \frac{1}{N}\right)\, .
\label{eq:omega-fact}
\eea
The power of $N$ in each transformed function is determined by whether the corresponding weight $w_i$ appears with 
$\hat{s}$ or $\sqrt{\hat{s}}$ in the momentum-space refactorized cross section, Eq.~(\ref{threshconv}).
On the left-hand side, we have suppressed the argument describing the factorization scale dependence for simplicity. 
For the initial-state jets the moment variables $N_i$ are shifted by simple kinematic factors from the value $N$ in the definition 
of the moment:
\bea
N_a\ &=&\ N\, \left(\frac{-\hat u}{\hat s}\right)\, ,
\nn\\[2mm]
N_b\ &=&\ N\, \left(\frac{- \hat t}{\hat s}\right)\, ,
\label{eq:Ni-defs}
\eea
which, as noted above, then depend on the center of mass rapidity. Here and below, it is convenient to use the common notation
\bea
\bar N\ \equiv \ N{\mathrm{e}}^{\gamma_E}\ =\ \exp\left[-\; \int_0^1 dz\, \frac{z^N-1}{1-z}\right ]\ +\ {\cal O}\left( \frac{1}{N}\right)  \, .
\eea
By analogy to the methods employed in Refs.\ \cite{Almeida:2009jt} and \cite{Hinderer:2014qta} for dihadron production, we  treat the hard-scattering function in Mellin moment space as an intermediate step.  Once resummed in Mellin space, we can invert its combination with the fragmentation function to momentum space, as in Eq. (\ref{eq:Omega-inverse}). 
To derive the single-particle inclusive cross section, we may then do the resulting integration over $x_a$ and $x_b$ with parton distributions, as in Eq.\ (\ref{taufac}).   We will return to this and other possible prescriptions in Sec.\ \ref{sec:zspace}.

We next review the resummed forms that we will use for the functions appearing in the moment-space hard scattering function (\ref{eq:omega-fact}), starting with the jet functions. We note that the hard functions $H_{ab\to cr}$ in Eq.~(\ref{eq:omega-fact})
were already given to one-loop for all partonic subprocesses in Ref. \cite{Kelley:2010fn} and in our previous paper~\cite{Hinderer:2014qta}
on dihadron production, and we do not present them again here. 
The same functions appear in single-inclusive production. 
Very recently, also the two-loop corrections to the $H_{ab\to cr}$ have become
available~\cite{Broggio:2014hoa}, and we will make use of this information in our exploratory phenomenological results. 

\subsection{Evolution in the refactorized cross section}

As usual, the functions appearing in refactorized hard scattering cross sections satisfy evolution equations, which control dependence on logarithms of the moment variable.  
The logarithmic corrections in the moment variable $N$ in each of the functions in the refactorized inclusive hard scattering function, Eq.\ (\ref{eq:omega-fact}) can be resummed at leading power in $N$ \cite{Sterman:1986aj,Catani:1989ne}.   Such resummations have been carried out for various cross sections and in different notations and normalizations, and we shall not try to present a full history here.   
For our purposes, it is convenient to observe that when the refactorization and factorization scales are taken as equal, each of our jet functions satisfies a specific evolution equation. 
 We will give here the relevant evolution equations for the ``in'' and ``fragmentation'' jets (the same equation for all three), for the recoil jet and then the soft function.

First, consider the ``in", and ``fragmentation" jets, where for consistency of notation we keep an argument $1=\mu_F/\mu_{rf}$.   The equation is of a standard form, with $\gamma_{J^{(i)}}$ the ``Drell-Yan (DY)" anomalous dimension \cite{Belitsky:1998tc},
\bea
\mu_{rf} \frac{d}{d\mu_{rf}}\, 
\ln \tilde{J}_{\rm in,\, fr}^{(i)} \left( \frac{\hat s}{\bar{N}^2\mu_{rf}^2}, 1,\as(\mu_{rf}) \right)\ =\ 
 A_i\big(\as(\mu_{rf})\big)  \ln\left( \frac{\bar N^2\mu_{rf}^2}{\hat s}\right) - \gamma_{J^{(i)}} \big(\as(\mu_{rf})\big)\, .
\label{eq:F-evolve}
\eea
When $\mu_F$ varies at fixed $\mu_{rf}$ these same functions satisfy an equation that generates the ``cusp" part of collinear evolution,
\bea
\mu_F\frac{\partial }{\partial \mu_F}\, \ln \tilde{J}_{\rm in,\, fr}^{(i)} \left( \frac{\hat s}{\bar{N}^2\mu_{rf}^2}, \frac{\mu_F}{\mu_{rf}},\as(\mu_{rf}) \right)
\bigg|_{\mu_{rf}}\ =\ 
\ln \bar N^2\, A_i\big(\as(\mu_F)\big) \, .
\label{eq:j-in-out-muF}
\eea
We show how to satisfy this pair of equations in the following section on jet functions.   We already see here, however, that the $\ln N$-dependence in the first of these equations is due entirely to the change in the factorization scale.  

Next, for the recoil jet, we will show in the following section that this function depends on two ratios involving the moment variable $N$.
The result that we use here is that its dependence on $\mu=\mu_{rf}$ is independent of $N$, although the function itself still depends on $N$,
\bea
\mu_{rf}\frac{d}{d\mu_{rf}}\, \ln \tilde{J}_{\rm rec}^{(r)} \left( \frac{\hat s}{\bar{N}\mu_{rf}^2},\frac{\hat s}{\bar{N}^2\mu_{rf}^2},
\as(\mu) \right) &=&
A_r\big(\as(\mu_{rf})\big)  \ln\left( \frac{\mu_{rf}^2}{\hat s}\right) \ -\ \gamma_{J_{\rm rec}^{(r)}} \big(\as(\mu_{rf})\big) \, ,
\label{eq:rec-evolve}
\eea
with $r$ the flavor of the parton that initiates the recoil jet.   We will find the single-log anomalous dimension $\gamma_{J^{(r)}_{\rm rec}}$ below.   
Note that the recoil jet is independent of the factorization scale $\mu_F$ since it does not involve any
parton distributions or fragmentation functions.

The soft function satisfies a familiar equation, with a matrix $\Gamma$ 
of anomalous dimensions and without an explicit logarithmic term, \cite{KS}
\bea
\mu\frac{d}{d\mu}\, \tilde S\left(\frac{ \hat {s}}{\bar{N}^2 \mu^2},\as(\mu), \hat\eta \right)
\ &=&\
-\; \Gamma^\dagger \left( \as(\mu), \hat\eta \right)\ \tilde S\left(\frac{ \hat {s}}{\bar{N}^2 \mu^2},\as(\mu), \hat\eta \right)
\nn\\[2mm]
&\ & \ -\; 
\tilde S\left(\frac{ \hat {s}}{\bar{N}^2 \mu^2},\as(\mu), \hat\eta \right)\ \Gamma\left( \as(\mu), \hat\eta \right)\, .
\label{eq:S-evolve}
\eea
The soft function knows nothing about the factorization scale, and the scale $\mu$ here is again the refactorization/renormalization scale.

Given the above, we can  derive equations for the hard functions $H$.   First, we demand that the short-distance inclusive
hard-scattering function be independent of the refactorization scale $\mu_{rf}$ at fixed factorization scale, $\mu_F$, 
\bea
\mu_{rf}\frac{d}{d\mu_{rf}}\,  \tilde \omega^{ab\to c} \left (\hat \eta,N \right )\, \big|_{\mu_F} \ =\ 0\ =\
\mu_{rf}\frac{d}{d\mu_{rf}}\,  \tilde \omega^{ab\to c} \left (\hat \eta,N \right )\, \big|_{\mu_F=\mu_{rf}}\ -\ \mu_F\frac{d}{d\mu_F}\,  \tilde 
\omega^{ab\to c} \left (\hat \eta,N \right )\, \big|_{\mu_{rf}}\, .
\label{eq:muR-fixed-muF}
\eea
Second, at fixed $\mu_{rf}$,  the function must obey a standard evolution in $\mu_F$, which at leading power in $N$ we can write as
\bea
\mu_F\frac{d}{d\mu_F}\, \ln \omega^{ab\to c} \left (\hat \eta,N \right )\, \big|_{\mu_{rf}} \ =\  \sum_{i=a,b,c} 
\left[ \ln \bar{N}^2\, A_i(\as(\mu_F))  \ -\  P_{i,\delta}(\as(\mu_F)) \right ]\, .
\label{eq:muF-fixed-muR}
\eea
Here the terms $P_{i,\delta}$ are the coefficients of $\delta(1-x)$ in the diagonal evolution kernels for parton $i$.

The evolution equations for the hard function now follow from the $\mu_{rf}$-independence of $\omega^{ab\to c}$, Eq.\ (\ref{eq:muR-fixed-muF}), combined with the evolution of the jet functions and soft matrix, Eqs.\  (\ref{eq:F-evolve}), (\ref{eq:rec-evolve}) and (\ref{eq:S-evolve}).   In particular, $\ln N$ dependence cancels in the derivative of the hard function $H$ with respect to the refactorization scale at fixed factorization scale, as it must, since the hard scattering function is purely virtual at partonic threshold.   This
leaves for the $\mu_{rf}$ evolution equation for $H$,
\bea
\mu_{rf}\frac{d}{d\mu_{rf}}\,H \left ( \as(\mu_{rf}),\frac{\sqrt{\hat{s}}}{\mu_{rf}} ,\frac{\mu_F}{\mu_{rf}},  \hat\eta  \right)\, \bigg|_{\mu_F} \ &=&\ 
-\ \sum_{i=a,b,c,r}\, \big [\, A_i\big(\as(\mu_{rf})\big)  \ln\left( \frac{\mu_{rf}^2}{\hat s}\right)\  
-\  \gamma_{J^{(i)}} \big(\as(\mu_{rf})\big)\,  \big ]\ 
\nn\\[2mm]
&\ & \hspace{20mm} \times \ H \left ( \as(\mu_{rf}),\frac{\sqrt{\hat{s}}}{\mu_{rf}} ,\frac{\mu_F}{\mu_{rf}},  \hat\eta  \right)  
\nn\\[2mm]
&\ &\ +\ \Gamma \left( \as(\mu_{rf}), \hat\eta \right) H \left ( \as(\mu_{rf}),\frac{\sqrt{\hat{s}}}{\mu_{rf}} ,\frac{\mu_F}{\mu_{rf}},  \hat\eta  \right) 
\nn\\[2mm]
&\ &\ +\
H \left ( \as(\mu_{rf}),\frac{\sqrt{\hat{s}}}{\mu_{rf}} ,\frac{\mu_F}{\mu_{rf}},  \hat\eta  \right) \, \Gamma^\dagger \left( \as(\mu_{rf}), \hat\eta \right)\, ,
\eea
where we have dropped the subscript $ab\to cr$ labeling the process.
To find $\mu_{rf}$-dependence in fixed order computations of $H$, an equivalent evolution equation is
\bea
\mu_{rf}\frac{\partial }{\partial \mu_{rf}}\,H \left ( \as(\mu_{rf}),\frac{\sqrt{\hat{s}}}{\mu_{rf}} ,\frac{\mu_F}{\mu_{rf}},  \hat\eta  \right)\, \bigg|_{\mu_F,\alpha_s} \ &=&\ 
-\ \sum_{i=a,b,c,r}\, \big [\, A_i\big(\as(\mu_{rf})\big)  \ln\left( \frac{\mu_{rf}^2}{\hat s}\right)\  
-  \gamma_{J^{(i)}}\big(\as(\mu_{rf})\big)\,  \big ]\ 
\nn\\[2mm]
&\ & \hspace{20mm} \times \ H \left ( \as(\mu_{rf}),\frac{\sqrt{\hat{s}}}{\mu_{rf}} ,\frac{\mu_F}{\mu_{rf}},  \hat\eta  \right)  
\nn\\[2mm]
&-&\ \as(\mu_{rf})\,\beta\left(\alpha_s(\mu_{rf})\right)\, \frac{\partial }{\partial \as(\mu_{rf})}  H \left ( \as(\mu_{rf}),\frac{\sqrt{\hat{s}}}{\mu_{rf}} ,\frac{\mu_F}{\mu_{rf}},  \hat\eta  \right)  
\nn\\[2mm]
&+&\ \Gamma \left( \as(\mu_{rf}), \hat\eta \right) H \left ( \as(\mu_{rf}),\frac{\sqrt{\hat{s}}}{\mu_{rf}} ,\frac{\mu_F}{\mu_{rf}},  \hat\eta  \right) 
\nn\\[2mm]
&+&\ H \left ( \as(\mu_{rf}),\frac{\sqrt{\hat{s}}}{\mu_{rf}} ,\frac{\mu_F}{\mu_{rf}},  \hat\eta  \right) \, \Gamma^\dagger \left( \as(\mu_{rf}), \hat\eta \right)\, .
\label{eq:H-evolve-2}
\eea
Here we have  exhibited the $\mu_{rf}$ dependence of the coupling through 
\beq
\beta(\as) \equiv \frac{1}{\as} \frac{d\as}{d\ln(\mu^2)}=-\beta_0\frac{\alpha_s}{4\pi}
+{\cal O}(\alpha_s^2)\, ,
\label{eq:beta-def}
\eeq
with $\beta_0=(11C_A-2N_f)/3$. 
Meanwhile the dependence of the hard function on the factorization scale is determined by
\be
\mu_F\frac{d}{d\mu_F}\,H \left ( \as(\mu_{rf}),\frac{\sqrt{\hat{s}}}{\mu_{rf}} ,\frac{\mu_F}{\mu_{rf}},  \hat\eta  \right) \, \bigg|_{\mu_{rf}}\ =\ 
-\sum_{i=a,b,c} P_{i,\delta}(\as(\mu_F))\, H \left ( \as(\mu_{rf}),\frac{\sqrt{\hat{s}}}{\mu_{rf}} ,\frac{\mu_F}{\mu_{rf}},  \hat\eta  \right) \, ,
\label{eq:H-evovle}
\ee
so that 
\be\label{HmuF}
H \left ( \as(\mu_{rf}),\frac{\sqrt{\hat{s}}}{\mu_{rf}} ,\frac{\mu_F}{\mu_{rf}},  \hat\eta  \right)\,=\,
H \left ( \as(\mu_{rf}),\frac{\sqrt{\hat{s}}}{\mu_{rf}} ,1,  \hat\eta  \right)\,
\exp\left[\,-\sum_{i=a,b,c}\int^{\mu_F}_{\mu_{rf}}\f{d\mu}{\mu} \, P_{i,\delta}\big(\as(\mu_F)\big) \,\right]\,.
\ee
That is, we must associate the part of evolution due to parton self-energies with the hard function.   Taken together with the purely eikonal evolution equations for the incoming and fragmenting jets, (\ref{eq:j-in-out-muF}), this ensures that  the full $\mu_F$ dependence of the complete perturbative function, $\omega^{ab\to c}$, is given by Eq.\ (\ref{eq:muF-fixed-muR}). 

The relationship of this kind between factorization and evolution equations was explored in Ref.\ \cite{Contopanagos:1996nh}, and equations of precisely this form are a familiar feature of factorized jet and soft functions in direct analyses in QCD \cite{Belitsky:1998tc,Sterman:2013nya} and in soft-collinear effective theory (SCET) \cite{Becher:2009th,Becher:2007ty}.   Technical comparisons of the formalisms were given in 
\cite{Sterman:2013nya}, \cite{Lee:2006nr}, \cite{Bonvini:2012az}, and \cite{Almeida:2014uva}.    Solutions to Eqs.\ (\ref{eq:F-evolve}) - (\ref{eq:S-evolve}) will provide our basic results, including factorization scale dependence where needed.
  
 The solutions to Eqs.\ (\ref{eq:F-evolve}) -- (\ref{eq:S-evolve}) that we will use below relate the jet and soft functions evaluated at a ``hard" scale, $\mu_h$, of order $\hat s$ to its value at a ``soft" scale, $\mu_s$, which can, but need not, be chosen to control $\bar N$ dependence.
The hard scale is normally chosen at a value that characterizes the hard scattering.    Whatever the choices of $\mu_h$ and $\mu_s$ introduced in moment space, it is possible to invert the transform, so long as $\mu_s$ is large enough that the integral over $\mu$ avoids any singularities associated with the running coupling, the ``Landau poles".   In SCET the conventional choice for the lower limit is a fixed scale, $\mu_{\rm jet}$ or $\mu_{\rm soft}$, chosen to best approximate the combination of the hard-scattering functions with parton distributions,
although other choices are possible; see Ref.\ \cite{Czakon:2018nun}. 
The conventional choice for many direct QCD analyses has been, for example, $\mu_s=\sqrt{\hat s}/\bar{N}$ for the in and fragmentation jets, which  resums all logarithms of $\bar{N}$ in moment space, to any fixed order in the coupling.  The integral now captures all logarithms of $\bar{N}$, but has a branch cut beginning at $\bar{N}=(\sqrt{\hat{s}}/\Lambda_{\rm QCD})$ in the complex $N$ plane.    Nevertheless, we can use a minimal procedure to invert the transform \cite{Catani:1996yz}, by evaluating a contour in moment space that is to the left of the branch cut.   This procedure keeps all the logs, at the cost of unphysical, but power suppressed,
contributions~\cite{Abbate:2007qv}. The anomalous dimensions, of course, do not depend on this solution to the evolution equation for our jet and soft functions.   We give the application
of this method to incoming jets first, closely following the discussion of Ref.\ \cite{Hinderer:2014qta}, and then turn to recoil and fragmentation jets, which require a different treatment.   For definiteness, we provide solutions that organize all $\ln \bar N$ dependence, and turn later to a discussion of other choices of hard and soft scales.

\section{Partonic Jet Functions for Single-Particle Cross Sections}
\label{sec:resum}
      
      \subsection{Initial-state jets\label{sec:inijets}}

The initial-state jets, $\tilde{J}_{\mathrm{in}}^{(a)}$ and $\tilde{J}_{\mathrm{in}}^{(b)}$ 
in Eq.\ (\ref{eq:omega-fact}) are normalized in terms of threshold resummation for the Drell-Yan cross section, which involves only initial-state jets at partonic threshold. When taken at a common Mellin moment $N$, the squares of the $\tilde{J}_{\rm in}^{(i)}$, $i=a,b$  give the threshold-resummed  Drell-Yan and Higgs production hard-scattering functions to leading power in $N$.   
We define them as a function of a single scale, again labeled $\mu_{rf}$ here, as
\bea
\tilde{J}_{\rm in}^{(i)} \left( \frac {\hat s}{\bar N_i^2 \mu_{rf}^2},1, \as(\mu_{rf}) \right)\ =\ \left[\, W^{(i)}_{\rm DY}\left( \frac {\hat s}{\bar N_i^2 \mu_{rf}^2}, \as(\mu_{rf}) \right)\, \right]^{1/2}\, ,
\label{eq:J-sqrtW}
\eea
in terms of $W^{(i)}_{\rm DY}$, the vacuum expectation value of the ``cusp" color singlet product of Wilson lines in the color representation of parton $i$.   For incoming jets, the moment variable $N_i$ is defined in Eq.\ (\ref{eq:Ni-defs}).
Equating the factorization and refactorization scales, $\mu_F=\mu_{rf}$, this function obeys the renormalization group 
equation Eq.\ (\ref{eq:F-evolve}). Its evolution 
has been discussed widely \cite{Belitsky:1998tc,Sterman:2013nya,Becher:2007ty,Becher:2009th}.   

   To be explicit, the single-log anomalous dimension in Eq.\ (\ref{eq:F-evolve}) is given by~\cite{Belitsky:1998tc}
\bea\label{eq:Dgmma-J-del}
\gamma_{J^{(i)}}(\as)
&=&\ \frac{1}{2}\; \left(\frac{\as}{\pi}\right)^2\, C_i\, \left[C_A\left(-\f{101}{27}+\f{7}{2}\,\zeta(3)\right) + \f{14}{27}\,N_f   + \frac{1}{4}\, \beta_0 \zeta(2) \right]\,+\,
{\cal O}(\alpha_s^3)\, .
\label{eq:gamW-def}
\eea
Because of (\ref{eq:J-sqrtW}), this is one-half of the full anomalous dimension associated with the function $W_{\rm DY}^{(i)}$.

The relevant solution of Eq.\ (\ref{eq:F-evolve}) relates the jet function at infrared and ultraviolet scales, $\mu_{rf}=\mu_2$ and $\mu_{rf}=\mu_1$, respectively,
\begin{eqnarray}
\tilde J_{\rm in}^{(i)}  \left( \frac {\hat s}{\bar N_i^2 \mu^2_1},1,\as(\mu_1)\right) &=& 
\tilde J_{\rm in}^{(i)}  \left(  \frac {\hat s}{\bar N_i^2 \mu^2_2},1,\as(\mu_2)\right)
\nn\\[2mm]
&\times&
\exp \left[ \int_{\mu_2}^{\mu_1} \frac{d\mu}{\mu}\ \left(A_i\big(\as(\mu)\big)  
\, \ln\left( \frac{\mu^2 \bar N_i^2}{\hat{s}}\right)\, -\, \gamma_{J^{(i)}}  
\big(\as(\mu)\big) \right) \right ]\, .
\label{eq:soln-gen}
\end{eqnarray}
We will return to the use of fixed scales below, but here we develop the resummed expressions of direct QCD, in which the infrared scale is chosen to generate all logarithmic dependence in the moment variable, $N$.   We carry out the resummation for $\mu_F=\mu_{rf}$, choosing $\mu_2=\sqrt{\hat s}/\bar N_i$ as the infrared scale and leaving $\mu_1=\mu_{rf}\sim \sqrt{\hat s}$ as the ultraviolet scale in (\ref{eq:soln-gen}).   The solution of Eq.\ (\ref{eq:soln-gen}) then becomes
\bea
&&\hspace*{-1.5cm} \tilde{J}_{\rm in}^{(i)} \left ( \frac {\hat s}{\bar N_i^2\mu_{rf}^2 },1,\as(\mu_{rf}) \right) 
\nn\\[2mm]
\hspace*{-0.25cm} 
&=& \tilde J_{\rm in}^{(i)}  \left(  1,1,\as(\sqrt{\hat s}/\bar N_i)\right)\, 
\exp \left[  \int^{\mu_{rf}}_{\sqrt{\hat s}/\bar{N}_i} 
\frac{d\mu}{\mu} \left(A_i\big(\as(\mu)\big) \, \ln\left( \frac{\mu^2\bar N_i^2}{\hat{s}}
\right) - \gamma_{J^{(i)}} \big(\as(\mu)\big) \right)\right ]\, .
\label{eq:J-in-soln1-pre-muF}
\eea
We then use Eq.\ (\ref{eq:j-in-out-muF}) to reintroduce an independent factorization-scale dependence by extending the integration limit for the integral over the ``cusp" anomalous dimension $A_i$,
\bea
&&\hspace*{-1.7cm} \tilde{J}_{\rm in}^{(i)} \left ( \frac {\hat s}{\bar N_i^2\mu_{rf}^2 },\frac{\mu_F}{\mu_{rf}},\as(\mu_{rf}) \right)
\,=\, \tilde J_{\rm in}^{(i)}  \left(1,  1,\as(\sqrt{\hat s}/\bar N_i)\right)\,  \nn\\[2mm]
\hspace*{0.5cm} 
&&\hspace*{-13mm}\times\,\exp \left[ \int^{\mu_{rf}}_{\sqrt{\hat s}/\bar{N}_i} 
\frac{d\mu}{\mu} \left(A_i\big(\as(\mu)\big)  \ln\left( \frac{\mu^2\bar N_i^2}{\hat{s}}
\right) - \gamma_{J^{(i)}} \big(\as(\mu)\big) \right)+ \ln({\bar N}^2_i)\int^{\mu_F}_{\mu_{rf}} 
\frac{d\mu}{\mu}  \, A_i \big(\as(\mu)\big) \right ] .
\label{eq:J-in-soln1}
\eea
In this form, the prefactor still generates logarithms of $N$ through the running coupling, but we can promote this dependence to the exponent \cite{Sterman:2013nya}, thereby generating all NNLL logarithms from the resulting anomalous dimension.   First, we introduce the notation,
\bea
\widehat R_i\big(\as\big)\ &\equiv&\ \tilde J_{\rm in}^{(i)}  \left(  1,1,\as \right)
\nn\\ [2mm]
&=&\ 1+ \frac{\as}{4\pi} A_i^{(1)}\zeta(2)+{\cal O}(\alpha_s^2)\, ,
\label{eq:R-hat}
\eea
where the explicit value given in the second line can be found in \cite{Hinderer:2014qta,Belitsky:1998tc}, for example.
For the expansions of $A_i$ and all other functions, we will use the conventions
\beq
A_i(\alpha_s)\,=\,\frac{\alpha_s}{\pi}\,A_i^{(1)} + \left(\frac{\alpha_s}{\pi}\right)^2 A_i^{(2)}+{\cal O}(\alpha_s^3)\, .
\eeq
We now note that
\bea
\frac{\widehat R_i\big(\as(\sqrt{\hat s}/\bar N)\big)}{\widehat R_i\big(\as(\mu_{rf})\big)}\ = \ \exp\left[ \, -\ \int_{\sqrt{\hat s}/\bar N}^{\mu_{rf}}\, \frac{d\mu}{\mu}  \times  \mu \frac{\partial}{\partial \mu} \ln \widehat R_i\big(\as(\mu)\big)\right ]\, .
\label{eq:Rrunning}
\eea
This enables us to write the resummed in-jet, Eq.\ (\ref{eq:J-in-soln1}) in a form where all $N$ dependence is generated in the exponent,
\bea
\tilde{J}_{\rm in}^{(i)} \left ( \frac {\hat s}{\bar N_i^2\mu_{rf}^2 },\frac{\mu_F}{\mu_{rf}},\as(\mu_{rf}) \right) 
&=&\widehat R_i\big(\as(\mu_{rf})\big)\, \exp \left[ \int^{\mu_{rf}}_{\sqrt{\hat s}/\bar{N}_i} 
\frac{d\mu}{\mu} \left(A_i\big(\as(\mu)\big)  \ln\left( \frac{\mu^2\bar N_i^2}{\hat{s}}
\right) - \frac{1}{2}\,\hat D_i \big(\as(\mu)\big) \right)\right.\nn\\[2mm]
&+& \left. \ln(\bar{N}_i)\int^{\mu_F^2}_{\mu_{rf}^2} 
\frac{d\mu{}^2}{\mu{}^2}  \, A_i \big(\as(\mu)\big) \right ]\, .
\label{eq:J-in-soln2}
\eea
Compared to (\ref{eq:J-in-soln1}), the coupling in the prefactor is evaluated at scale $\mu_{rf}\sim\sqrt{\hat s}$, and we introduce a new function, $\hat D_i$ defined by
\bea
\hat D_i\big(\as(\mu)\big) &=& 2 \gamma_{J^{(i)}}\big(\as(\mu)\big)   + 4\as(\mu)\, \beta\big(\as(\mu)\big) 
\,\frac{d}{d\as}\ln \widehat R_i\big(\as(\mu)\big)\, ,
\label{eq:hatD-def} 
\eea
in terms of the QCD $\beta$-function, Eq.\ (\ref{eq:beta-def}).
 The function $\hat D_i$, which differs from $\gamma_{J^{(i)}}$ in Eq.\ (\ref{eq:gamW-def}) by the term proportional to $\beta_0$, also begins at order at $\as^2$,
\bea\label{eq:Dhat2}
\hat D_i\big(\as(\mu)\big)
&=&\left(\frac{\as}{\pi}\right)^2\,C_i\,
\left[C_A\left(-\f{101}{27}+\f{7}{2}\,\zeta(3)\right) + \f{14}{27}\,N_f   \right]\,+\,
{\cal O}(\alpha_s^3)\, .
\eea
Interestingly, the $\zeta(2)$ term in $\gamma_{J^{(i)}}$ is fully cancelled by exponentiating the $N$-dependence of the jet prefactor.
Appendix~\ref{AppB} collects the explicit low-order expansions of 
all other anomalous dimensions needed for the NNLL resummed jet functions $\tilde{J}_{\rm in}^{(i)}$.
Closed expressions for NNLL expansions of the integrals in Eq.~(\ref{eq:J-in-soln2}) are given in Appendix~\ref{AppC}.

For completeness, we go on to link the solution, Eq.\ (\ref{eq:J-in-soln2}) to another standard expression, with an explicit Mellin transform in the exponent.  This form is particularly natural when there is only a single hard scale, as in threshold resummation for electroweak annihilation, or for single-particle inclusive cross sections in electron-positron annihilation.
 Details of the transformation to all logarithmic accuracy were given in \cite{Catani:2003zt} and reviewed in \cite{Sterman:2013nya} and elsewhere.   For this discussion, we identify $\mu_{rf}=\sqrt{\hat s}$.
We then define two shifted $R$ and $D$ functions,
\bea
R_i(\as)\ =\  \widehat R_i(\as)\ - \frac{\as}{\pi} A_i^{(1)}\zeta(2)\ =\ 1\ -\ \frac{3\as}{4\pi} A_i^{(1)}\zeta(2)+{\cal O}(\alpha_s^2)
\label{eq:R-def}
\eea
and
\bea\label{eq:Drel}
D_i(\as)&=&\hat D_i(\as)\  +\left(\frac{\as}{\pi}\right)^2\zeta(2)\,\beta_0\, A_i^{(1)} \, .
\label{eq:D-def}
\eea
In these terms, and staying at the level of NNLL, we have 
\bea
&&\hspace*{-12mm}\tilde{J}_{\rm in}^{(i)} \left (\frac {1}{\bar N_i^2 },\frac{\mu_F}{\sqrt{\hat s}}, \as(\sqrt{\hat s}) \right)  \nn\\[2mm]
&&\hspace*{-10mm}=\, R_i\big(\as(\sqrt{\hat s})\big) \exp \Bigg \{ \int_0^1dy\ \frac{y^{N_i-1}-1}{1-y} \left[ \int_{\mu_F^2}^{(1-y)^2 {\hat s}} \f{d\mu 
{}^2}{\mu {}^2} \,
A_i \big(\as(\mu)\big)+\frac{1}{2}\,D_i\big(\as((1-y)\sqrt{\hat s})\big)\right]\Bigg\} .
\label{eq:in-jet}
\eea
This expression, with a moment integral in the exponent, is discussed in detail in Ref.\ \cite{Catani:2003zt} for Higgs production (see particularly 
appendix\ A there); the only difference here is in the factor $R_i(\as)$, which we normalize to Drell-Yan cross sections.    

\subsection{Recoil and fragmentation jets}
\label{sec:rec-frag-jets}

The recoil jet and fragmentation jet functions in the factorized expressions Eq.\ (\ref{threshconv}) or (\ref{eq:omega-fact}) can be extracted  from the singular $z\to 1$ behavior of single-inclusive cross sections in electroweak annihilation.    We will describe this procedure, giving some details of how the refactorization that leads to (\ref{threshconv}) and (\ref{eq:omega-fact}) can be carried out.

The cross sections in electroweak annihilation, ${\rm e^+e^-}\to hX$, 
can be put in factorized form starting from Eq.\ (\ref{eq:taufac-rewrite}), simply replacing the parton distributions by two delta functions, $\delta(1-x_{a,b})$, giving,
\bea
p_T^3\,\frac{d^2\sigma^{{\rm e^+e^-}\to hX}}{dp_Td\eta}\,&=&\, \sum_{c}\
\int_z^1 {dz_c}\, {z^2_c} \,D_c^h(z_c,\mu_F)\,\omega^{{\rm e^+e^-}\to c}\left(\hat\eta,\frac{z}{z_c},\frac{\mu_F^2}{\hat s}\right)\, .
\label{eq:epem-fac}
\eea
For ease of comparison to the general form in Eq.\ (\ref{threshconv}), we have kept the rapidity dependence relative to the beam direction.  
For $\rm e^+e^-$ annihilation, $z$, $\hat\eta=\eta$ and $x_T$ are related by  $z=x_T\cosh\hat \eta$ and $z_c=x_T/\hat x_T$.    Equation (\ref{eq:epem-fac}) applies  to cross sections for both hadrons and partons, and to analyze the partonic hard scattering we consider the cross section for an observed parton $c$ in the (dimensionally) infrared-regulated theory. 

After subtraction of collinear poles from radiated and virtual partons in the fragmentation direction, the inclusive hard-scattering function 
$\omega^{{\rm e^+e^-} \to c}$ in (\ref{eq:epem-fac}) is infrared finite for $z_c>z$, but with singularities in the limit $z/z_c\to 1$.    It can be a function only of the remaining 
invariant mass of the radiated state, which is given by $\hat{s}(1-z/z_c)$ (note that for ${\rm e^+e^-}$ collisions $\hat{s}=s$).  The hard-scattering function for this cross section then (re-)factorizes at partonic threshold into a short-distance function and two jets, one in the direction of the observed parton and the other of its recoiling partner, with additional wide-angle radiation.    As we shall see, for this particular case we will not need a soft function \cite{Moch:2009my}.

As an intermediate step in bringing out this structure, we refactorize the inclusive hard-scattering function $\omega^{{\rm e^+e^-}\to c}$ 
into a true short-distance non-radiative factor, denoted $H^{{\rm e^+e^-}\to c}$ and the function that contains all soft and collinear radiation, denoted $\Sigma_{\bar c}$,
\be
\label{eq:Sigma-def1}
\omega^{{\rm e^+e^-}\to c} \left (\hat \eta,y,\frac{\mu_F^2}{\hat{s}}\right) \,=\,
H^{{\rm e^+e^-}\to c}\left (\as(\mu_{rf}) , \frac{\sqrt{\hat s}}{\mu_{rf}}, \frac{\mu_F}{\mu_{rf}},\hat \eta\right) 
\  \Sigma_{\bar c} \left (\bar y, \frac{\bar y\hat s}{\mu_{rf}^2},\frac{\mu_F}{\mu_{rf}},\as(\mu_{rf}) \right) \, ,
\ee
where $\bar{y}=1-y$ with $y=1-\hat s_4/\hat s=z/z_c$ (see Eq.~(\ref{eq:z-def})).  This is analogous to Eq.\ (\ref{threshconv}), but without separation of jets in the final state.   As usual, the function $\Sigma_{\bar c}$ will itself require renormalization, so that an additional ``refactorization scale'', labelled $\mu_{rf}$ as above will appear in both it and in the short distance ``hard part", $H^{{\rm e^+e^-}\to c}$.     All $y$-dependence, associated with the final states, is in the function $\Sigma_{\bar c}$, and at leading power in $1/\bar y$, the function $H^{{\rm e^+e^-}\to c}$ has contributions only from virtual corrections.  
The $\bar y$ dependence in $\Sigma_{\bar c}$ links its evolution to that of the fragmentation function $D_c^h(z_c,\mu_F)$ in Eq.\ (\ref{eq:epem-fac}).  At the same time, $H^{{\rm e^+e^-}\to c}$ has a residual dependence on the factorization scale $\mu_F$, which will be determined as in the general case, Eq.\ (\ref{eq:H-evovle}).   
We are going to give a further factorization of the function $\Sigma_{\bar c}$ to define normalizations for the recoil and fragmentation jets introduced above.   

The virtual hard function $H^{{\rm e^+e^-}\to c}$ in (\ref{eq:Sigma-def1}) is normalized so 
that at lowest order the final-state function $\Sigma_{\bar{c}}$ is a simple delta function setting $y$ to one,
\bea
\omega^{{\rm e^+e^-}\to c, (0)}(\hat \eta,y)\ =\ H^{{\rm e^+e^-}\rightarrow c, (0)}(\hat \eta)\; \delta(\bar y)\, .
\eea
At higher orders, the function $\Sigma_{\bar c}$ contains all long-distance dynamics, associated with soft gluons and collinear enhancements in the recoil and fragmentation directions.  In the threshold limit, all radiation is forced to be soft, except for particles emitted in the recoil jet direction.  In electroweak annihilation, the flavor associated with this jet is naturally $\bar c$, as indicated by the subscript on $\Sigma$.   The singular, long-distance behavior in $\omega^{{\rm e^+e^-}\to c}$ near partonic threshold is given in terms of a sum over states consisting of all radiation except for the observed parton, as produced by the field of the other high energy parton, which initiates the recoil jet. This field, which we will denote by
$\bar{\phi}_r$, is joined locally to a lightlike Wilson line that extends to infinity in the direction opposite to the recoiling jet.
  
To construct  the function $\Sigma_{\bar c}$ in a manner that reproduces all singular behavior, we sum over a phase space that matches the phase space of the 1PI cross section near partonic threshold, including all singular regions, and is weighted by squared amplitudes that generate all soft and collinear enhancements.   For this analysis, we  take equal scales, $\mu_F=\mu_{rf} \sim \sqrt{\hat s}$.    As for the initial-state jets, independent factorization scale dependence will be reintroduced in the cross section by standard evolution equations.  For arbitrary flavor, $\bar c\equiv r$, the function $\Sigma_r$  is now given in terms of a single scale by  \cite{Becher:2006qw,Berger:2003iw}
\bea
\Sigma_r  \left ( \bar y, \frac{\bar y\hat s}{\mu_{rf}^2},1,\as(\mu_{rf})  \right)\ 
&=& {\cal N}_0\;  \sum_{\ket{\xi}}\  \delta \left (1 - \frac{p^2_\xi}{\bar y\hat s} \right) \nn\\[2mm]
&\times&
  {\rm Tr}_{\{j\}} \, \langle 0|\, \bar T \left( \Phi_{\bar \beta_r}^{(\bar r)\, \dagger} \phi_r(0)_{\{j\}} \right )
|\xi\rangle \, \langle \xi | \,T \left(  \Phi_{\bar \beta_r}^{(\bar r)} \bar \phi_r(0)_{\{j\}} \right) |0\rangle\, ,
\label{eq:Sigma-def}
\eea
with the normalization factor ${\cal N}_0$ chosen to set $\Sigma_r$ to $\delta_{c\bar r}\delta(\bar y)$ at zeroth order.
In (\ref{eq:Sigma-def}), 
$\bar \phi_r$ is the partonic field that produces the recoil jet; the trace is over color indices and spin. 
The familiar Wilson lines $\Phi_{\beta}^{(f)}$ are 
\beq
\Phi_{\beta}^{(f)}\ = \ {\cal P}\exp\left(-ig\int_0^\infty d{\eta}\; {\beta}{\cdot} A^{(f)} ({\eta}{\beta})\right)\, .
\label{eq:wilson} 
\eeq
As defined in (\ref{eq:Sigma-def}), the renormalization of $\Sigma_r$ includes the removal of virtual collinear singularities associated with the presence of the Wilson line, in addition to the collinear singularties resulting from real radiation.   This renormalization is equivalent to the subtraction of counterterms defined by the perturbative fragmentation function in (\ref{eq:epem-fac}).   

The function $\Sigma_r$ in (\ref{eq:Sigma-def}) is the imaginary part of a renormalized ``jet function", as defined in Ref.\  \cite{Becher:2006qw} with a light-like Wilson line (and in Ref.\ \cite{Berger:2003iw} with a Wilson line off the light cone).   Because only collinear poles have been subtracted in the factorized expression (\ref{eq:epem-fac}), the renormalized function $\Sigma_r$ can still have singular plus distributions associated with radiation collinear to the $\bar r$ direction.   For 1PI cross sections in $\rm e^+e^-$ annnihilation this is the direction of the observed particle ($\bar r=c$ at partonic threshold).   This will motivate us to make a further factorization below, into recoil and fragmentation jet functions, as in the general form of Eq.\ (\ref{threshconv}).   
 
The sum over states in Eq.\ (\ref{eq:Sigma-def}) matches the phase space for radiation  in the limit of $y\to 1$ for single-particle annihilation with a parton observed in direction $\bar \beta_r$.    In the center of mass frame defined by the timelike momentum $p_\xi$ and a lightlike vector $p_{\bar r}^\mu= \sqrt{{\hat s}/2}\,\bar\beta_r^\mu$, states $|\xi\rangle$ consist of soft radiation in the $\bar\beta_r$ direction and collinear radiation in the direction $\beta_r$.   Note that $\hat s$ enters only in the combination $\bar y \hat s$.
Collinear-soft radiation is generated by the path-ordered exponential in representation $\bar r$ with constant  velocity $\bar{\beta}_r$, originating at the origin, as above.
In Eq.\ (\ref{eq:Sigma-def}) applied to $\rm e^+e^-$ 1PI cross sections,  $\Phi_{\bar \beta_r}$ plays the dual roles of representing the outgoing observed parton and of serving as a ``gauge link" for the partonic field, $\phi_r$, rendering the matrix elements in Eq.\ (\ref{eq:Sigma-def}) gauge invariant.

At this stage, it is convenient to take moments, under which the 1PI convolution in Eq.\ (\ref{eq:epem-fac}) breaks into a simple product of the moments of the parton-to-parton fragmentation function (taking $h=c$ in Eq.\ (\ref{eq:epem-fac})) times the moments of 
$\Sigma_r$,
\bea
\tilde \Sigma_r \left( \frac{\hat s}{\bar{N}\mu_{rf}^2} ,1, \as(\mu_{rf}) \right)\ &=&\
\int_0^1 d y\;  y^{N-1}\ \Sigma_r \left ( \bar y, \frac{\bar y\hat s}{\mu_{rf}^2},1,\as(\mu_{rf}) \right)
\nn\\[2mm]
\ &=&\ \int_0^\infty d y\;  {\mathrm{e}}^{-N\, \bar y}\ \Sigma_r \left ( \bar y, \frac{\bar y\hat s}{\mu_{rf}^2},1,\as(\mu_{rf}) \right) \ +\ {\cal O}(1/N) \, .
\label{eq:C-moment}
\eea
At leading power, $\Sigma_r$ has an overall behavior of $1/\bar y=\hat s/p^2_\xi$ times logarithms of 
$\bar y\hat s/\mu_{rf}^2$, so that, as usual in threshold resummation, the function has one less argument in moment space.

We will eventually refactorize the function $\tilde{\Sigma}_r$ into perturbative recoil and fragmention jet functions, as in  Eq.\ (\ref{threshconv}).   This procedure is based on the evolution equation satisfied by 
$\tilde{\Sigma}_r$, which is known.  Because $\tilde\Sigma_r$  is built from the composite parton-Wilson line vertex in Eq.\ (\ref{eq:Sigma-def}), with $\mu_F=\mu_{rf}$, it again obeys an evolution equation  \cite{Korchemsky:1994jb,Becher:2006qw} of the general ``cusp" form~(\ref{eq:F-evolve}), 
with another single-logarithmic anomalous dimension, which we label $\gamma_\Sigma^{(r)}$,
\bea
\mu_{rf}\frac{d}{d\mu_{rf}}\, \ln \tilde{\Sigma}_r \left( \frac{\hat s}{\bar{N}\mu_{rf}^2}, 1,\as(\mu_{rf}) \right)\ =\ 2\,A_r(\as(\mu_{rf}))\ln\left( \frac{\bar{N}
\mu_{rf}^2}{\hat s}\right) - \gamma^{(r)}_\Sigma \big( \as(\mu_{rf})\big)\, .
\label{eq:C-evolve}
\eea
The value of $\gamma_\Sigma^{(r)}$ reflects renormalization of the composite operator linking the parton field $\phi_r$ with Wilson line $\Phi^{(\bar r)}_{\bar \beta_r}$ in (\ref{eq:Sigma-def}).   
We will determine it by comparison precisely to the threshold-resummed single-particle annihilation cross section, Eq.\ (\ref{eq:epem-fac}) \cite{Moch:2009my}.   

The solution to Eq.\ (\ref{eq:C-evolve}) organizes logarithms of $N$ as an exponentiated integral from scale $\hat s/N$ to a hard scale, $\mu_{rf} \sim \sqrt{\hat s}$,
\bea
\tilde{\Sigma}_r \left( \frac{\hat s}{\bar{N}\mu_{rf}^2},1, \as(\mu_{rf}) \right)\ &=&\ \tilde{\Sigma}_r \left( 1, 1,\as\big(\sqrt{\hat s/\bar{N}}\,\big) \right)
\nn \\[2mm]
& \times &
 \exp \left [ \int_{\sqrt{\hat s/\bar{N}}}^{\mu_{rf}}  
   \frac{d\mu }{\mu }\ \left(  2\, A_r(\as(\mu))\ \ln \left( \frac{\mu{}^2{\bar{N}}}{\hat s} \right)\,
   -\, \gamma^{(r)}_\Sigma \big(\as(\mu) \big) \right) \right]\, .\;\;\;\;\;\;
   \label{eq:C-soln}
   \eea
We now go through the same steps as for the incoming jets of the previous subsection, starting by promoting the $N$-dependence of the coupling of the evolved coefficient function to the exponent \cite{Sterman:2013nya}
\bea
\label{eq:Sigma-r-muF}
\tilde{\Sigma}_r \left( \frac{\hat s}{\bar{N}\mu_{rf}^2},\frac{\mu_F}{\mu_{rf}}, 
\as(\mu_{rf}) \right)\  &=& \tilde{\Sigma}_r \left( 1,1, \as(\sqrt{\hat s}\,) \right)\ \exp \left [ \int_{\sqrt{\hat s/\bar{N}}}^{\mu_{rf}}  
\frac{d\mu}{\mu }\ \left(  2\, A_r(\as(\mu))\ \ln \left( \frac{\mu^2{\bar{N}}}{\hat s} \right)\,
\right.\right.
\nn \\[2mm]
 & \ & \left. \left. \hspace{-1mm} 
-\,2 \,\hat B_r \big(\as(\mu) \big)\right)\  +\ \ln(\bar{N}_i)\int^{\mu_F^2}_{\mu_{rf}^2} 
\frac{d\mu{}^2}{\mu{}^2}  \, A_i \big(\as(\mu)\big)  \right]\, ,\label{eq:C-soln2}
\eea
where we introduce factorization scale dependence, as in Eq.\ (\ref{eq:J-in-soln1}) for the in jets, and where we define
\bea
\hat B_r \big(\as(\mu) \big)\ =\ \frac{1}{2} \gamma_\Sigma^{(r)}\big(\as(\mu)\big) + \as(\mu)\, \beta\big(\as(\mu)\big) 
\,\frac{d}{d\as}\ln \tilde{\Sigma}_r\big(1,1,\as(\mu)\big)\, ,
\label{eq:hat-Br-def}
\eea
by analogy to the shift in the function $\hat D$ for initial state jets (see Eq.~(\ref{eq:hatD-def})).

At this point, we can determine $\hat B$ and then $\gamma_\Sigma$  from the literature.   To do so, we further rewrite (\ref{eq:C-soln}) with moments explicit in the exponent, again as for incoming jets,
\bea
\tilde{\Sigma}_r \left( \frac{\hat s}{\bar{N}\mu_{rf}^2}, 1,\as(\mu_{rf}) \right)&=& \tilde{\Sigma}_r \left( 1,1, \as\big(\mu_{rf}\,\big) \right)\ \left( 1 - \frac{\as(\mu_{rf})}{\pi}\, \frac{\zeta(2)}{2}\, A_r^{(1)} \right)
\nn \\[2mm]
&\ & \hspace{-35mm} \times\ \exp \left [ \int_0^1 dy \,\frac{y^{N-1}-1}{1-y}
 \left( \int_{\mu_{rf}^2}^{(1-y)\hat{s}} \frac{d\mu^2}{\mu^2}\ A_r(\as(\mu))+ B_r\Big(\as(\sqrt{(1-y)\,\mu_{rf}^2}\,)\Big) \right) \right] .\;\;\;
 \label{eq:Sigma-resum}
\eea
Reexpressing the exponent in the explicit Mellin form leads, as above, to a new prefactor at the hard scale, as shown explicitly,  and to a 
shift from $\hat B_r$ to $B_r$.   The latter depends on the kinematic range 
$\mu_{rf}^2< \mu^2<(1-y)\hat s$ in (\ref{eq:Sigma-resum}), and turns out to be  one-eighth of the corresponding shift in the ``Drell-Yan" function $D(\as)$, 
which appeared in the discussion of incoming jets above.   
Again following the steps taken in Refs.\ \cite{Catani:2003zt} and \cite{Sterman:2013nya}, we find 
\bea
 B_r^{(1)}\,&=&\, \hat B_r^{(1)}\, , \nn\\[2mm]
 B_r^{(2)}\,&=&\, \hat B_r^{(2)}\ +\ \frac{1}{8}\,\zeta(2)\, \beta_0\, A_r^{(1)}\, . 
\label{eq:B-Bhat}\
\eea
By taking $\mu_{rf}^2=\hat s$, we can compare to the explicit expression for $B_r$ as it appears in single-particle inclusive resummation at center-of-mass energy $\hat{s}$, as  given \cite{Moch:2009my}.
At ${\cal O}(\alpha_s^2)$, we find for $\hat B_r$,
\bea
&&\hspace*{-1cm}
\hat B_r^{(2)}\,=\, \left\{ \begin{array}{cc} \frac{C_F^2}{2} \left( -\frac{3}{16}+\frac{3}{2}\zeta(2) -3 \zeta(3)\right)+\frac{C_FC_A}{2}
\left( -\frac{3155}{432} + \frac{11}{12}\zeta(2)+5\zeta(3)\right)+\frac{C_FN_f}{2} \left( \frac{247}{216}-\frac{1}{6}\zeta(2)\right)
& (r=q)\,,
\\[3mm]
\frac{C_A^2}{2} \left( -\frac{611}{72}+\frac{11}{4}\,\zeta(2)+2 \zeta(3)\right)+
\frac{C_AN_f}{2}\left(\frac{107}{54}-\frac{1}{2}\zeta(2)\right)+\frac{C_F N_f}{8}-
\frac{5N_f^2}{108} & (r=g)\,.
\end{array}
\right.\nn\\[2mm]
\label{eq:Br-def}
\eea
This is the anomalous dimension we will use below.
We recall the values of the $B_r^{(2)}$ in~(\ref{eq:B-Bhat}) as given in the literature \cite{Moch:2009my} in Appendix~\ref{AppB}.

We can now solve Eq.\ (\ref{eq:hat-Br-def}) to find an explicit expression for $\gamma_\Sigma^{(r)}(\as)$.  This requires the one-loop Mellin space prefactors 
$\tilde\Sigma_r(1,1,\as)$, which we can compute directly (see Sec.~\ref{sec:1loop-jets} below) to find
\bea
\label{Sigfactor}
\tilde{\Sigma}_q\left( 1,1, \as\right)&=&1 + \frac{\as}{\pi} \, C_F \left(\frac{7}{4}\, -\, \zeta(2)\right)+{\cal O}(\as^2)\nn\\[2mm]
\tilde{\Sigma}_g\left( 1,1,\as\right)&=&1 + \frac{\as}{\pi} \, \left\{
C_A\left(\frac{67}{36}\, -\, \zeta(2)\right)-\frac{5}{18}N_f\right\} +{\cal O}(\as^2)\,.
\eea
Using these explicit forms in Eq.\ (\ref{eq:hat-Br-def}), we have for the two-loop term of $\gamma_\Sigma^{(q)}$,
\bea
\gamma^{(q),(2)}_\Sigma  &=& 2\hat{B}_r^{(2)}\ +\ \frac{1}{2}\,\beta_0 C_F \left(\frac{7}{4}\, -\, \zeta(2)\right)\, .
\label{eq:gamma-Sigma-solve}
\eea
At this point we may recall the relation observed in Ref.\ \cite{Vogt:2000ci,Moch:2005ba}, valid to two loops,   
\bea\label{eq:BDPrel}
B_q(\as)\ =\ \frac{1}{2}\, D_q(\as) -P_{q,\delta}(\as)  - \left( \frac{\as}{2\pi}\right)^2 \frac{7}{4}\,\beta_0 C_F +\ {\cal O}(\as^3)\, ,
\eea
where $P_{q,\delta}$ is the coefficient of $\delta(1-x)$ in the quark DGLAP splitting function (see App.~\ref{AppB}), and as above
 $D_i$ (see Eq.~(\ref{eq:Drel}))
is the single-logarithmic anomalous dimension for Drell-Yan resummation 
\cite{Sterman:1986aj,Catani:1989ne,Vogt:2000ci,Catani:2003zt}. In terms of the anomalous dimensions $\gamma_{J^{(q)}}(\as)$ 
of Eq.~(\ref{eq:gamW-def}) and the above $\gamma^q_\Sigma(\as)$, relation~(\ref{eq:BDPrel}) becomes
\be
\frac{1}{2}\,\gamma^q_\Sigma\,=\,\gamma_{J^{(q)}}-P_{q,\delta}\,.
\ee
A similar result holds for the gluon.   We also note that at order $\as$, the single-loop anomalous dimension is given by the delta function piece of the splitting function, which is consistent with jet anomalous dimensions in direct QCD and soft-collinear effective theory.

In the form of Eq. (\ref{eq:C-soln}), the  integral over scales $\mu$ that appear in the running coupling can be reorganized in a trivial fashion to separate the recoil jet from the fragmentation jet, as presented, for example, in Refs.\ \cite{Vogt:2000ci,Moch:2005ba},
\be
\tilde{\Sigma}_r \left( \frac{\hat s}{\bar{N}\mu_{rf}^2},\frac{\mu_F}{\mu_{rf}}, \as(\mu_{rf}) \right)\ 
=\ \tilde J_{\rm fr}^{(\bar r)} \left ( \frac {\hat s}{\bar N^2\mu_{rf}^2 },\frac{\mu_F}{\mu_{rf}},\as(\mu_{rf}) \right) \,
\tilde J_{\rm rec}^{(r)} \left( \frac{\hat s}{\bar{N}\mu_{rf}^2},\frac{\hat s}{\bar{N}^2\mu_{rf}^2},\as(\mu_{rf}) \right)\, ,
\label{eq:Sigma-refactor}
\ee
where as shown, we introduce factorization scale dependence into the fragmentation jet as we did for the incoming jets.   Note that dependence on the ratio $\hat s/\bar N^2\mu_{rf}^2$ cancels in the product of the recoil and fragmentation jets.

The factorization in Eq.\ (\ref{eq:Sigma-refactor}) is, of course, not unique.  For application to QCD scattering, we normalize the outgoing jets so that they preserve the one-loop structure of the soft anomalous dimension matrix at two loops \cite{Aybat:2006mz}.   We can do this by choosing the fragmentation jet to be equal to the incoming jets of Drell-Yan as a function of $N$.   This is the natural choice for double-inclusive annihilation, as noted in \cite{Sterman:2006hu}.
It is also the manner in which the soft anomalous dimension matrices were normalized and computed in \cite{Aybat:2006mz}.
The formal definition of the eikonal fragmentation jet functions differs from that for the incoming jet function in being defined 
by the eikonal double-inclusive annihilation cross section, rather than the eikonal Drell-Yan cross section.  Soft radiation is thus emitted from an outgoing color-singlet pair of Wilson lines, rather than the incoming color singlet pair 
of  $\tilde{J}_{\rm in}^{(i)}$, Eq.\ (\ref{eq:J-sqrtW}).   
The equality of such eikonal fragmentation functions and parton distributions defined at fixed energy is shown in Ref.\ \cite{Sterman:2006hu}.

With this normalization, the fragmentation jet is given by the expression for incoming jets, Eq.\ (\ref{eq:J-in-soln2}), involving both the cusp anomalous dimension and the function $D$.  We will choose partonic subscript $c$ for the fragmentation jet to emphasize that our choice here extends beyond $\rm e^+e^-$ annihilation,
\bea
\tilde J_{\rm fr}^{(c)} \left ( \frac {\hat s}{\bar N^2\mu_{rf}^2 },\frac{\mu_F}{\mu_{rf}},\as(\mu_{rf}) \right) \ &=&\  
J_{\rm in}^{(c)} \left ( \frac {\hat s}{\bar N^2\mu_{rf}^2 },\frac{\mu_F}{\mu_{rf}},\as(\mu_{rf}) \right) \, 
\nn\\[2mm]
&\ & \hspace{-25mm} =\
\widehat R_c(\as(\mu_{rf}))\, \exp \left [ \int^{\mu_{rf}}_{\sqrt{\hat s}/\bar{N}} 
\frac{d\mu}{\mu} \left(A_c\big(\as(\mu)\big) \, \ln\left( \frac{\mu^2\bar N^2}{\hat{s}}
\right) - \frac{1}{2}\,\hat D_c \big(\as(\mu)\big) \right)\right.
\nn\\[2mm]
&&\hspace*{12mm}+\ \ln(\bar{N})\int^{\mu_F^2}_{\mu_{rf}^2} 
\frac{d\mu{}^2}{\mu{}^2}  \, A_c \big(\as(\mu)\big) \bigg]\, .
  \label{eq:out-jet-def}
\eea
This definition is the same as  in Ref.\ \cite{Hinderer:2014qta}. 

Using Eq.\ (\ref{eq:Sigma-resum}) for 
$\tilde\Sigma_r$ and (\ref{eq:out-jet-def}) for the fragmentation jet in (\ref{eq:Sigma-refactor}), we find that the recoil jet includes 
the full 
$\hat{B}_r$ term in the exponent, while the terms involving $A_r(\as)$ change,
\bea
\tilde J_{\rm rec}^{(r)} \left( \frac{\hat s}{\bar{N}\mu_{rf}^2},\frac{\hat s}{\bar{N}^2\mu_{rf}^2},
\as(\mu_{rf}) \right) \ &=& \frac{\tilde{\Sigma}_r \left( 1, 1, \as(\mu_{rf}\,) \right)}{\widehat{R}_{\bar r}(\as(\mu_{rf}))}\,  
\exp \left [ \, -\, \int_{\sqrt{\hat s}/\bar{N}}^{\sqrt{\hat s/\bar{N}}}
\frac{d\mu}{\mu }\ \, A_r(\as(\mu))\ \ln \left( \frac{\mu^2{\bar{N}}^2}{\hat s} \right)\, \right. \nn\\[2mm]
& &\left. \hspace{-45mm}
+\,  \int^{\mu_{rf}}_{\sqrt{\hat s/\bar{N}}}
\frac{d\mu}{\mu }\,\left( A_r(\as(\mu))\ \ln \left( \frac{\mu^2}{\hat s} \right)\ - \, 2\,
  \hat B_r \big(\as(\mu) \big)\right)
\, +\,  \frac{1}{2} \int_{\sqrt{\hat s}/\bar N}^{\mu_{rf}} \frac{d\mu}{\mu} \hat D_r \big(\as(\mu) \big)  \right]\, .
\label{eq:J-rec-MV}
\eea
We see here that $\tilde{J}^{(r)}_{\rm rec}$ obeys the evolution equation given in (\ref{eq:rec-evolve}), with a single-log anomalous dimension that is a combination of the functions $\hat{B}_r$ and $\hat{D}_r$.
We note that although the function $\hat{D}_r$ cancels in the 1PI electroweak annihilation cross section, these expressions can be chosen as the factorizing jets in QCD cross sections, where the outgoing partons are not necessarily of the same flavor.   Alternative choices for the jets are possible, but in general will lead to soft anomalous dimensions with additional terms proportional to the identity matrix at two loops.  Although the separation of recoil from 
fragmentation in Eqs.\ (\ref{eq:out-jet-def}) and (\ref{eq:J-rec-MV}) is a simple reshuffling of the integral, it has a direct physical interpretation because of the relation of the scale that appears in $A_r(\as)$ to the momenta of ``web functions'', which, as discussed for example in Ref.\ \cite{Laenen:2000ij}, reflect the transverse momentum and hence direction of soft radiation.
Closed expressions for NNLL expansions of the jet functions in Eqs.~(\ref{eq:out-jet-def}) and~(\ref{eq:J-rec-MV}) 
are again given in Appendix~\ref{AppC}.

In summary, for the short-distance coefficient function for electroweak annihilation, we reexpress Eq.\ (\ref{eq:Sigma-def1}) as a special case of the general factorized cross section, which exhibits the separation of the fragmenting and recoil jets in the simplest case,
\bea
\label{eq:omega-resum-epem}
\tilde\omega^{{\rm e^+e^-}\to c} \left (\hat \eta,N, \frac{\mu_F^2}{\hat{s}}\right) &=&
H^{{\rm e^+e^-}\to c}\left (\as(\mu_{rf}) , \frac{\sqrt{\hat s}}{\mu_{rf}},\frac{\mu_F}{\mu_{rf}},\hat \eta\right) 
\nn\\[2mm]
&\times&\tilde J_{\rm fr}^{(c)} \left ( \frac {\hat s}{\bar N^2\mu_{rf}^2 },\frac{\mu_F}{\mu_{rf}},\as(\mu_{rf}) \right) \,
\tilde J_{\rm rec}^{(\bar c)} \left( \frac{\hat s}{\bar{N}\mu_{rf}^2},\frac{\hat s}{\bar{N}^2\mu_{rf}^2},\as(\mu_{rf}) \right)\, ,
\eea
where the $\mu_F$ dependence of the function $H$ is determined by Eq.\ (\ref{eq:H-evovle}), in this case with only a single term, for the fragmenting parton.
It is straightforward to verify the consistency of this expression with the explicit results for single-inclusive annihilation given in Ref.\ \cite{Moch:2009my}.

\subsection{One-loop partonic jets}
\label{sec:1loop-jets}

For later reference, we now expand each of the jet functions 
to order $\alpha_s$ and transform the result back to $z$ space. For each, we write the result in the form
\be
J(z)\,=\,\delta(1-z)+\frac{\alpha_s}{\pi}\,J^{(1)}(z)\,.
\ee
For the initial-state functions discussed in Sec.~\ref{sec:inijets}, we use the explicit moments given by Eq.~(\ref{eq:Ni-defs}). 
If the underlying Born process is $ab\to cr$, with parton $c$ fragmenting, we have
\bea\label{Jab}
J^{(a),(1)}_{{\mathrm{in}}}(z)&=&C_a\left[ 
2 \left(\frac{\ln(\bar{z})}{\bar{z}}\right)_+ - 2\ln(v)\left(\frac{1}{\bar{z}}\right)_++
\delta(\bar{z}) \left( \ln^2(v) - \frac{3}{4}\zeta(2)\right)\right]\,,\nn\\[2mm]
J^{(b),(1)}_{{\mathrm{in}}}(z)&=&C_b\left[ 
2 \left(\frac{\ln(\bar{z})}{\bar{z}}\right)_+ - 2\ln(1-v)\left(\frac{1}{\bar{z}}\right)_++
\delta(\bar{z}) \left( \ln^2 (1-v) - \frac{3}{4}\zeta(2)\right)\right]\,,
\eea
where $C_q=C_F$ and $C_g=C_A$ and $\bar{z}=1-z$, and where
\beq\label{eq:v-def}
v\,\equiv\,1+\frac{\hat{t}}{\hat{s}}\,=\,\frac{{\mathrm{e}}^{\hat\eta}}{{\mathrm{e}}^{\hat\eta}+{\mathrm{e}}^{-\hat\eta}}\,.
\eeq
In Eq.~(\ref{Jab}) we have neglected contributions that are suppressed near threshold. 
Note that we have also suppressed the dependence on the factorization  scale,
which may be easily reconstructed. 

The construction of jet functions for the outgoing fragmenting partons was presented in the previous subsection.   In summary, we compute the renormalized function $\Sigma_r$ for each parton flavor, and separate the fragmentation and recoil jets as in Eqs.\ (\ref{eq:Sigma-resum})-(\ref{eq:out-jet-def}).  The $\Sigma_r$ for quarks and gluons are not related to each other by simple exchange
$C_F\leftrightarrow C_A$.   At one loop, we find for quarks and gluons
 from~(\ref{eq:Sigma-def})
\bea\label{eq:sigma1loop}
\Sigma_q(\bar{z},\alpha_s)&=&\delta(\bar{z})+\frac{\alpha_s}{\pi}\,C_F\left\{\left(\frac{\ln(\bar{z})}{\bar{z}}\right)_+-\frac{3}{4}\left(\frac{1}{\bar{z}}\right)_+ 
+\left(\frac{7}{4}-\frac{3}{2}\,\zeta(2)\right) \delta(\bar{z})\right\}\,,
\nn\\[2mm]
\Sigma_g(\bar{z},\alpha_s)&=&\delta(\bar{z})+\frac{\alpha_s}{\pi}\,\left\{ C_A \left(\frac{\ln(\bar{z})}{\bar{z}}\right)_+-
\pi b_0\, \left(\frac{1}{\bar{z}}\right)_+ 
+\left(C_A \left( \frac{67}{36}-\frac{3}{2}\,\zeta(2)\right) -\frac{5}{18}N_f \right) \delta(\bar{z})\right\}\,.
\nn\\
\eea
The jet function associated with the fragmenting parton is
given by Eq.\ (\ref{eq:out-jet-def}), and is found at one loop by a simple expansion and inverse transform,
\be
J^{(c),(1)}_{{\mathrm{fr}}}(z)\,=\,C_c\, \left[\, 2\left(\frac{\ln(\bar{z})}{\bar{z}}\right)_+\ -\ \frac{3\zeta(2)}{4}\,\delta(\bar z) \right]\, .
\label{eq:Jc-1loop}
\ee
The jet functions for the recoiling quarks and gluons are defined as the remaining part of the final-state function $\Sigma_r$, Eq.\ (\ref{eq:Sigma-def}), and are found from~(\ref{eq:sigma1loop}) by removal of the fragmenting jet (\ref{eq:Jc-1loop}), 
\be
J^{(q),(1)}_{{\mathrm{rec}}}(z)\,=\,C_F\left[
- \left(\frac{\ln(\bar{z})}{\bar{z}}\right)_+ -\frac{3}{4}\left(\frac{1}{\bar{z}}\right)_+ 
+\left(\frac{7}{4}\,-\,\frac{3}{4} \zeta(2)\right) \delta(\bar{z})\,\right]\,,
\ee
and
\be\label{Jrecg}
J^{(g),(1)}_{{\mathrm{rec}}}(z)\,=\,
-C_A \left(\frac{\ln(\bar{z})}{\bar{z}}\right)_+-\pi b_0\left(\frac{1}{\bar{z}}\right)_+ 
+\left( C_A\left(\frac{67}{36}-\,\frac{3}{4} \zeta(2)\right)-\frac{5}{18}N_f\right) 
\delta(\bar{z})\, .
\ee
Here we have introduced $b_0=\beta_0/(4\pi)=(11C_A-2N_f)/(12\pi)$.   We note that the logarithmic terms are the same as those found from the expansion of the resummed expression for the recoil jets, Eq.\ (\ref{eq:J-rec-MV}).

It is useful here to make again contact to ${\rm e^+e^-}\to hX$, which we used in Sec.~\ref{sec:rec-frag-jets}
to define the outgoing jets.  
The hard function in this case is obtained from the known~\cite{Altarelli:1979ub} one-loop virtual correction:
\bea
H(\hat\eta,\alpha_s)&=&H^{(0)}(\hat\eta)+\frac{\alpha_s}{\pi}\,
H^{(1)}(\hat\eta)+{\cal O}(\alpha_s^2)\nn\\[2mm]
&=&H^{(0)}(\hat\eta)\,\bigg\{ 1 + \frac{\alpha_s}{2\pi}\,C_F
\big[ -8+7\zeta(2) \big]+{\cal O}(\alpha_s^2)\bigg\}\,,
\eea
where we have suppressed the label $e^+e^-\to q\bar{q}$ for the process, and where
\be
H^{(0)}\,(\hat\eta)\,=\,4\pi C_A\alpha^2 \big(v(1-v)\big)^2 \,\big(v^2+(1-v)^2\big)\,.
\ee
We thus find
\bea
H(\hat\eta,\alpha_s)\,\Sigma_q(\bar{z},\alpha_s)&=&
H^{(0)}(\hat\eta)\,\delta(\bar{z})+\frac{\alpha_s}{\pi}\,H^{(1)}(\hat\eta)\,\delta(\bar{z})+
\frac{\alpha_s}{\pi}\,H^{(0)}(\hat\eta)\left(
J^{(q),(1)}_{{\mathrm{fr}}}(z)+J^{(q),(1)}_{{\mathrm{rec}}}(z)\right)\nn\\[2mm]
&=&H^{(0)}(\hat\eta)\left[ \delta(\bar{z})+\frac{\alpha_s}{\pi}\,
C_F\left\{\left(\frac{\ln(\bar{z})}{\bar{z}}\right)_+-\frac{3}{4}\left(\frac{1}{\bar{z}}\right)_+ 
+\left(2\,\zeta(2)-\frac{9}{4}\right) \delta(\bar{z})
\right\}\right]\nn\\[2mm]
&=&\omega^{{\rm e^+e^-}\to c,(1)}(z)\,,
\eea
where $\omega^{{\rm e^+e^-}\to c,(1)}$ is the one-loop inclusive hard-scattering function of Eq.~(\ref{eq:epem-fac}),
calculated to NLO in~\cite{Altarelli:1979kv}. As designed, our formalism thus reproduces the full NLO correction near partonic 
threshold, without the need for any additional soft function.    
For general single-particle cross sections in hadronic scattering, a soft function is of course necessary.

\section{Soft Functions for Single-Particle Cross Sections}
\label{sec:define-S1pi}

Soft radiation in the central region between the jets is well approximated by light-like Wilson lines in the color representations 
of the partons participating in the hard scattering   \cite{KS,KOS,Catani:1996yz,Ferroglia:2013awa}.     We begin our discussion of the necessary soft functions by briefly reviewing their resummation.

\subsection{Resummation for soft functions}

The soft functions in the refactorized expressions, Eqs.\ (\ref{threshconv}) and (\ref{eq:omega-fact}) are matrices in the space of color exchange tensors for the partonic process $ab\to cr$ \cite{KS,Laenen:1998qw}.    We will specify their definitions in the next subsection, and here introduce some notation, since we will use both momentum and moment space.   
The soft function scales as an overall $w_S^{-1}$, with additional logarithms in the combination 
$w_S\sqrt{\hat s}/\mu_{rf}$.   The soft function is constructed to have at most one such logarithm per loop.  Higher logarithms are associated with collinear enhancements, which are universal, and are factored into the jet functions \cite{KOS,Laenen:1998qw}.   To identify fixed-order terms, we introduce the momentum-space notation,
\bea
S\left ( w_S,\frac{w_S\sqrt{\hat s}}{\mu_{rf}},\as(\mu_{rf}),\hat \eta \right)\ &=& \sum_{n=0}^\infty\ \left ( \frac{\as(\mu_{rf})}{\pi} \right)^n\ S^{(n)} \left( w_S,\frac{w_S\sqrt{\hat s}}{\mu_{rf}},\hat \eta \right)\,,
\nn\\[2mm]
S^{(n)} \left(w_S, \frac{w_S\sqrt{\hat s}}{\mu_{rf}},\hat \eta \right)\
&=&\   S^{(n)}_0 (\hat \eta)\, \delta(w_S) + 
\sum_{i=0}^{n-1} S^{(n)}_{i+1} (\hat \eta)\, \left[ \frac{\ln^i \big[w_S\sqrt{\hat s}/\mu_{rf}\big]}{w_S} \right]_+ \, ,
\label{eq:S-expand}
\eea
where here and in the following we usually suppress the explicit labeling of the underlying partonic process, $ab\to cr$. 
Note that from the refactorization expression, Eq.\ (\ref{threshconv}) the argument of each $S^{(n)}_i$ appears in a convolution, whose range is $0\le w_S \le \hat s_4/\hat s$.   The corresponding moment-space expansion 
of the last equation will be denoted by
\bea
\tilde S^{(n)}\left ( \frac{\hat s}{\bar{N}^2\mu_{rf}^2},\hat \eta \right)\  =\  \tilde S^{(n)}_0(\hat \eta) + \sum_{j=1}^{n} \tilde S^{(n)}_j(\hat \eta)\,  
\ln^j \left(\frac{\sqrt{\hat s}}{\bar{N}\mu_{rf}}\,\right) \, ,
\label{eq:tilde-S-expand}
\eea
where as before $\bar N\equiv N {\mathrm{e}}^{\gamma_E}$.
In the following section, we will provide field-theoretic definitions of these soft functions, and give an example of the calculation of $\tilde S^{(1)}_0$.   
The coefficient $\tilde S^{(1)}_0$ itself provides a series of NNLL logarithms, as we now review,
in addition to $\tilde S_1^{(1)}$, which is proportional to the anomalous dimension matrix $\Gamma$, and contributes at NLL.

The resummed soft factor in moment space is the solution to Eq.\ (\ref{eq:S-evolve}) \cite{KS,KOS,Catani:1996yz} in terms of the refactorization scale,
\be
\tilde S\left( \frac{\hat s}{\bar{N}^2\mu_{rf}^2}, \alpha_s(\mu_{rf}) , \hat{\eta}\right)  \ = \ {\cal{S}}^\dagger   \left(\bar{N},
\as(\mu_{rf}), \frac{\mu_{rf}}{\sqrt{\hat s}} , \hat{\eta} \right) \,
\tilde {S}\left(1, \as \Big(\sqrt{\hat s}/\bar{N} \,\Big),\hat{\eta} \right)\, 
 {\cal{S}}  \left(\bar{N},\as(\mu_{rf}), \frac{\mu_{rf}}{\sqrt{\hat s}} , \hat{\eta} \right).
 \label{eq:ressoft}
\ee
The second factor on the right-hand side of this expression is the soft function in moment space, Eq.\ (\ref{eq:tilde-S-expand}), with a scale choice that makes all its logarithmic terms vanish.   A full NNLL resummation takes into account logarithms due to the expansion of the running coupling for the one-loop soft function at scale 
$\sqrt{\hat s}/\bar{N}$, 
in much the same way as it includes logarithms from the function $R_i(1,\as)$ from incoming and fragmentation jets, as in 
Eqs.\ (\ref{eq:J-in-soln1}),(\ref{eq:Rrunning})
and from $\tilde{\Sigma}_r \left( 1,1, \as\right)$ from the recoil jet, in Eq.\ (\ref{eq:J-rec-MV}).
This factor serves as the boundary condition for the evolution of the soft function from infrared to ultraviolet scales.
The factor that resums the evolution logarithms of the moment variable in Eq.\ (\ref{eq:ressoft}) is given by the ordered exponential \cite{Laenen:1998qw}
\beeq
\label{GammaSoft}
{\cal S}_{ab\to cr} \left(\bar{N},\as(\mu_{rf}), \frac{\mu_{rf}}{\sqrt{\hat s} } , \hat{\eta}  \right) 
&=& 
{\cal P}\exp\left[  \int^{\sqrt{\hat s}/\bar{N}}_{\mu_{rf}} 
\frac{d\mu}{\mu} \ \Gamma_{ab\to cr} 
\left(\hat{\eta},\as(\mu) \right)\right] \, ,
\eeeq
with ${\cal P}$ denoting path ordering.
The process-dependent anomalous dimension matrix, $\Gamma_{ab\to cr}$ is determined entirely from virtual corrections. 
As observed in Ref.~\cite{Aybat:2006mz}, the two-loop anomalous dimension matrix, $\Gamma^{(2)}_{ab\to cr}$  is proportional to the one-loop 
matrix,
\be
\Gamma^{(2)}_{ab\to cr}(\hat\eta) = \frac{K}{2}\, \Gamma^{(1)}_{ab\to cr}(\hat\eta)\; ,
\ee
with $K=C_A(67/18-\pi^2/6)-5 N_f/9$.  This specific relation does not extend to three loops, as was 
demonstrated by explicit calculation in Ref.\ \cite{Almelid:2015jia}, but this does not lead to qualitative differences in the analysis beyond NNLL.  As in Refs.~\cite{Almeida:2009jt,Hinderer:2014qta},   exponentiation of these matrices is readily carried out numerically, by iterating the exponential series to an adequately high order.

\subsection{Operator definitions}

As in the dihadron case, the elements $\tilde{S}_{LI}$ of the soft matrix of Eq.\ (\ref{eq:omega-fact}) are computed using the method described in Ref.\ \cite{KOS},  which, however, we must adapt in a significant way to single-particle inclusive cross sections.  The all-orders form is clearest in moment space, 
where it is
given as the ratio of the moments of a fully eikonal cross section $\hat \sigma^{ab\to cr}_{LI}$ and four factorized jets, two to absorb the factorizing collinear singularities of the incoming parton lines, and two to absorb the collinear singularities of outgoing lines, 
all in eikonal approximation:
\bea
&&\hspace*{-20mm}
\Bigg(\tilde{S}_{ab\to cr}\bigg( \frac{\hat s}{\bar{N}^2\mu_{rf}^2}, \alpha_s(\mu_{rf}) , \hat{\eta} \bigg)\Bigg)_{LI}\ \,=\,
\nn\\
&\ & \hspace{-5mm} 
\frac{\hat \sigma^{ab\to cr}_{LI} \left( \bar{N},\frac{\hat s}{\mu_{rf}^2},\as(\mu_{rf}), \hat \eta, 
\varepsilon\right)}{\prod\limits_{i=a,b} \tilde j_{\rm in}^{(i)}\left(\frac{\hat s}{\bar{N}_a^2\mu_{rf}^2},\as(\mu_{rf}), \varepsilon \right)\
 \tilde j_{\rm fr}^{(c)}\left( \frac{\hat s}{\bar{N}^2
\mu_{rf}^2},\as(\mu_{rf}), \varepsilon \right)\ \tilde j_{\rm rec}^{(r)}\left( \frac{\hat s}{\bar{N}
\mu_{rf}^2}, \frac{\hat s}{\bar{N}^2
\mu_{rf}^2},\as(\mu_{rf}), \varepsilon \right)}\, ,
\label{eq:soft-matrix}
\eea
where as above $\hat \eta$ is the rapidity of the fragmenting jet in the partonic center of mass frame, and $\hat s$ sets the scale of the invariant mass of the partonic system in the single-particle inclusive case we are considering. As
indicated, dimensional regularization with $D=4-2\varepsilon$ dimensions is used to regulate divergences.

In order to formulate operator definitions for the quantities on the right-hand side of Eq.~(\ref{eq:soft-matrix}),
we introduce a slight generalization of the definition of the Wilson line in Eq.~(\ref{eq:wilson}), 
\beq
\Phi_{\beta}^{(f)}({\lambda}_2,{\lambda}_1;x)\ \equiv \
{\cal P}\exp\left(-ig\int_{{\lambda}_1}^{{\lambda}_2}d\lambda\; {\beta}{\cdot} A^{(f)} (\lambda\cdot\beta+x)\right)\, ,
\label{eq:wilson1} 
\eeq
which will appear both for the numerator and  for the ``eikonal jet" functions in the denominator of~(\ref{eq:soft-matrix}). 
For $2\rightarrow 2$ scattering, the ends of two incoming and two outgoing Wilson lines are coupled locally by a 
constant color tensor ${\cal C}_I$, and we define
\beqa
w_I^{(ab\to cr)}(x)_{\{j\}}
&\equiv&
\sum_{\{i\}}
\Phi_{\beta_r}^{(r)}(\infty,0;x)_{j_r,i_r}\; 
\Phi_{\beta_c}^{(c)}(\infty,0;x)_{j_c,i_c}\nn \\[2mm]
&\times& \left( {\cal C}_I^{(ab\to cr)}\right)_{i_ri_c,i_bi_a}\; 
\Phi_{\beta_a}^{(a)}(0,-\infty;x)_{i_a,j_a}
\Phi_{\beta_b}^{(b)}(0,-\infty;x)_{i_b,j_b}\, .
\label{eq:wivertex}
\eeqa 
These operators will produce radiation at all scales and directions, including collinear radiation.   
As described in Refs.\ \cite{KS,KOS} and below, the 
incoming and fragmentation
jets, 
$\tilde{j}_{\rm in}$ and $\tilde{j}_{\rm fr}$, respectively are constructed to match collinear 
singularities and radiation phase space in the partonic threshold limit, avoiding double counting with 
the partonic jet functions in the re-factorized hard function, Eqs.\ (\ref{threshconv}) and (\ref{eq:omega-fact}).
In the same way, we also define singlet operators that link two lines in conjugate representations, extending from 
the infinite past and joined at the origin by  a color singlet tensor:
\beqa
w_0^{(a\bar a)}(x)_{\{j\}} &=& 
\sum_{\{i\}}
 \delta_{i_a,i_{\bar{a}}}\; 
\Phi_{\beta_{\bar a}}^{(\bar a)}(0,-\infty;x)_{i_a,j_a}\,
\Phi_{\beta_a}^{(a)}(0,-\infty;x)_{i_{\bar a},j_{\bar a}}\, .
\label{eq:w0vertex}
\eeqa
We will use these to construct the incoming eikonal jets.

In these terms,
for the incoming eikonal jets, we construct the eikonal analogs of partonic (Drell-Yan) annihilation.
Unlike the case of the final state jets below, the phase space for the initial state jets is defined by a total energy, and is hence finite.
The kinematics of the process 
is reflected in the rescaled Mellin
moment variables, as in Eq.\ (\ref{eq:Ni-defs}) \cite{Laenen:1998qw}.
The ``in'' jets 
constructed in this way are found in Ref.\ \cite{Hinderer:2014qta} on dihadron production, and are given by
\bea
\left( \tilde j_{\rm in}^{(a)}\left( \frac{ \hat s}{\bar{N}^2\mu_{rf}^2},\as(\mu_{rf}),\varepsilon \right) \right)^2\
&=& \   \int_0^1 dy\  y^{N-1}\ \sum_{\ket{\xi}}\ \delta \left ( 1-y- \frac{p_\xi^0}{\sqrt{\hat s}} \right)
\nn \\[2mm]
& \times & {\rm Tr}_{\{j\}} \;  \langle 0|\, \bar T\left( w_0^{(a\bar a)}{}^\dagger 
\left(0\right)_{\{j\}} \right)  |\xi\rangle \, \langle \xi| T \left(  w_0^{(a\bar a)}(0)_{\{j\}} \right) |0\rangle\, .
\label{eq:eik-in-jet}
\eea
With this choice, $\big(\tilde{j}^{(a)}_{\rm in}\big)^2$ is the eikonal Drell-Yan cross section, computed at two 
loops in \cite{Belitsky:1998tc}.   In fact, the eikonal jets that remove collinear singularities from the eikonal cross section are the same as the incoming jet functions that appear in the re-factorized hard-scattering functions, defined as in Eq.\ (\ref{eq:J-sqrtW}).   This is simply a reflection of the multiplicative nature of factorization in moment space.

For 
the outgoing jets, we turn to eikonal
single-particle inclusive ${\rm e^+e^-}$ annihilation.   Here, the eikonal cross section is defined at fixed values of the invariant mass of all radiation recoiling against the observed particle \cite{Laenen:1998qw,Catani:1998tm,Catani:2013vaa}, as in the definition of the recoil and fragmentation jets, derived
above
 from functions $\Sigma_r$, Eq.\ (\ref{eq:Sigma-def}).   In particular, because our cross section is defined at fixed momentum fraction $y$, a light cone fraction in the direction of the observed momentum $p_c$, we
 must
  incorporate the limitation on the energy of radiation in the direction of the recoil jet in the definition of the eikonal cross section and its eikonal jet subtractions.   
  This is because 
  fixing $y$ alone does not result in a finite phase space in $\hat \sigma^{ab\to cr}_{LI}$.   
Specifically, fixing 
$y$ and hence $s_4$ still allows collinear divergences in the recoil direction from arbitrarily large energies.   The situation is to be contrasted to pair invariant mass threshold resummation, where a fixed energy automatically imposes a limited phase space.    In the present case, we must truncate the sum over radiation collinear to the recoil direction.

To this end, we match the physical phase space for partons in the soft function to the phase space near 1PI partonic threshold to cut off unbounded collinear behavior.   The details of the collinear truncation will cancel in the ratio of Eq.\ (\ref{eq:soft-matrix}), because collinear partons factor from soft gluons emitted at fixed angles both in the numerator and denominator.    In defining the space of states over which to sum, we can thus replace the partonic recoiling jet by a single particle, whose momentum we will denote by $p_4^{(h)}$.   
The complete ``recoil" momentum, $p_R$ is then the sum of $p_4^{(h)}$ and the momenta of partons emitted by the eikonals, 
\bea
p_R &=&\sum_{i=1}^n k_i + p^{(h)}_4 \nonumber\\[2mm]
p_R^2 &=& \hat s(1-y)\ =\ \hat s_4\, ,
\label{eq:phase space}
\eea
where we integrate over the $n$-particle phase space, requiring $\big(p_4^{(h)}\big)^2=0$.   

In the $p_c / p_R$ c.m.\ frame, with $p_c$ in the plus direction, $p_R$ has only plus and minus components, which we denote as
\bea
p_R &=& \left( p_R^+ , p_R^-, p_R^\perp \right)\,=\,
\left( \sqrt{ \frac{\hat s}{2}} \,(1-y), \,\sqrt{\frac{\hat s}{2}}, \,0_\perp \right)
\nn \\[2mm]
&=&\ \sqrt{ \frac{\hat s}{2}}\, (1-y)\, \beta_c\ +\ \sqrt{ \frac{\hat s}{2}}\, \beta_r\, ,
\label{eq:pR-def}
\eea
where in the second equality $\beta_c$ is the lightlike vector in the direction of the observed particle, and $\beta_r$ is the ``opposite-moving" lightlike vector in the direction of the recoiling jet at partonic threshold in this frame, with $\beta_c\cdot \beta_r =1$.   This definition will enable us to evaluate integrals over partonic phase space in other frames, and this is the form we will use below.

In terms of the operators $w_I$, the eikonal cross section is defined by sums over states $\ket{\xi}=
\ket{\{k_i\}}$ radiated freely by the Wilson lines, 
subject only to the momentum constraint of Eq.\ (\ref{eq:phase space}) involving $p_R$, Eq.\ (\ref{eq:pR-def}):
\bea
\hat \sigma^{ab\to cr}_{LI} \left(\bar N, \frac{\hat s}{
\mu_{rf}^2}, \as(\mu_{rf}), \hat \eta,\varepsilon\right) &=&  
\int_0^1 d y\;  y^{N-1}\ \sum_{\ket{\xi}}\  \delta \left (1-y - \frac{2p_\xi\cdot p_R-p_\xi^2}{\hat s} \right)
\nn\\[2mm]
&  \times & {\rm Tr}_{\{j\}} \;  \langle 0|\, \bar T\left( w_L^{(ab\to cr)}{}^\dagger 
\left(0\right)_{\{j\}} \right)  |\xi \rangle \, \langle \xi | T \left(  w_I^{(ab\to cr)}(0)_{\{j\}} \right) |0\rangle\, ,\;\;\;\;\;\;
\label{eq:eik-cross section}
\eea
where $p_\xi$ is the momentum of state $|\xi\rangle$. Thus we define the eikonal soft function with exact 
eikonal matrix elements, but as a sum over the states $\ket{\xi}$, consisting of radiated gluons,
and at high orders, quark pairs. At NLO, for the soft function, $S^{(1)}$, we need only the single-gluon final state, $n=1$, for which the
condition (\ref{eq:phase space}) is equivalent to
\bea
k\cdot p_R \ =\ p_R^2/2\, ,
\eea
with $k$ the momentum of the radiated gluon.  Again, the momentum $p_4^{(h)}$ is introduced only to define the phase space; in the evaluation of the eikonal cross section at one loop below, only the recoilless ``eikonal" vector $p_4$ and the total recoil momentum $p_R$ appear.

The definition of the eikonal 
fragmentation 
jets is designed to match the phase space of the partonic recoil jets.  For $\beta_c$ the fragmentation direction, it is given by the 
same 
eikonal fragmentation function as for
the full jet functions in Eq.\ (\ref{eq:out-jet-def}) 
\cite{Hinderer:2014qta}
\bea
\tilde j_{\rm fr}^{(c)}\ =\  \tilde{j}_{\rm in}^{(c)}\, .
\label{eq:jout-c}
\eea
Similarly, the eikonal recoil jet is the eikonal analog of the partonic recoil jet extracted from Eq.~(\ref{eq:Sigma-def}), using the same phase space as in the eikonal cross section (\ref{eq:eik-cross section}) and is given by
\bea
  \tilde j_{\rm rec}^{(r)}\left( \frac{\hat s}{\bar{N}\mu_{rf}^2}, \frac{\hat s}{\bar{N}^2
\mu_{rf}^2},\as(\mu_{rf}),\varepsilon \right) 
&=&  \frac{1}{ \tilde j_{\rm fr}^{(\bar r)}\left( \frac{\hat s}{\bar{N}^2\mu_{rf}^2},
\as(\mu_{rf}),\varepsilon \right)\,}\nn\\[2mm]
&\times&\int_0^1 d y\;  y^{N-1}\ \sum_{\ket{\xi}}\  \delta \left (1-y - \frac{2p_\xi\cdot p_R-p_\xi^2}{\hat s}\, \right )
\nn\\[2mm]
&\times& {\rm Tr}_{\{j\}} \;  \langle 0|\, \bar T\left(w_0^{(r\bar r)}{}^\dagger \left(0\right)_{\{j\}} \right)  
|\xi \rangle \, \langle \xi |\, T \left( w_0^{(r\bar r)}(0)_{\{j\}} \right) |0\rangle\, ,
\label{eq:eik-out-jet}
\eea
with  $\tilde{j}_{\rm fr}^{(\bar r)}$ the same $\rm \overline{MS}$ distribution as in Eq.\ (\ref{eq:jout-c}), whose flavor is defined by the parton initiating the recoil jet.   

\subsection{One-loop soft functions}
\label{sec:calc-S1pi}

We are now ready to determine the finite soft function at one loop and beyond, taking the simplest case of the scattering of quarks of different flavors.   The calculation of soft matrices for the other partonic processes follows exactly the same pattern.  
As we discussed above, and as we shall see explicitly, 
the soft matrices for single-particle inclusive cross sections reflect the phase space of this process and differ from related cross sections with the same partonic channels, such as dihadron production, treated in a similar fashion in \cite{Hinderer:2014qta}.

The explicit calculation of $(S_{ab\to cr})_{LI}$ at one loop as given here is equivalent to the procedure described in Ref.\ \cite{Hinderer:2014qta}. The functions on the right side of (\ref{eq:soft-matrix}) are are written to one loop as
\bea
\hat \sigma^{ab\to cr}_{LI}  &=&
\big(\tilde{S}_0^{(0)}\big)^{ab\to cr}_{LI}\ +\ \frac{\alpha_s}{\pi}\,
\hat \sigma^{ab\to cr ,(1)}_{LI}\ + \ {\cal O}\big(\alpha_s^2\big)\, , \nn\\[2mm]
\tilde{ j} &=& 1\ +\ \frac{\alpha_s}{\pi}\,
\tilde{ j}^{(1)}\ +\ {\cal O}\big(\alpha_s^2\big)\, ,
\label{eq:lo-SLI-and-jets}
\eea
where $\tilde{S}^{(0)}_0=S^{(0)}_0$ is the moment-space tree-level soft matrix,  Eq.\ (\ref{eq:tilde-S-expand}) corresponding to the particular partonic process \cite{KS}, and where $\tilde{j}$ can be any of our eikonal jet functions $\tilde{j}_{\mathrm{in}},\tilde{j}_{\mathrm{frag}},
\tilde{j}_{\mathrm{rec}}$. 
In moment space, the collinear singularities of the eikonal cross section $\hat \sigma_{LI}$ are cancelled 
by
those of the incoming 
and outgoing jet functions, constructed as above. Expanding the soft function to first order, we have in moment space
\bea
\left(\tilde S_0^{(1)}\right)^{ab\to cr}_{LI}\ = \hat \sigma^{ab\to cr, (1)}_{LI}\ - \ \big(\tilde S_0^{(0)}\big)^{ab\to cr}_{LI} 
\left[ \,\sum_{i=a,b}\ \tilde j_{\rm in}^{(i),(1)}\ +\  \tilde{j}_{\rm fr}^{(c),(1)}\ +\ \tilde{j}_{\rm rec}^{(r),(1)} \,\right ]\, .
\label{eq:eik-diff}
\eea
   At one loop, this will result in a finite soft function by simple cancellation in Eq.\ (\ref{eq:eik-diff}),
   after renormalization of both $\hat \sigma^{ab\to cr}_{LI}$ and the jets.  
   That is, division by the  jet functions plays the role of the collinear factorization of the soft function.   
It also provides finite, factorizing corrections to the soft function, which depend in turn on the definitions of the jet functions.    
The eikonal cross section, and its in- and out-jet subtractions, are given by 
Eqs.\ (\ref{eq:eik-in-jet}), (\ref{eq:eik-cross section}), (\ref{eq:jout-c}) and (\ref{eq:eik-out-jet}).
The diagrammatic content of each of these functions at one loop is given by gluons emitted by eikonal lines in the directions that define the Wilson lines.   

The eikonal cross section at order $\as$ may be determined from the phase space integrals (defined here in Feynman gauge)
\bea
\frac{d I_{ij}}{d \bar y} \left (\bar y, v, \frac{\hat s}{\mu_{rf}^2} \right )\ =\ 
g_s^2\mu_{rf}^{2\varepsilon} \int \frac{d^D k}{(2\pi)^D}\ (2\pi)\,\delta_+(k^2) \; 
\delta\left (\bar{y} -\frac{2p_R\cdot k}{\hat s}\right)\ \frac{ \beta_i^\mu\,(-g_{\mu\nu}) \,\beta_j^\nu}{(\beta_i \cdot k)(\beta_j\cdot k)}\, ,
\label{eq:beta-ij-def}
\eea
with the momentum $p_R$ defined in Eq.\ (\ref{eq:pR-def}), $\bar y \equiv 1-y$, and the kinematic variable $v$ 
as defined in Eq.~(\ref{eq:v-def}). These integrals appear times color coefficients, which we label as ${\cal R}_{ij}$, that are 
color tensors and depend on which Wilson lines, $i,j=1, \dots ,4$ are connected by the emitted gluon.  
 Similar integrals are encountered in the treatment of threshold resummation for dihadron cross sections \cite{Hinderer:2014qta}, which extend over a different phase space.  
 
The momentum-space eikonal integrals $I_{ij}$ are, of course, the same in any representation for the Wilson lines, and are  given in $D=4-2\varepsilon$ dimensions by
\ba\label{eq:integrals1}
\frac{d I_{12}}{d \bar y}&=&-\frac{\as}{\pi}\left(\frac{4\pi\,{\mathrm{e}}^{-\gamma_E}\mu_{rf}^2}{\hat{s}}\right)^\varepsilon\left[\left(\frac{1}{\varepsilon^2}-\frac{\zeta(2)}{2}+\frac{1}{
\varepsilon}\ln(v(1-v)) +\frac{1}{2}\ln^2(v(1-v))\right) \delta(\bar y)\right. \nn\\[2mm]
&+& \left. \left(-\frac{2}{\varepsilon}-2\ln(v(1-v)) \right)\left(\frac{1}{\bar y}\right)_+ + 
4 \left(\frac{\ln(\bar y)}{\bar y}\right)_+\,\right]\,, \nn \\[2mm]
\frac{d I_{13}}{d \bar y}&=& -\frac{\as}{\pi}\left(\frac{4\pi\,{\mathrm{e}}^{-\gamma_E}\mu_{rf}^2}{\hat{s}}\right)^\varepsilon
\left[\left(\frac{1}{\varepsilon^2}-\frac{\zeta(2)}{2}-\frac{1}{
\varepsilon}\ln\left(\frac{1-v}{v}\right) +\frac{1}{2}\ln^2\left(\frac{1-v}{v}\right)\right) \delta(\bar y)\right. \nn\\[2mm]
&+& \left. \left(-\frac{2}{\varepsilon}+2\ln\left(\frac{1-v}{v}\right) \right)\left(\frac{1}{\bar y}\right)_+ + 
4 \left(\frac{\ln(\bar y)}{\bar y}\right)_+\,\right] \,,\nn \\[2mm]
\frac{d I_{23}}{d \bar y} &=& \frac{d I_{13}}{d \bar y} \big|_{v\leftrightarrow 1-v}\,,\nn\\[2mm]
\frac{d I_{14}}{d \bar y} &=&\frac{d I_{24}}{d \bar y} =\frac{d I_{34}}{d \bar y} \,=\,-\frac{\as}{\pi} 
\left(\frac{4\pi\,{\mathrm{e}}^{-\gamma_E}\mu_{rf}^2}{\hat{s}}\right)^\varepsilon
\left[\left(\frac{1}{\varepsilon^2}-\f{3}{2}\zeta(2)\right)\delta(\bar y)-\frac{1}{\varepsilon}\left(\frac{1}{\bar y}\right)_+ + 
\left(\frac{\ln(\bar y)}{\bar y}\right)_+ \,\right]\, .
\nn\\[0mm]
\ea
We note here that these integrals involve only logarithms of the kinematic variable $v$, rather than the dilogarithms found for dihadron threshold resummation \cite{Hinderer:2014qta}.  The explicit forms found here are, of course, necessary to reproduce singular threshold behavior at one loop and beyond.

We now illustrate the calculation of the soft matrix using these ingredients, specializing to our example for 
the partonic channel with quarks of different flavor, $qq'\to qq'$.
In this case, the Wilson lines are in the fundamental representation, and  the color-space matrices ${\cal R}_{ij}$ 
describe how single-gluon exchange mixes the couplings of the Wilson lines that represent soft radiation \cite{KOS}.   For definiteness, we choose a basis of $t$-channel color exchange between quarks of distinct flavors. The lowest-order soft matrix is independent 
of $\hat\eta$ and in this basis given by \cite{Kidonakis:2000gi}
\be\label{S0qqp}
\big(S_0^{(0)}\big)_{qq'\to qq'}\,=\,\left( \begin{array}{cc} \frac{C_A^2-1}{4} & 0 \\[2mm] 
0 & C_A^2 \end{array} \right) \,.
\ee
We note that the $S_0^{(0)}$ for all other partonic channels are collected in Appendix~\ref{sec:soft-matrices-explicit}.

With lines in the fundamental representation, the ${\cal R}_{ij}$ mix singlet and octet exchange.
Acting on the amplitude represented by vectors with the $t$-channel color singlet coupling of the Wilson lines in the first entry 
and the $t$-channel octet in the second,
they are given by
\ba
{\cal R}_{12}&=&{\cal R}_{34}\,=\,\frac{C_F}{2}
\left(
   \begin{array}{cc}
   -1& N_c \\[2mm]
 N_c & 0
   \end{array}
   \right)\;,\nn\\[2mm]
   {\cal R}_{13}&=&  {\cal R}_{24}\,=\, -\,\frac{C_F}{2}
 \left(
   \begin{array}{cc}
   -\frac{1}{2}& 0 \\[2mm]
0 & 2 N_c^2
   \end{array}
   \right)\;,\nn\\[2mm]
     {\cal R}_{14}&=&   {\cal R}_{23}\,=\, -\, \frac{C_F}{2}
\left(
   \begin{array}{cc}
   \frac{1}{2}(N_c^2-2)& N_c \\[2mm]
N_c & 0
   \end{array}
   \right)\;.
   \label{eq:R-matrices}
\ea
In these eikonal factors, the interference between initial- and final-state emission has a relative minus
sign which we 
exhibit here, changing the notation slightly from that of Ref.\ \cite{Hinderer:2014qta}.
Notice that the sum of the ${\cal R}_{ij}$ is proportional to $S^{(0)}$.   A corresponding result holds for all channels, and
follows from the gauge theory Ward identities. This ensures that at one loop double poles factorize and
cancel in Eq.\ (\ref{eq:eik-diff}).

Together with the $dI_{ij}/d \bar{y}$, these matrices define the eikonal cross 
section (\ref{eq:eik-cross section}) at one loop. Taking moments, we find 
\ba\label{R3mellinnew}
\hspace*{0.1cm}
\hat \sigma^{qq'\to qq' ,(1)}\ = \ \int_0^1 dy\,y^{N-1}\sum_{ij}\,{\cal R}_{ij}\, \frac{d I_{ij}}{d \bar y}
\nonumber\\[2mm]
&&\hspace{-6.4cm} = \ \left(\frac{4\pi\,{\mathrm{e}}^{-\gamma_E}\mu_{rf}^2}{\hat{s}}\right)^\varepsilon\,
\Bigg[\frac{2C_F}{\varepsilon^2}\,\tilde{S}_0^{(0)}-\f{1}{2\varepsilon}\,\Big[(\Gamma^{(1)})^{\dagger}\tilde{S}_0^{(0)}+
\tilde{S}_0^{(0)}\Gamma^{(1)}\Big]\nn\\[2mm]
&&\hspace*{-6.4cm}+\ \frac{C_F}{\varepsilon}
\,\big(\ln\bar N_a+\ln\bar N_b+\ln\bar N \big)\tilde{S}_0^{(0)} \nn \\[2mm]
&&\hspace*{-6.4cm} + \ C_F\left[\ln^2\bar N_a +  \ln^2\bar N_b+ \ln^2\bar N-\f{1}{2}
\ln^2\bar N\right]\tilde{S}_0^{(0)}-\, \ln\bar N\left(\big(\Gamma^{(1)}\big)^{\dagger}\tilde{S}_0^{(0)}
+\tilde{S}_0^{(0)}\Gamma^{(1)}\right) \nn\\[2mm]
&&\hspace*{-6.4cm}+\ \frac{C_F}{2}
    \left\{\left(
   \begin{array}{cc}
   -\frac{N_c^2-1}{4} \ln^2\left(v(1-v)\right)+2 \ln(1-v)\ln(v) & -2 N_c\ln(1-v)\ln(v) \\[2mm]
  -2 N_c\ln(1-v)\ln(v) & -N_c^2\ln^2\left(v(1-v)\right)
   \end{array}
      \right)+\zeta(2)\tilde{S}_0^{(0)}\right\}\Bigg],\nn\label{full1}\\
\ea
where as before $N_a=N v$, $N_b=N (1-v)$.  On the right-hand side of this relation, we have suppressed the subscript denoting the process, $qq'\to qq'$ for the soft function $\tilde{S}^{(0)}_0$ and the anomalous dimension matrix $\Gamma^{(1)}$, which is the 
first-order term of the matrix defined by Eq.\ (\ref{GammaSoft}). For the process at hand (and in fact for 
all quark-quark scattering processes), we have~\cite{Kidonakis:2000gi}
\be\label{eq:Gqqp}
\Gamma^{qq'\to qq',(1)}(\hat\eta)\,=\,\left( \begin{array}{cc}
-\frac{1}{C_A}\big(T+U\big)+2C_F U & 2 U \\[2mm]
\frac{C_F}{C_A}\,U & 2 C_F T \end{array}\right)\,,
\ee
where $T=\ln(1-v)+i\pi$, $U=\ln(v)+i\pi$.

Following Eq.\ (\ref{eq:soft-matrix}), to derive the soft function, we next need to divide 
$\hat \sigma^{qq'\rightarrow qq'}_{LI}$ by the eikonal jet functions. 
In computing the ratio, we need consider only real-gluon contributions.   
Virtual corrections to these scaleless integrals can be defined as pure counterterms, which cancel the infrared poles of the real contributions, as in Ref.\ \cite{Czakon:2009zw} for example.   All double poles also cancel in the ratio, leaving only a single pole, which can be considered part of the renormalization of the soft function. To order $\as$ the real-gluon contributions to the incoming 
eikonal jets defined in~(\ref{eq:eik-in-jet}) are, in momentum ($y$) space,
\bea
\tilde{j}_{\rm in,  real}^{(a)}\ = \ 1&+&\frac{\alpha_s}{2\pi}\left(\frac{4\pi\,{\mathrm{e}}^{-\gamma_E}\mu_{rf}^2}{\hat{s}}\right)^\varepsilon 
C_a\left[ \,4 \left(\frac{\ln(\bar y)}{\bar y}\right)_+ -\frac{2}{\varepsilon} \left(\frac{1}{\bar y}\right)_+
- 4\ln(v)\left(\frac{1}{\bar y}\right)_+ \right.\nn\\[2mm]
&+&\left.\delta(\bar y) \left(\frac{1}{\varepsilon^2}+2 \ln^2(v) - \frac{3}{2}\zeta(2)+\frac{2}{\varepsilon}\ln (v)\right)\right]\,,\nn\\[2mm]
\tilde{j}_{\rm in,  real}^{(b)}\ =\ 1&+&\frac{\alpha_s}{2\pi}\left(\frac{4\pi\,{\mathrm{e}}^{-\gamma_E}\mu_{rf}^2}{\hat{s}}\right)^\varepsilon 
C_b\left[ \,4 \left(\frac{\ln(\bar y)}{\bar y}\right)_+ -\frac{2}{\varepsilon} \left(\frac{1}{\bar y}\right)_+- 4\ln(1-v)\left(\frac{1}{\bar y}\right)_+ \right.\nn\\[2mm]
&+&\left.\delta(\bar y) \left(\frac{1}{\varepsilon^2}+2 \ln^2 (1-v) - \frac{3}{2}\zeta(2)+\frac{2}{\varepsilon}\ln (1-v)\right)\right]\, ,
\eea
with the kinematic variable $v$ defined as in Eq.\ (\ref{eq:v-def}), and as usual, $C_q=C_{q'}=C_F$, $C_g=C_A=N_c$.

For the observed (fragmenting) parton, defined as in Eq.\ (\ref{eq:jout-c}), we have at order $\as$
for the real-gluon contributions
\be\label{Jqeikobs}
\tilde{j}_{\rm fr,real}^{(c)}\,=\,1+\frac{\alpha_s}{2\pi}\left(\frac{4\pi\,{\mathrm{e}}^{-\gamma_E}\mu_{rf}^2}{\hat{s}}\right)^\varepsilon
C_c \left[\,   
4 \left(\frac{\ln(\bar y)}{\bar y}\right)_+   
-\frac{2}{\varepsilon}\left(\frac{1}{\bar y}\right)_+ +  \delta(\bar y) \left( \frac{1}{\varepsilon^2}\, -\, \frac{3}{2} \zeta(2) \right ) \,\right]\, ,
\ee
with $c=q$ or $g$.
Finally, to obtain the one-loop jet function for the unobserved recoiling parton, $r$ as defined in (\ref{eq:eik-out-jet}),
we subtract $\tilde{j}_{\rm fr,real}^{(q')}$ from  $dI_{34}/d\bar{y}$ in~(\ref{eq:integrals1}), which results in
\be\label{Jqeikrec}
\tilde{j}_{\rm rec,real}^{(r)}\,=\,1+\frac{\alpha_s}{2\pi}\left(\frac{4\pi\,{\mathrm{e}}^{-\gamma_E}\mu_{rf}^2}{\hat{s}}\right)^\varepsilon
C_r\left[\, -2 \left(\frac{\ln(\bar y)}{\bar y}\right)_+\, +\, \delta(\bar y)\left(\frac{1}{\varepsilon^2}-\frac{3}{2}\zeta(2)\right)
\,\right]\, ,
\ee
with $r=q$ or $g$.
Taking moments and dividing out the jet functions from~(\ref{R3mellinnew}) cancels all collinear and infrared poles, and we obtain
at ${\cal O}(\alpha_s)$ 
\ba\label{R3mellinnew2}
\int_0^1 dy\,y^{N-1}\sum_{ij}\,{\cal R}_{ij}\,dI_{ij}/ d\bar y-   \tilde{S}^{(0)}_0
\left(  \tilde{j}^{(q)}_{{\mathrm{in}}}\,\tilde{j}^{(q')}_{{\mathrm{in}}  }\,
\tilde{j}^{(q)}_{{\mathrm{fr}}}\,\tilde{j}^{(q')}_{\mathrm{rec}} \right )_{\mathrm{real}}^{(1)}
&=&\  -\left(\f{1}{2\varepsilon}+\ln\bar{N}\right) \Big[(\Gamma^{(1)})^{\dagger}\tilde{S}_0^{(0)}+\tilde{S}_0^{(0)}\Gamma^{(1)}\Big]
\nn\\[2mm]
&&\hspace*{-9.3cm}+\ \frac{C_F}{2}
    \Bigg \{ \left(
   \begin{array}{cc}
   -\frac{N_c^2-1}{4} \ln^2\left(v(1-v)\right)+2 \ln(1-v)\ln(v) & -2 N_c\ln(1-v)\ln(v) \\[2mm]
  -2 N_c\ln(1-v)\ln(v) & -N_c^2\ln^2\left(v(1-v)\right) 
   \end{array} \right)
      +2\zeta(2)\tilde{S}^{(0)}\Bigg \} \,.\nn\\[0mm]\label{full2}
\ea
After $\overline{{\mathrm{MS}}}$ renormalization, the terms proportional to the anomalous dimension matrix result from the evolution of the zeroth order soft function, and the remaining expression is the one-loop finite term of the soft function, 
as in Eq.\  (\ref{eq:tilde-S-expand}):
\beq\label{R3mellinnewqqp}
\tilde{S}^{(1)}_0\,=\,
\frac{C_F}{2}
    \left\{\left(
   \begin{array}{cc}
   -\frac{N_c^2-1}{4} \ln^2\left(v(1-v)\right)+2 \ln(1-v)\ln(v) & -2 N_c\ln(1-v)\ln(v) \\[2mm]
  -2 N_c\ln(1-v)\ln(v) & -N_c^2\ln^2\left(v(1-v)\right)
   \end{array}
      \right)+2\zeta(2)\tilde{S}_0^{(0)}\right\}\;,
\eeq
for $qq'\to qq'$.
According to Eq.\ (\ref{eq:ressoft}), to NNLL accuracy and beyond, this function will appear in the resummed hard scattering, Eq. (\ref{eq:omega-fact})  times $\as$ evaluated at scale 
$\sqrt{\hat s}/\bar{N}$. We note that
the first-order soft matrix may be cast into the form
\ba\label{Sid}
\tilde{S}^{(1)}_0&=&\tilde{S}_0^{(0)}\Bigg[\frac{1}{4}(C_a+C_b+C_c-C_r) \, \left( \ln^2\left(\frac{1-v}{v}\right)+2\zeta(2)
\right)-\,C_a\ln^2(v)-C_b\ln^2(1-v)\Bigg]\nn\\[2mm]
&&\hspace*{9mm}-\,2 \ln(1-v)\ln(v) \,{\cal R}_{12}\,,
\ea
where $C_a=C_b=C_c=C_r=C_F$, and with ${\cal R}_{12}$ as in Eq.~(\ref{eq:R-matrices}). 

The soft matrix constructed here, along with its matrix anomalous dimension, bears a close relation to the 
``wide-angle" soft function computed in the context of top production in Ref.\ \cite{Ferroglia:2013awa}.   
In fact, non-diagonal entries in the soft function are the same (in the same basis), but the soft matrices are not identical. 
Our evolved soft function acquires only a 
single logarithm of the moment variable per loop, while the soft function in \cite{Ferroglia:2013awa} has up to 
two logarithms per loop.   These double logs are color-transfer independent, however, and the difference is due to the different choices in soft-collinear factorization commonly made in SCET, compared to direct QCD.  This comparison is discussed, for example, in Ref.\ \cite{Lee:2006nr}.  
In particular, in the direct QCD treatment chosen here, all double logarithmic behavior is factorized into jet functions.   
The consistency of the non-trivial single-logarithmic evolution is another confirmation of the underlying consistency of the two treatments.

Soft functions for all partonic subprocesses are constructed in the same fashion, starting from eikonal cross sections, and dividing out eikonal jets.   Results are presented in 
Appendix~\ref{sec:soft-matrices-explicit}. We found that in each case, the result takes the form
given by Eq.~(\ref{Sid}), except that for processes with a $q\bar{q}^{\,(}{}'{}^{)}$ initial or final 
state, the sign of the last term needs to be reversed. 

\section{The Resummed Short-Distance Function and Moment Inversion \label{sec4}}
\label{sec:zspace}

\subsection{The resummed inclusive hard-scattering function in moment space}
\label{sec:hard-resum}

We are now ready to combine our previous results and present the resummed hard-scattering function, Eq.\ (\ref{eq:omega-fact}) in transform space, to leading power in the transform variable $N$.  

We will base inverse transforms of the expressions above on a hard scattering function 
written almost entirely in terms of logarithmic integrals. Indeed, this is the form on which our derivation in Sec.\ \ref{sec:resum} of the resummation is based.   
The expressions we need are Eqs.\ (\ref{eq:R-hat}) and (\ref{eq:J-in-soln2})  for the incoming jets, Eq.\ (\ref{eq:out-jet-def}) for the fragmentation jet, 
(\ref{Sigfactor}) and (\ref{eq:J-rec-MV}) for the recoil jet, and (\ref{eq:ressoft}),(\ref{GammaSoft}) for the soft matrix.  Explicit expressions for the anomalous dimensions are given in App.~\ref{AppB}, and for the one-loop soft functions and their anomalous dimension matrices in Appendix \ref{sec:soft-matrices-explicit}.

 We give the result for a specific choice of refactorization scale, $\mu_{rf}=\sqrt{\hat s}$.   This is a scale that simplifies existing expressions for the hard-scattering function \cite{Broggio:2014hoa}.   We comment below on alternative refactorization scale choices.

 In these terms, the full expression with $\mu_{rf}=\sqrt{\hat s}$, to NNLL, is a product of exponentials  associated with the jets, multiplied by a trace that links the hard and resummed soft matrices,  
\bea
\tilde{\omega}^{ab\to c}(\hat\eta,N)&=& \frac{\tilde{\Sigma}_r \left( 1, 1,\as\big(\sqrt{\hat s}\,\big) \right)}{ \hat R_{\bar r}\big(\as(\sqrt{\hat s})\big)}\,
\prod_{i=a,b,c}  \hat R_i\big(\as(\sqrt{\hat s})\big)\,  \nn
\\[2mm]
&&\hspace*{-2.4cm}\times\,
\prod_{i=a,b,c}\ \exp \left [ \int^{\sqrt{\hat{s}}}_{\sqrt{\hat s}/\bar{N}_i} 
\frac{d\mu}{\mu} \left(A_i\big(\as(\mu)\big)\, \ln\frac{\mu^2\bar N^2_i}{\hat{s}}\ -\, \frac{1}{2}\,
\hat D_i\big(\as(\mu)\big) \right) +\, \ln(\bar{N}_i)\int^{\mu_F^2}_{\hat s} 
\frac{d\mu^2}{\mu^2}  \, A_i (\as(\mu)) \right]
\nn\\[2mm]
&&\hspace*{-2.4cm} \times  \ \exp\ \Bigg [ -\, \int_{\sqrt{\hat s}/\bar{N}}^{\sqrt{\hat s/\bar{N}}}
\frac{d\mu}{\mu }\ \, A_r(\as(\mu))\ \ln \left( \frac{\mu^2{\bar{N}}^2}{\hat s} \right)\
  \nn\\[2mm]
& & \hspace{-8.2mm}
+\,  \int^{\sqrt{\hat s}}_{\sqrt{\hat s/\bar{N}}}
\frac{d\mu}{\mu } \, \left(A_r(\as(\mu))\ \ln \left( \frac{\mu^2}{\hat s} \right)\ - \, 2\,   \hat B_r \big(\as(\mu) \big)\right)
\, +\,  \frac{1}{2} \int_{\sqrt{\hat s}/\bar N}^{\sqrt{\hat s}} \frac{d\mu}{\mu} \hat D_r \big(\as(\mu) \big)   \Bigg]\nn
\\[2mm]
&&\hspace{-2.4cm}\times \
{\mathrm{Tr}} \left[ H  \left(\as(\sqrt{\hat s}),1, \frac{\mu_F}{\sqrt{\hat s}},\hat\eta
\right)\, \left\{ {\cal P}\exp\left[  \int^{\sqrt{\hat s}/\bar{N}}_{\sqrt{\hat{s}}}
\frac{d\mu}{\mu} \ \Gamma_{ab\to cr} 
\left(\hat{\eta},\as(\mu) \right)\right]  \right\}^\dagger\right.
\nn\\[2mm]
&& \left.\hspace{-1.2cm} \times\ \tilde {S}\left(1, \as \Big(\sqrt{\hat s}/\bar{N} \Big),\hat{\eta} \right)\, 
 {\cal P}\exp\left[ \int^{\sqrt{\hat s}/\bar{N}}_{\sqrt{\hat{s}}}
\frac{d\mu}{\mu} \ \Gamma_{ab\to cr} 
\left(\hat{\eta},\as(\mu) \right)\right]\, \right] \, .
\label{eq:omega-fact-res-2}
\eea
Here, the kinematically-rescaled moment variables $N_a$ and $N_b$ for the incoming partons are given by 
$N_a=N(-\hat u/\hat s),\, N_b=N(-\hat t/ \hat s)$ (see Eq.\ (\ref{eq:Ni-defs})), and we define $N_c\equiv N$. 

In practical applications of the result in Eq.\ (\ref{eq:omega-fact-res-2}) we will typically need an 
expansion in terms of the coupling at a fixed scale, $\mu_R$. For instance, matching to fixed-order calculations,
we will write the hard function in the form
\bea\label{HCexp}
H \left ( \as(\mu_{rf}),\frac{\sqrt{\hat{s}}}{\mu_{rf}} ,\frac{\mu_F}{\mu_{rf}},  \hat\eta  \right)
\ &=&\ 
C^{(2)} \left(\hat \eta \right) \left( \frac{\as(\mu_R)}{\pi} \right)^2 \ +\ 
\sum_{n \ge 3}  C^{(n)} \left ( \hat \eta, \frac{\mu_R}{\sqrt{\hat{s}}}, \frac{\mu_{rf}}{\sqrt{\hat s}},\frac{\mu_F}{\sqrt{\hat s}} \right)\, 
\left( \frac{\as(\mu_R)}{\pi} \right)^n\, ,
\nn\\
\eea
in which the refactorization scale appears in coefficients.   These coefficients have up to two logs 
in $\mu_{rf}/\sqrt{\hat s}$ per loop, because of the renormalization group equation satisfied by the hard function, Eq.\ (\ref{eq:H-evolve-2}).  Additional single-logarithmic terms appear from the reexpansion of the running coupling.   Dependence on the renormalization scale $\mu_R$ now begins, as usual, at the next uncalculated order.   As noted above, the factorization scale dependence of the hard matrix is known order-by-order 
through Eq.\ (\ref{HmuF}).  For our choice of $\mu_{rf}=\sqrt{\hat s}$, we may use the notation,
\bea\label{Hformnew}
H \left ( \as(\mu_{rf}),\frac{\sqrt{\hat{s}}}{\mu_{rf}} ,\frac{\mu_F}{\mu_{rf}},  \hat\eta  \right) \bigg|_{\mu_{rf}=\sqrt{\hat s}}
&\equiv& H \left(\as(\mu_R),\frac{\mu_R}{\sqrt{\hat s}}, \frac{\mu_F}{\sqrt{\hat s}},\hat\eta \right )\, .
\eea
The general form of these hard functions is, of course, similar to 
that given in Ref.\ \cite{Hinderer:2014qta} for dihadron cross sections, but with different final-state jets and soft functions, as developed above.   Specialized to the case of prompt photon production ($c=\gamma$), these results extend the NLL treatments given 
previously in the literature~\cite{Catani:1998tm,Kidonakis:1999hq,Sterman:2000pt}.
At NNLL (and beyond) they are also consistent with the soft-collinear analysis for prompt photons in Ref.\ \cite{Becher:2009th}, as we discuss in Sec.\ \ref{sec:N-indep}  below, with similarities in moment space as explored for Drell-Yan in Refs.\ \cite{Forte:2002ni,Sterman:2013nya}.  
 
In much the same way as Eq.\ (\ref{HCexp}), the resummed exponents in Eq.\ (\ref{eq:omega-fact-res-2}) will
also start to depend on a renormalization scale $\mu_R$ in fixed-order expansions. In fact, even when keeping the 
all-order resummed form of Eq.\ (\ref{eq:omega-fact-res-2}), dependence on a scale $\mu_R$ will arise in the usual 
``minimal expansions'' \cite{Catani:1996yz} of the exponents, despite the fact that the full exponents are strictly
independent of such a scale. This is because the perturbative expansion of the resummed exponents to a desired 
logarithmic accuracy necessarily truncates the perturbative series. The dependence of the resummed exponents on $\mu_R$ is explicitly 
seen in the expansions given in Eqs.~(\ref{EE}) and~(\ref{JJ}) in App.~\ref{AppC}. It is also the dependence that 
we will explore numerically in Sec.~\ref{secnumres}.

Equation (\ref{eq:omega-fact-res-2}) is the central result of this paper.   As observed above, we have given it for a specific choice of refactorization scale, $\mu_{rf}=\sqrt{\hat s}$, because this choice simplifies logarithms in the hard function, $H$.   Other choices are, of course, possible, leading to changes in the expression, and presumably, in numerical results.   The simplest modification is to vary $\mu_{rf}$ by a factor around $\sqrt{\hat s}$, say 
$\mu_{rf} = \zeta \sqrt{\hat s}$, with $1/2\le \zeta \le 2$.   In Eq.\ (\ref{eq:omega-fact-res-2}), this can be implemented by replacing $\sqrt{\hat s}$ by $\zeta\sqrt{\hat s}$ wherever it appears explicitly {\it without} a factor of $\bar N$.   Logarithms of $\zeta$ will also appear in the explicit expansion of the hard matrix $H$.    
Of course, variations can  also be implemented independently in both the hard and soft endpoints of each scale
integration in Eq.\ (\ref{eq:omega-fact-res-2}) \cite{Czakon:2018nun}.

A common choice for the factorization scale is $p_T$ itself.   The simplest implementation of this choice is to replace $\sqrt{\hat s}$ by $p_T$ in the definitions of the jet and soft functions in Sec.\ \ref{sec:refact-mtm-sp}, and in the phase space delta functions of Eq.\ (\ref{threshconv}).   The scale $p_T$ would then replace $\sqrt{\hat s}$ everywhere in Eq.\ (\ref{eq:omega-fact-res-2}).   Again, the fixed-order hard function would change as logarithms of $\sqrt{\hat s}$ are shifted to logs of $p_T$.   These shifts would be simple functions of the rapidity $\hat \eta$, or equivalently of $\ln v$ and $\ln (1-v)$.   For the purposes of this resummation formalism, none of these terms is taken to be parametrically large, although this depends to some extent on the process- and energy-dependent kinematics.   
A formalism with applicability to large partonic rapidity, as treated in electroweak annihilation in Ref.\ \cite{Banerjee:2018vvb}, will require further development for QCD hard scattering.

\subsection{Comparison to NLO}
\label{sec:compare}

With the complete short-distance functions in hand, we have all the ingredients necessary to compare to
the full one-loop calculations available in the literature~\cite{Aversa:1988vb,Jager:2002xm}.
Expanding the resummed partonic cross sections to NLO, we should recover 
all leading contributions that arise near partonic threshold in the full NLO cross
sections. These are all terms with distributions of the form  $(\ln(\bar{y})/\bar{y})_+$,
$1/\bar{y}_+$, and $\delta(\bar{y})$ where, as before, $\bar{y}=1-y$. This is a powerful
test of our resummation procedure. It is worth pointing out that we do not actually {\it use}
the NLO cross section to determine the $\delta(\bar{y})$ pieces in our resummed cross
section, but that we independently predict these pieces from our expressions
given above. We note that in~\cite{Hinderer:2014qta} we showed that the resummation 
formalism fully reproduces also the NLO hadron pair cross section near threshold. We now extend the comparison to 
single-inclusive production. This primarily tests our final-state recoil jet function and
our new one-loop soft function for this case. We use the simplest process $qq'\rightarrow qq'$
as an example and present the corresponding results for all other partonic channels in Appendix~\ref{otherC}.

Expanding the inclusive hard-scattering function in Eq.~(\ref{eq:omega-fact}) to first order we have,
again transforming to momentum space, and using the  form of the one-loop soft function, Eq.\ (\ref{R3mellinnew2}), after 
factorization
\ba\label{expandarb}
\omega^{ab\to c}(\hat\eta,y)&=&{\mathrm{Tr}}\left\{ H^{(0)} \, S_0^{(0)} \right\}\,
\left[\delta(\bar{y})\ + \ \frac{\alpha_s}{\pi}\left(J^{(a),(1)}_{{\mathrm{in}}}(y)+
J^{(b),(1)}_{{\mathrm{in}}}(y)+J^{(c),(1)}_{{\mathrm{fr}}}(y)+J^{(r),(1)}_{{\mathrm{rec}}}(y)\right)\right]\nn\\[2mm]
&+&\frac{\alpha_s}{\pi}\,
\left(\frac{1}{\bar{y}}\right)_+ {\mathrm{Tr}}\left\{ H^{(0)} \,\left(\Gamma^{(1)}\right)^\dagger\, S_0^{(0)}+
H^{(0)} \, S_0^{(0)}\,\Gamma^{(1)}  \right\}\nn\\[2mm]
&+&\frac{\alpha_s}{\pi}\,\delta(\bar{y})\,
{\mathrm{Tr}}\left\{ H^{(1)} \, S_0^{(0)} +H^{(0)} \, S_0^{(1)}  \right\}\,,
\ea
where on the right we have dropped the label $ab\to cr$ of the subprocess.
We now present explicit results for the various 
contributions in the above equation for the process $qq'\to qq'$. Using the results 
in~(\ref{Jab})--(\ref{Jrecg}) and
in~(\ref{S0qqp}), (\ref{R3mellinnewqqp}), and defining
\ba
L&=&\ln(v)\,,\nn \\[2mm]
\bar{L}&=&\ln(1-v)\,,
\ea
we have:
\bea
J^{(a),(1)}_{{\mathrm{in}}}(y)+
J^{(b),(1)}_{{\mathrm{in}}}(y)+J^{(c),(1)}_{{\mathrm{fr}}}(y)+J^{(r),(1)}_{{\mathrm{rec}}}(y)
&=&C_F \left[5 \left(\frac{\ln(\bar{y})}{\bar{y}}\right)_+
-\left(2 (L+\bar{L})+\frac{3}{4}\right)\left(\frac{1}{\bar{y}}\right)_+ \right.\nn\\[2mm]
&+&\left. \delta(\bar{y})\left(L^2+\bar{L}^2+
\frac{7}{4}-3\zeta(2)\right)\right]\,,
\eea
and
\bea\label{qqpexp}
&&{\mathrm{Tr}}\left\{ H^{(0)} \, S_0^{(0)}  \right\}\,=\,\frac{C_F}{C_A}\,\frac{1+v^2}{(1-v)^2}\,,
\nn\\[2mm]
&&{\mathrm{Tr}}\left\{ H^{(0)} \,\left(\Gamma^{(1)}\right)^\dagger\, S_0^{(0)}+
H^{(0)} \, S_0^{(0)}\,\Gamma^{(1)}  \right\}\,=\,\frac{2C_F}{C_A^2}\,\frac{1+v^2}{(1-v)^2}\,
\Big(-\bar{L}+(C_A^2-2)L\Big)\,,\nn\\[2mm]
&&{\mathrm{Tr}}\left\{ H^{(1)} \, S_0^{(0)} +H^{(0)} \, S_0^{(1)}  \right\} \,=\,\frac{C_F}{C_A}\,\Bigg[
\frac{C_F}{(1-v)^2}\Big(- L^2+v^2\bar{L}^2+(1-v) L-2(1+2 v^2) L\,\bar{L}\nn\\[2mm]
&&\hspace*{5.2cm}\left.+\,\frac{13}{9}(1+v^2) -\frac{1}{3}(5-3v+2 v^2)\bar{L}+\frac{\pi^2}{6}  (5+11v^2)\right)\nn\\[2mm]
&&\hspace*{5.2cm}+\,\frac{1}{4C_A(1-v)^2}\Big((1-v^2) L^2+(7+v^2)\bar{L}^2+2(5+7 v^2) L\,\bar{L}\nn\\[2mm]
&&\hspace*{5.1cm}\left.
-\,2(1-v) L+ (3+v^2)\pi^2+\frac{170}{9}(1+v^2)-\frac{2}{3}(8-3v+17 v^2)\bar{L}\right)\nn\\[2mm]
&&\hspace*{5.2cm}+\,\frac{N_f}{9}\,\big(-5+3 \bar{L}\big)\,\frac{1+v^2}{(1-v)^2}\Bigg]\,.
\eea
With these, Eq.~(\ref{expandarb}) reproduces the full NLO~\cite{Aversa:1988vb,Jager:2002xm} for 
$qq'\to qq'$ near threshold, including all $\delta(\bar{y})$ contributions. Appendix~\ref{otherC} collects
the expansions for all other partonic processes; we have checked that in each case the resummed formulas 
reproduce NLO near threshold.

\subsection{Resummed cross section with $N$-independent scales}
\label{sec:N-indep}

As noted above, in Ref.\ \cite{Becher:2009th} and other effective theory resummation studies, it is customary to use fixed, rather than $N$-dependent, infrared and ultraviolet scales in solutions to the evolution equations that generate threshold resummation.   We can do this here as well, which leads to an expression that generalizes the results of Ref.\ \cite{Becher:2009th} for prompt photons to final-state hadrons.
Following this approach, we replace $\sqrt{\hat s}/\bar{N}$ by a soft scale, $\mu_s$ for the
incoming and fragmenting jets and the soft function, and $\sqrt{\hat s/\bar{N}}$ by a jet scale, $\mu_j$ for the  recoil jet.     In the solutions, dependence on the moment variable $N$ is retained in the prefactors for the various jet functions, so that the jet anomalous dimensions $\gamma_J$ and $\gamma_\Sigma$ appear in the exponents rather than the alternative anomalous dimensions $D$ and $B$. 
 
The solutions for the incoming and fragmentation jets can be found from
Eq.\ (\ref{eq:soln-gen}),
and  then the  recoil jet from  (\ref{eq:C-soln}) and (\ref{Sigfactor}).
Using these for the resummed cross section we obtain, instead of Eq.\ (\ref{eq:omega-fact-res-2}),
\bea
\tilde{\omega}^{ab\to c}(\hat\eta,N) &=&
\prod_{i=a,b,c}  \tilde J_{\rm in}^{(i)}  \left( \frac {\hat s}{\bar N_i^2 \mu^2_s},1,\as(\mu_s)\right)\, 
\nn\\[2mm]
&&\hspace*{-2.6cm}\times\,
\exp \left \{ \,\sum_{i=a,b,c} \left[ \int^{\sqrt{\hat{s}}}_{\mu_s} 
\frac{d\mu}{\mu} 
\left(A_i\big(\as(\mu)\big)\, \ln\frac{\mu^2\bar N^2_i}{\hat{s}}\ -\, 
\gamma_{J^{(i)}}\big(\as(\mu)\big) \right) +\, \ln(\bar{N}_i)\int^{\mu_F^2}_{\hat s} 
\frac{d\mu^2}{\mu^2}  \, A_i \big(\as(\mu)\big) \right] \right\} 
\nn\\[2mm]
&\ & \hspace{-2.6cm} 
\times\ \frac{\tilde{\Sigma}_{r} \left( \frac{\hat s}{\bar{N}\mu_j^2}, 1,\as(\mu_j) \right) }{\tilde{J}_{\rm in}^{(\bar r)}  \left( \frac {\hat s}{\bar N^2 \mu^2_s},1,\as(\mu_s)\right)} \, 
\exp \Bigg[  
 \int_{\mu_j}^{\sqrt{\hat{s}}}\ \frac{d \mu}{\mu} \left( 2 A_r\big(\as(\mu)\big)\, \ln \frac{\mu^2\bar{N}}{\hat{s}}\,-\,\gamma_\Sigma^{(r)} \big(\as(\mu) \big) 
\right)
\nn\\[2mm]
&\ & \hspace{3 cm}
-\, \int_{\mu_s}^{\sqrt{\hat{s}}}\ \frac{d \mu}{\mu} \,
\left( A_r\big(\as(\mu)\big)\, \ln \frac{\mu^2\bar N^2}{\hat{s}}\ -\ \gamma_{J^{(r)}} \left(\as(\mu)\right) \right )\,  \Bigg]
\nn\\[2mm]
&\ &\hspace{-2.6cm}\times \
{\mathrm{Tr}} \left[ \,H  \left(\as(\sqrt{\hat s}),1, \frac{\mu_F}{\sqrt{\hat s}},\hat{\eta} 
\right)\, \left\{ {\cal P}\exp\left[  \,\int^{\mu_s}_{\sqrt{\hat{s}}} 
\frac{d\mu}{\mu} \ \Gamma_{ab\to cr} 
\left(\hat{\eta},\as(\mu) \right)\right]  \right\}^\dagger\right.
\nn\\[2mm]
&& \hspace{-1.37cm} \left. \times\ \tilde {S}\left(\frac {\hat s}{\bar N_i^2 \mu^2_s},\as(\mu_s),\hat{\eta} \right)\, 
 {\cal P}\exp\left[ \, \int^{\mu_s}_{\sqrt{\hat{s}}}
\frac{d\mu}{\mu} \ \Gamma_{ab\to cr} 
\left(\hat{\eta},\as(\mu) \right)\right] \,\right] \, .
\label{eq:omega-fact-res-3}
\eea
 For $N$-independent choices of $\mu_j$ and $\mu_s$, logarithms of $N$ do not all appear from the exponents \cite{Sterman:2013nya}.   Again, we recall that $\bar{N}_c=\bar{N}$ here.

For compactness and ease of comparison to the literature, we introduce notation in the spirit, if not the exact letter, of that found in the treatment of prompt photon cross sections in Ref.\ \cite{Becher:2009th},
\bea
C_i\, S(\mu_1,\mu_2)\ &\equiv&\ -\  \int^{\mu_2}_{\mu_1} \frac{d\mu}{\mu} \ A_i\big(\as(\mu)\big)\, \ln\frac{\mu}{\mu_1}\, ,
\nn\\[2mm]
C_i\, A_\Gamma(\mu_1,\mu_2)\ &\equiv&\ -\ \int^{\mu_2}_{\mu_1} \frac{d\mu}{\mu} \ A_i\big(\as(\mu)\big)\, ,
\nn\\[2mm]
A_{\gamma}(\mu_1,\mu_2)\ &\equiv&\ -\  \int^{\mu_2}_{\mu_1} \frac{d\mu}{\mu}  \ \gamma\big(\as(\mu)\big)\, ,
\label{eq:expon-defs}
\eea
where the final definition applies to $\gamma=\gamma_\Sigma^{(r)},\gamma_{J^{(i)}}$.
(As a notational point, the ``Sudakov" factor $S(\mu_1,\mu_2)$ should not be confused with the soft function.)
In this notation, the moment-space resummed hard-scattering function Eq.\ (\ref{eq:omega-fact-res-3}) becomes
\bea
\tilde{\omega}^{ab\to c}(\hat\eta,N)&=&
\prod_{i=a,b,c}  \tilde J_{\rm in}^{(i)}  \left( \frac {\hat s}{\bar N^2 \mu^2_s},1,\as(\mu_s)\right)\ \exp \Bigg [ -2 \left(\, \sum_{i=a,b,c}  C_i \ln \bar N_i \right ) A_\Gamma(\sqrt{\hat s},\mu_F)  \Bigg ]
\nn\\[2mm]
&&\hspace*{-2.5cm}\times \ 
\exp \Bigg[   \sum_{i=a,b,c} \left( - 2C_i \,S( \mu_s,\sqrt{\hat{s}}) - C_i \,A_\Gamma(\mu_s,\sqrt{\hat s}) \left( \ln \frac{\mu_s^2 \bar N^2}{\hat s} +  \ln \frac{\bar N_i^2}{\bar N^2}\right)  + 
A_{\gamma^{J^{(i)}}}(\mu_s,\sqrt{\hat s}) 
\right)  \Bigg ]
  \nn\\ [2mm]
&&\hspace*{-2.5cm}\times \ 
\exp \Bigg[ \  2C_r \,S( \mu_s,\sqrt{\hat{s}}) + C_r \,A_\Gamma(\mu_s,\sqrt{\hat s})  \ln \frac{\mu_s^2 \bar N^2}{\hat s} -  A_{\gamma^{J^{(r)}}}(\mu_s,\sqrt{\hat s})   \Bigg ]  
\nn\\ [2mm]
&&\hspace*{-2.5cm}\times \ 
\frac{\tilde{\Sigma}_{r} \left( \frac{\hat s}{\bar{N}\mu_j^2}, 1,\as(\mu_j) \right) }{ J_{\rm in}^{(\bar r)}  \left( \frac {\hat s}{\bar N^2 \mu^2_s},1,\as(\mu_s)\right)} \, 
\exp \Bigg[  \ - 4C_r \,S(\mu_j,\sqrt{\hat{s}}) -   2C_r \ln \frac{\mu_j^2\bar N}{\hat s} A_\Gamma(\mu_j,\sqrt{\hat s}) \,  
+ \, A_{\gamma_\Sigma}(\mu_j,\sqrt{\hat s}) \Bigg]
\nn\\[2mm]
&\ &\hspace{-2.5cm}\times \
{\mathrm{Tr}} \left[ \,H  \left(\as(\sqrt{\hat s}),1, \frac{\mu_F}{\sqrt{\hat s}},\hat{\eta} 
\right)\, \left\{ {\cal P}\exp\left[  \,\int^{\mu_s}_{\sqrt{\hat{s}}} 
\frac{d\mu}{\mu} \ \Gamma_{ab\to cr} 
\left(\hat{\eta},\as(\mu) \right)\right]  \right\}^\dagger\right.
\nn\\[2mm]
&& \hspace{-1.25cm} \left. \times\ \tilde {S}\left(\frac {\hat s}{\bar N_i^2 \mu^2_s},\as(\mu_s),\hat{\eta} \right)\, 
 {\cal P}\exp\left[ \, \int^{\mu_s}_{\sqrt{\hat{s}}}
\frac{d\mu}{\mu} \ \Gamma_{ab\to cr} 
\left(\hat{\eta},\as(\mu) \right)\right] \,\right] \, .
\label{eq:omega-fact-res-4}
\eea
When specialized to prompt photon production, this result 
is consistent with Ref.\ \cite{Becher:2009th}, with a different organization of logarithms in the moment variable compared to Eq.\ (\ref{eq:omega-fact-res-2}).
We now turn to inversions of these transforms to find the leading threshold behavior.

\subsection{The Mellin inverse \label{secmell}}

To apply any of the moment-space resummations given above for phenomenological applications, we must perform a Mellin inverse.   The most direct approach is found from Eq.\ (\ref{eq:Omega-inverse}),
\be
\Omega^{ab\to c,\mathrm{resum}}(\hat\eta,z)\,\equiv\,
\frac{1}{2\pi i}\int_{\cal C}dN \,z^{-N}\,\tilde{D}_c^h(N+3)\,\tilde{\omega}^{ab\to c,{\mathrm{resum}}}(\hat\eta,N)\, ,
\label{eq:Omega-again}
\ee
with a suitable contour ${\cal C}$.
This result is then convoluted with the parton distributions to give (see Eq.~(\ref{eq:Omega-eq-1}))
\be
p_T^3\,\frac{d^2\sigma^{\mathrm{resum}}}{dp_Td\eta}\,=\, \sum_{abc}\int dx_a \,f_a(x_a) \int dx_b \,f_b(x_b) \,
\Omega^{ab\to c,\mathrm{resum}}\left(\hat\eta,z=\frac{x_T}{\sqrt{x_ax_b}}\cosh\hat\eta\right)\,.
\label{eq:Omega-ff}
\ee
In principle, because the moments of the fragmentation functions fall off quite 
rapidly, the numerical inversion of the moments will allow the integration
against the parton distributions to be carried out numerically.    This is analogous to the procedure in Ref.\ \cite{Hinderer:2014qta} for the threshold resummation of cross sections at fixed invariant mass and rapidity difference, although in that case the roles of the fragmentation functions and parton densities  in Eqs.\ (\ref{eq:Omega-again}) and (\ref{eq:Omega-ff}) are reversed.    
For dihadrons at NNLL, this worked well by defining the inverse transforms like Eq.\ (\ref{eq:Omega-again}) following the ``minimal" definition of Ref.\ \cite{Catani:1996yz}. 

In the ``minimal" procedure, the contour ${\cal C}$  in the complex $N$-plane is chosen to cross the real axis between the origin and the branch cut associated with the leftmost Landau singularity, which occurs in the resummed exponent (\ref{eq:omega-fact-res-2}) when $\bar N=\sqrt{\hat s}/\Lambda_{\rm QCD}$.   The presence of this singularity allows for power-suppressed but nonzero contributions in the unphysical region $z>1$, where for dihadrons, $z=M^2_H/\hat{s}$, with $M_H$ the dihadron mass.   These contributions were indeed numerically small for the phenomenologically-relevant kinematics studied in Ref.\ \cite{Hinderer:2014qta}.

Sample evaluations, however, show that single-particle inclusive cross sections are much less stable against contributions with unphysical origin ($z>1$), even though they are still power-suppressed.   This increased sensitivity can enter because once $z$ is greater than unity, the partonic fractions $x_a$ and $x_b$ in Eq.\ (\ref{eq:Omega-ff}) can become unphysically small (see Eq.\ (\ref{eq:z-def})).   Through Eq.\ (\ref{eq:hat-eta-def}), this 
leads in turn to very large values of partonic rapidity $\hat \eta$, which affects the limits of integration in the resummed exponents, as well as the soft functions, through their anomalous dimensions.  We know of no arguments that limit the normalization of such unphysical contributions.

Another  approach to using the resummed inclusive hard-scattering function (\ref{eq:omega-fact-res-2}) avoids the Landau pole, and unphysical contributions to the inverse transform~\cite{Becher:2006nr}.   In this approach, usually adopted in soft-collinear effective theory treatments of resummation, we
 replace the $N$-dependent solutions to the evolution equations for the soft and jet functions by $N$-independent 
 jet and soft scales: $\mu_j$ and $\mu_s$,
as discussed in the previous subsection.
Given such choices, the moment inversion of $\tilde \omega^{ab\to c}$ 
can be done explicitly, using only the identity
\bea
\int_{\cal C} \frac{dN}{2\pi i}\, z^{-N}\ {\mathrm{e}}^{-\zeta\, \ln \bar{N}}\
=\ \frac{{\mathrm{e}}^{-\gamma_E \zeta}\,(1-z)^{\zeta-1}}{\Gamma(\zeta)} \, , 
\eea
where ${\cal C}$ is a again a contour in the $N$-plane that can now be taken to the right of all singularities, and where in this form 
there are $1/N$ or ${\cal O}(1-z)$ corrections.  
In this analysis, $\zeta$ represents a sum of integrals over any anomalous dimensions between the 
 chosen soft or jet and the hard scale.  
Explicit $N$-dependence in the soft function is accounted for through derivatives with respect to the variable $\zeta$.   In the general case, logarithms of $N$ appear in soft and jet functions, and one uses
\bea
 \int_{\cal C} \frac{dN}{2\pi i}\ z^{-N}\  
 \widetilde F \left( \ln \frac{M^2}{\bar N^2\mu_i^2}, \as(\mu_i) \right)\, {\mathrm{e}}^{-\eta \ln \bar N^2}
\,=\,
\widetilde F \left(\ln \frac {M^2}{ \mu_i^2} +\frac{\partial}{\partial \eta}, \as(\mu_i) \right)\  
\frac {{\mathrm{e}}^{-2\gamma_E \eta}\, (1-z)^{2\eta-1}} {\Gamma\left( 2\eta \right ) }\, .
\nn\\
\label{eq:inverse-fixedscales}
\eea
This is readily applied to Eq.\ (\ref{eq:omega-fact-res-4}).   Defining the 
negative of the sum of $\ln \bar{N}^2$ coefficients in the exponent as 
\bea
\eta'\ \equiv \  \sum_{i=a,b,c} C_i\, A_\Gamma(\mu_s,\mu_F)\ -\ C_r\, A_\Gamma(\mu_s,\mu_j)\, ,
\label{eq:eta-sum}
\eea
and recalling the definitions of the $\bar N_i$ in Eq.\ (\ref{eq:Ni-defs}), we find from Eq.\ (\ref{eq:omega-fact-res-4})
\bea
{\omega}^{ab\to c}(\hat\eta,y)&=&
\left( \frac{-\hat u}{\hat s}\right)^{-2C_aA_\Gamma(\mu_s,\sqrt{\hat s})}
\left( \frac{-\hat t}{\hat s}\right)^{-2C_bA_\Gamma(\mu_s,\sqrt{\hat s})}
\left( \frac{\mu_s^2}{\hat s}\right)^{-[ \sum\limits_{i=a,b,c}C_i - C_r] A_\Gamma(\mu_s,\sqrt{\hat s})}
\left( \frac{\mu_j^2}{\hat s}\right)^{ -\, 2 C_r A_\Gamma(\mu_j,\sqrt{\hat s})}
\nn\\ [2mm]
& \times& \exp \Bigg[  -2\left(\,\sum_{i=a,b,c} C_i \,-\, C_r\right) S(\mu_s, \sqrt{\hat{s}}) \, -\, 4 C_r S(\mu_j, \sqrt{\hat{s}}) \Bigg]\, 
  \nn\\ [2mm]
&\times&  \exp \Bigg[  \,     \sum_{i=a,b,c}\,   A_{\gamma_{J^{(i)}}}(\mu_s,\sqrt{\hat s})\, -\, A_{\gamma_{J^{(r)}}}(\mu_s,\sqrt{\hat s})\,+\, A_{\gamma_\Sigma}(\mu_j,\sqrt{\hat s})    \Bigg]
  \nn\\ [2mm]
&\times& \prod_{i=a,b,c}  \tilde J_{\rm in}^{(i)}  \left( \frac {\hat s}{\mu^2_s} +  \frac{\partial}{\partial \eta'} ,1,\as(\mu_s)\right)\ 
\tilde{\Sigma}_r \left( \frac{\hat s}{\mu_j^2} + \frac{\partial}{\partial \eta'}, 1,\as(\mu_j) \right)\, 
\tilde  J_{\rm in}^{(\bar r)}\,{}^{-1}  \left( \frac {\hat s}{\mu^2_s} +  \frac{\partial}{\partial \eta'} ,1,\as(\mu_s)\right)
\nn\\[2mm]
&\times&
{\mathrm{Tr}}  \Bigg[ H  \left(\as(\sqrt{\hat s}),1, \frac{\mu_F}{\sqrt{\hat s}},\hat{\eta} 
\right)\,  \left\{ {\cal P}\exp\left[  \,\int^{\mu_s}_{\sqrt{\hat{s}}} 
\frac{d\mu}{\mu} \ \Gamma_{ab\to cr} 
\left(\hat{\eta},\as(\mu) \right)\right]  \right\}^\dagger
 \,\nn\\[2mm]
&& \hspace*{0.68cm}\times \;\tilde {S}\left( \frac{\hat s}{ \mu_s^2} +  \frac{\partial}{\partial \eta'}, \as(\mu_s) ,\hat{\eta} \right)\, 
 {\cal P}\exp\left[ \, \int^{\mu_s}_{\sqrt{\hat{s}}}
\frac{d\mu}{\mu} \ \Gamma_{ab\to cr} 
\left(\hat{\eta},\as(\mu) \right)\right]  \Bigg]
 \nn\\[2mm]
&\times & \frac{{\mathrm{e}}^{-2\gamma_E \eta'}}{\Gamma(2\eta')}\, (1-y)^{2\eta'-1}
\ + \ {\cal O}\left((1-y)^0\right)\, .
\label{eq:omega-fact-res-5}
\eea
In this form, if we set
$\mu_F=\sqrt{\hat s}$, we can reduce to the case of prompt photon emission, $c=\gamma$, and compare to the effective theory treatment of Ref.\ \cite{Becher:2009th}, taking $\mu=\sqrt{\hat s}$ in their notation.   To do so, we only need the anomalous dimensions associated with the 
two partonic processes, pair annihilation and QCD Compton scattering, found in Ref.\ \cite{Laenen:1998qw}.   We find that the two expressions are consistent in explicit dependence on the soft and jet scales, Sudakov factors and other anomalous dimensions, the relevant sums of which are equal at least to two loops.

 In closing this discussion, we note that the advantages of both of these well-explored approaches may be incorporated by a hybrid choice, already suggested by the analysis presented in Ref.\ \cite{Shimizu:2005fp}, given, for example, by
    \bea
\mu_s(N) = \sqrt{\hat{s}}/\bar N + \mu' \, ,
\label{eq:mu-noz>1}
\eea
for the soft scale, with $\mu'$ larger than $\Lambda_{\rm QCD}$.
   This is just the sum of conventional direct QCD and SCET boundary conditions, differing significantly from a ``pure" $\sqrt{\hat{s}}/\bar{N}$ choice only when $N$ is very large, where it bottoms out at a ``jet" or ``soft" scale $\mu'$ in the SCET language.   
   The ``Landau'' branch point is now moved to a large negative position 
$\bar N_L = -\sqrt{\hat{s}}/(\mu' - \Lambda_{\rm QCD} )$.   Qualitatively, a singularity at this position affects the inverse transform to variable $z$ 
as an additive contribution that scales as
$z^{-\sqrt{\hat{s}}/\mu'}$.
The influence of this nonperturbative singularity is thus suppressed exponentially.
This is more or less the same power structure as the Landau pole in standard ``minimal resummation" \cite{Catani:1996yz}, but now without contributions for $z > 1$, and with the non-perturbative singularity far from the origin to the left of the contour.   
We shall not pursue phenomenological implementations of this or other methods to invert the transform of our resummed expressions in momentum space here.
We anticipate exploring a formalism using  Eq.\ (\ref{eq:mu-noz>1}) and other possibilities in future work.
For now we will restrict our numerical analysis to a brief exploration of fixed-order  expansions 
and scale dependence 
of the 
resummed cross section, setting aside $z>1$ contributions associated with the Landau pole. 

\subsection{Numerical results for fixed-order expansions \label{secnumres}}
 
We have seen in Sec.\ \ref{sec:compare} that the order $\alpha_s$ expansion of the resummed cross section reproduces the singular behavior at threshold for each partonic channel.  A natural first test is to compare the numerical result of the expanded resummation formula to the full NLO with realistic choices of kinematics and parton distributions and fragmentation functions. In addition, given that
resummation provides insight into the size of beyond-NLO effects, we will also explore expansions to NNLO.

In the following, we will consider neutral-pion production $pp\to \pi^0X$ at two center-of-mass energies,
$\sqrt{s}=31.5$~GeV and $\sqrt{s}=200$~GeV. The former corresponds to one of the energies used
in the Fermilab fixed-target experiment E706~\cite{Apanasevich:2002wt}, while the latter is relevant
for experiments at RHIC. Although we will not present any actual comparisons to data, we adopt
the proper kinematics for these two cases, integrating over pseudorapidity $|\eta|\leq 0.75$ in the 
E706 case and over $|\eta|\leq 0.35$ 
for RHIC, corresponding to measurements by the PHENIX collaboration~\cite{Adare:2007dg},
and considering the cross sections as functions of pion transverse momentum, $p_T$. We will use
the CT14 parton distribution functions~\cite{Dulat:2015mca} as implemented in the LHAPDF database~\cite{Buckley:2014ana}.
CT14 provides both NLO and NNLO sets of parton distributions, which is useful for our expansions. 
We compute LO and NLO cross sections with the NLO set of parton distributions, and the NNLO 
expansions with the NNLO ones. 
For the $\pi^0$ fragmentation functions we use the set of~\cite{deFlorian:2014xna}, which is at NLO.
As discussed above, we need Mellin moments of the fragmentation functions, whereas the set of~\cite{deFlorian:2014xna}
is available as a numerical code in $z$-space. Technically, we obtain the moments by performing a 
fit to each of the fragmentation functions for a given set of scales. The functional form of the fit is chosen such 
that one can easily take its Mellin moments. We have checked that this approach works to about $1\%$ accuracy
in the kinematic regimes of interest to us. 

As discussed after Eq.~(\ref{HCexp}) and shown by Eqs.~(\ref{EE}) and~(\ref{JJ}) in App.\ \ref{AppB}, the explicit NNLL expansions of the jet functions induce dependence on a renormalization scale $\mu_R$. 
As is usually done, we will choose $\mu_R$ and $\mu_F$ of order $p_T$.

\begin{figure}[t!]
\vspace*{-6cm}

\hspace*{-0.8cm}
\epsfig{figure=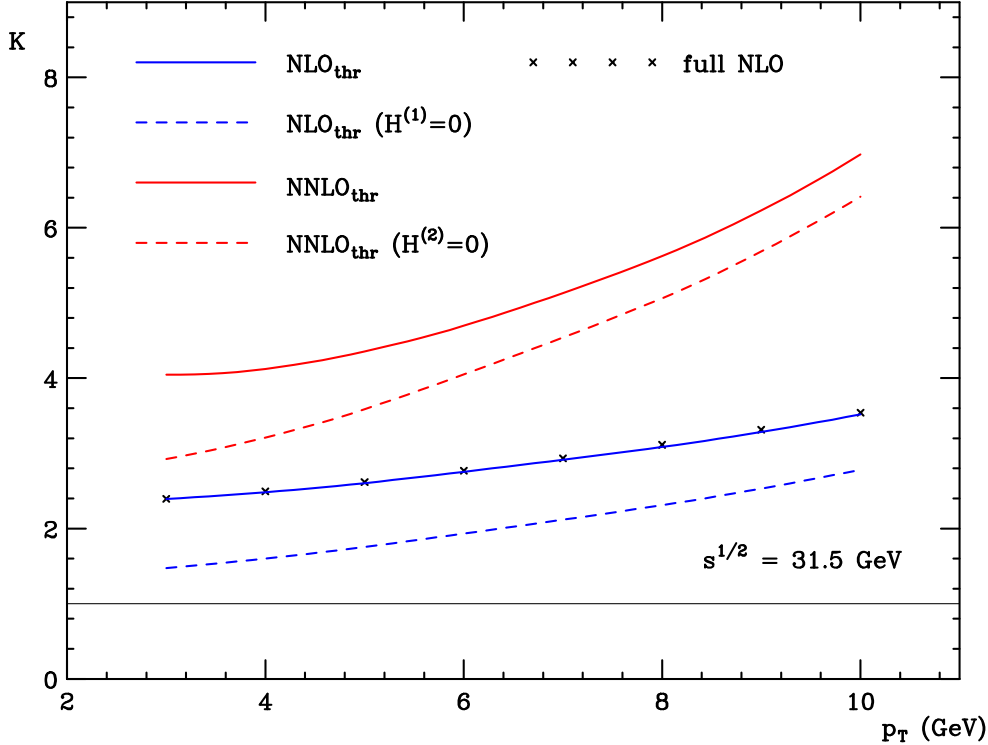,width=1\textwidth}

\vspace*{-6.4cm}
\caption{{\it NLO and NNLO $K$-factors for $pp\to \pi^0X$ for E706 kinematics with $\sqrt{s}=31.5$~GeV. 
All results are normalized to the LO cross section. We have chosen $\mu_F=\mu_R=p_T$.}}
\label{fig:fixed-order}
\end{figure}

Figure~\ref{fig:fixed-order} shows various ``$K$-factors'' for $pp\to \pi^0X$ with E706 kinematics with
$\mu_F=\mu_R=p_T$, where
\be
K\,\equiv\,\frac{d\sigma^{\mathrm{N^kLO}}}{d\sigma^{\mathrm{LO}}}\,,
\ee
with $k=1,2$. The crosses show the $K$-factor corresponding to the full NLO result of Ref.~\cite{Jager:2002xm}.
We see that the NLO $K$-factor is large, exceeding 
$2$ throughout the $p_T$ regime considered.  
The lower solid line shows the NLO expansion of the resummed cross section. 
The agreement of the NLO expansion with full NLO is excellent, 
at about $2\%$ or better. This is an improvement over
previous comparisons for the rapidity-integrated cross sections given in Ref.\ \cite{deFlorian:2005yj}. It provides
confidence that resummation indeed captures the dominant contributions to the cross section and
motivates the study of higher-order expansions of the resummed cross section as a method of
obtaining accurate results beyond NLO. 

The upper dashed line in Fig.~\ref{fig:fixed-order} presents such an expansion, to NNLO.
All the new ingredients at NNLL that we have derived in the previous sections become relevant here. 
For now, in the upper dashed line we truncate the expansion of the hard function $H_{ab\to cr}$ 
after its ${\cal O}(\alpha_s)$ term $H^{(1)}_{ab\to cr}$, although we include the scale logarithms in the hard 
function through second order. 
Evidently, the expansion to higher orders leads to further sizable enhancements, especially at
high transverse momentum where the threshold logarithms become more and more sizable. 
This result is in line with what was found in Ref.\ \cite{deFlorian:2005yj}. 

An interesting observation is that, although the threshold logarithms provide a large part of the 
enhancements seen at NLO, the one-loop hard functions $H^{(1)}_{ab\to cr}$ are numerically very important
as well. This is shown by the lower dashed line, which again presents the NLO expansion of the resummed cross section,
but this time without the contributions by the $H^{(1)}_{ab\to cr}$. Specifically, in the notation of Eq.~(\ref{Hformnew})
we set the ${\cal O}(\alpha_s)$ correction in $H_{ab\to cr}(\alpha_s,1,1,\hat\eta)$ to zero, but keep the scale logarithms of 
$H_{ab\to cr}$ that arise at that order. Clearly the result is much lower and falls
short of the full NLO result. We recall that in Mellin-space the $H^{(1)}_{ab\to cr}$ appear as constant pieces 
in the NLO cross section, corresponding to contributions $\propto \delta(\hat{s}_4)$ in the cross section in physical space.
The only other sources of such terms are the one-loop soft functions $S^{(1)}_{ab\to cr}$ and the normalizations
$R_i$ and $\tilde{\Sigma}_r$ in the jet functions in Eqs.~(\ref{eq:in-jet}) 
and~(\ref{eq:Sigma-resum}), respectively. These turn out to be numerically insignificant in comparison to 
the $H^{(1)}_{ab\to cr}$. 

Given the importance of the $\delta(\hat{s}_4)$ contribution for obtaining a good agreement between
the NLO expansion of the resummed cross section and full NLO, one may wonder how accurately the NNLO 
expansion shown by the upper dashed line in Fig.~\ref{fig:fixed-order} will really match the full NNLO cross section. 
Among the five ``towers'' of leading NNLO corrections near threshold, $\alpha_s^2 \big[\ln^3(\hat{s}_4/\hat{s})/\hat{s}_4\big]_+$,
$\alpha_s^2 \big[\ln^2(\hat{s}_4/\hat{s})/\hat{s}_4\big]_+$, $\alpha_s^2 \big[\ln(\hat{s}_4/\hat{s})/\hat{s}_4\big]_+$,
$\alpha_s^2 \big[1/\hat{s}_4\big]_+$, and $\delta(\hat{s}_4)$, the first four are fully accounted for by
our formalism described in the previous sections. 
The $\delta(\hat{s}_4)$ contribution, however, requires knowledge
of the two-loop hard functions $H^{(2)}_{ab\to cr}$, as well as of the presently 
unknown two-loop soft functions $S^{(2)}_{ab\to cr}$ 
and the ${\cal O}(\alpha_s^2)$ corrections to the $R_i$ and $\tilde{\Sigma}_r$. 
Recalling the dominance of the $H^{(1)}_{ab\to cr}$ in the NLO $\delta(\hat{s}_4)$ contribution seen above,
one might expect that the $H^{(2)}_{ab\to cr}$ are equally important for the corresponding NNLO terms. 
Fortunately, the complete set of $H^{(2)}_{ab\to cr}$ has been given in Ref.~\cite{Broggio:2014hoa},
so that we may include it in our studies. The color basis adopted in that paper differs from the one we use, 
but it is relatively straightforward to transform the results to our basis. Reference~\cite{Broggio:2014hoa}
also provides the results for the $H^{(1)}_{ab\to cr}$, and we have verified that after the change of 
basis all our $H^{(1)}_{ab\to cr}$ are correctly reproduced. 

The upper solid line in Fig.~\ref{fig:fixed-order} shows the NNLO expansion when the full two-loop terms 
$H^{(2)}_{ab\to cr}$ are included. 
Compared to the upper dashed line, this means that we also include now the 
${\cal O}(\alpha_s^2)$ correction in the $H_{ab\to cr}(\alpha_s,1,1,\hat\eta)$.
To the extent that the $H^{(2)}_{ab\to cr}$ are as dominant
in the NNLO $\delta(\hat{s}_4)$ contribution as the $H^{(1)}_{ab\to cr}$ are for the NLO one,
the result shown would include all five leading NNLO terms near threshold and hence be expected
to provide a faithful estimate of full NNLO. We observe that the $H^{(2)}_{ab\to cr}$ lead to a further
significant enhancement of the $K$-factor. As expected, this enhancement is moderated toward
higher $p_T$ where the plus distributions become more and more dominant in size.
 
\begin{figure}[t!]
\vspace*{-6cm}

\hspace*{-0.8cm}
\epsfig{figure=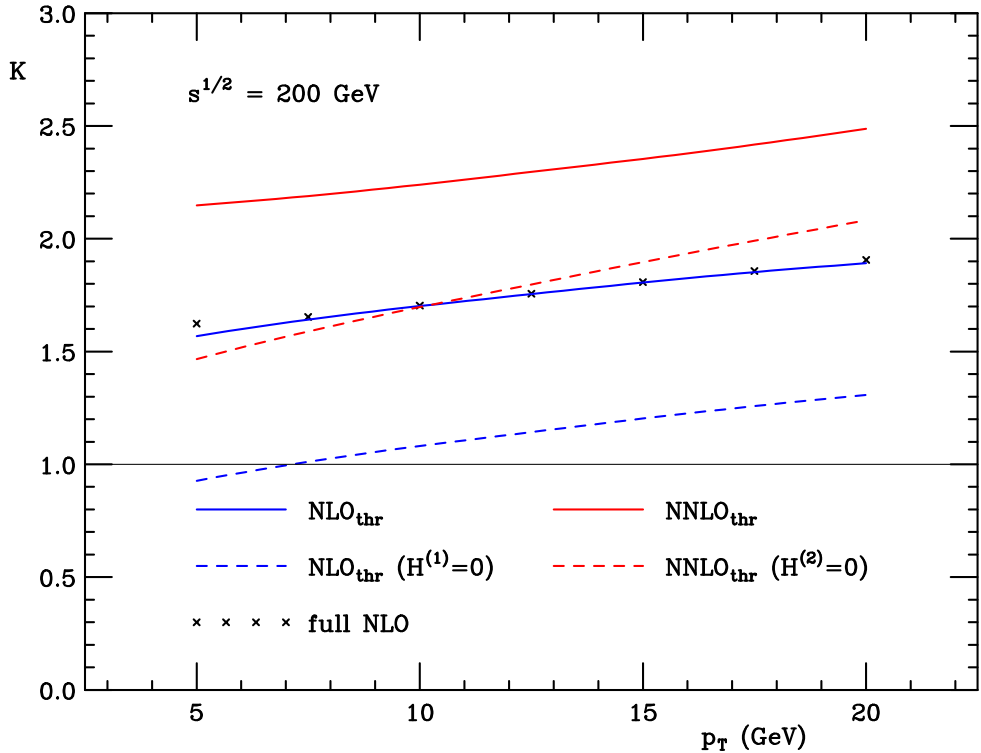,width=1\textwidth}

\vspace*{-6.4cm}
\caption{{\it NLO and NNLO $K$-factors for $pp\to \pi^0X$ for RHIC kinematics with $\sqrt{s}=200$~GeV. 
All results are normalized to the LO cross section.}}
\label{fig:collider}
\end{figure}

Figure~\ref{fig:collider} shows similar results, now for $pp$ collisions at RHIC energy $\sqrt{s}=200$~GeV,
again using $\mu_F=\mu_R=p_T$.
Some of the qualitative features from the previous figure carry over: Again the NLO expansion of
the resummed cross section matches the full NLO one very accurately -- despite the fact
that we are on average further away from partonic threshold due to the higher collision energy. 
The contribution by the $H^{(1)}_{ab\to cr}$ is again crucial in order to achieve this. In fact,
the $H^{(1)}_{ab\to cr}$ are relatively more important than in the fixed-target case, as one would
expect. The NNLO terms in the expansion again provide an enhancement of the cross section;
this time the enhancement becomes particularly pronounced only when the $H^{(2)}_{ab\to cr}$ 
are included as well. 

\begin{figure}[t!]
\vspace*{-6.4cm}

\hspace*{-0.8cm}
\epsfig{figure=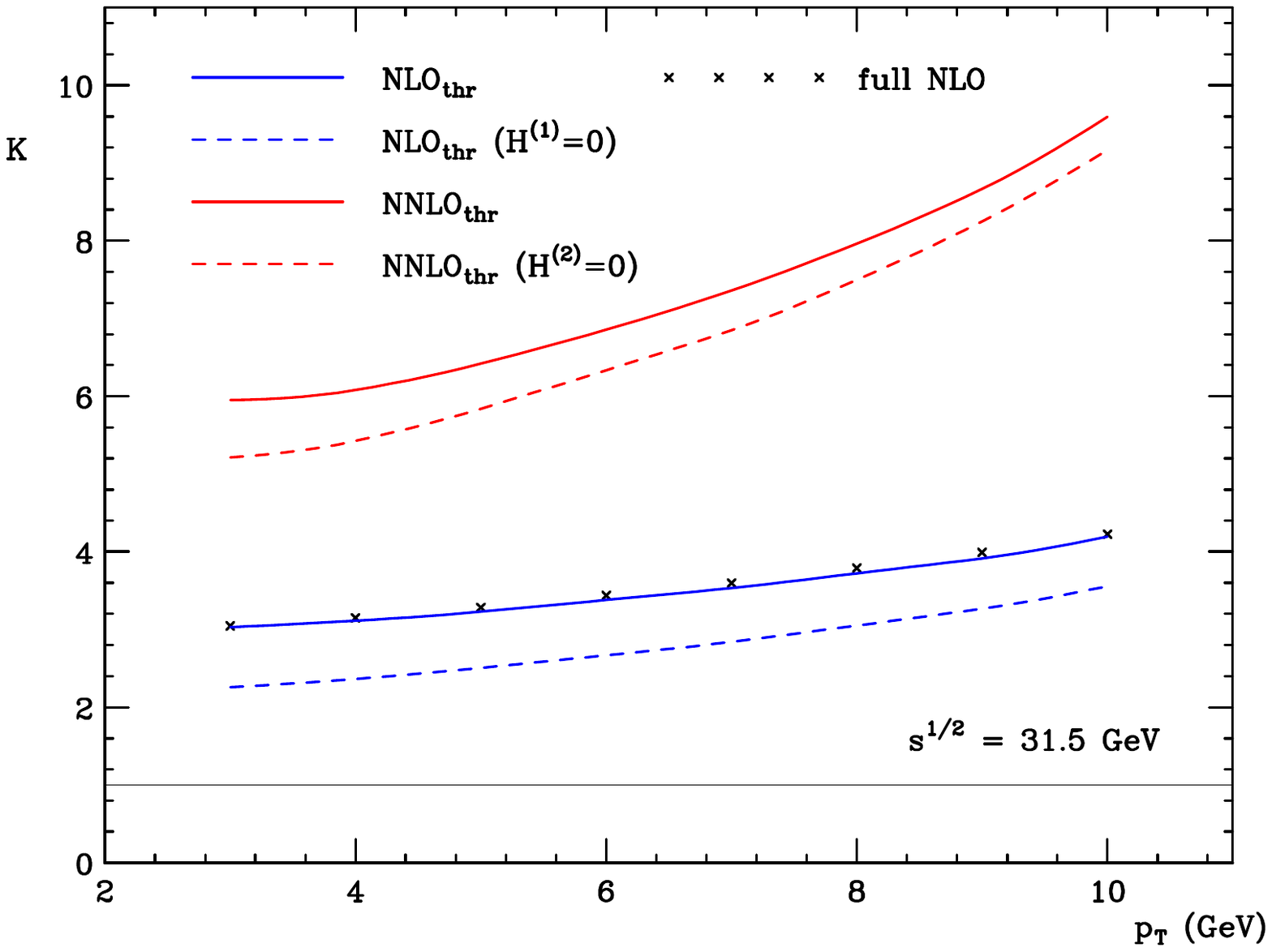,width=1\textwidth}

\vspace*{-6.4cm}
\caption{{\it Same as Fig.~\ref{fig:fixed-order}, but for scales $\mu_F=\mu_R=2p_T$.}}
\label{fig1a}
\vspace*{-4.8cm}

\hspace*{-0.8cm}
\epsfig{figure=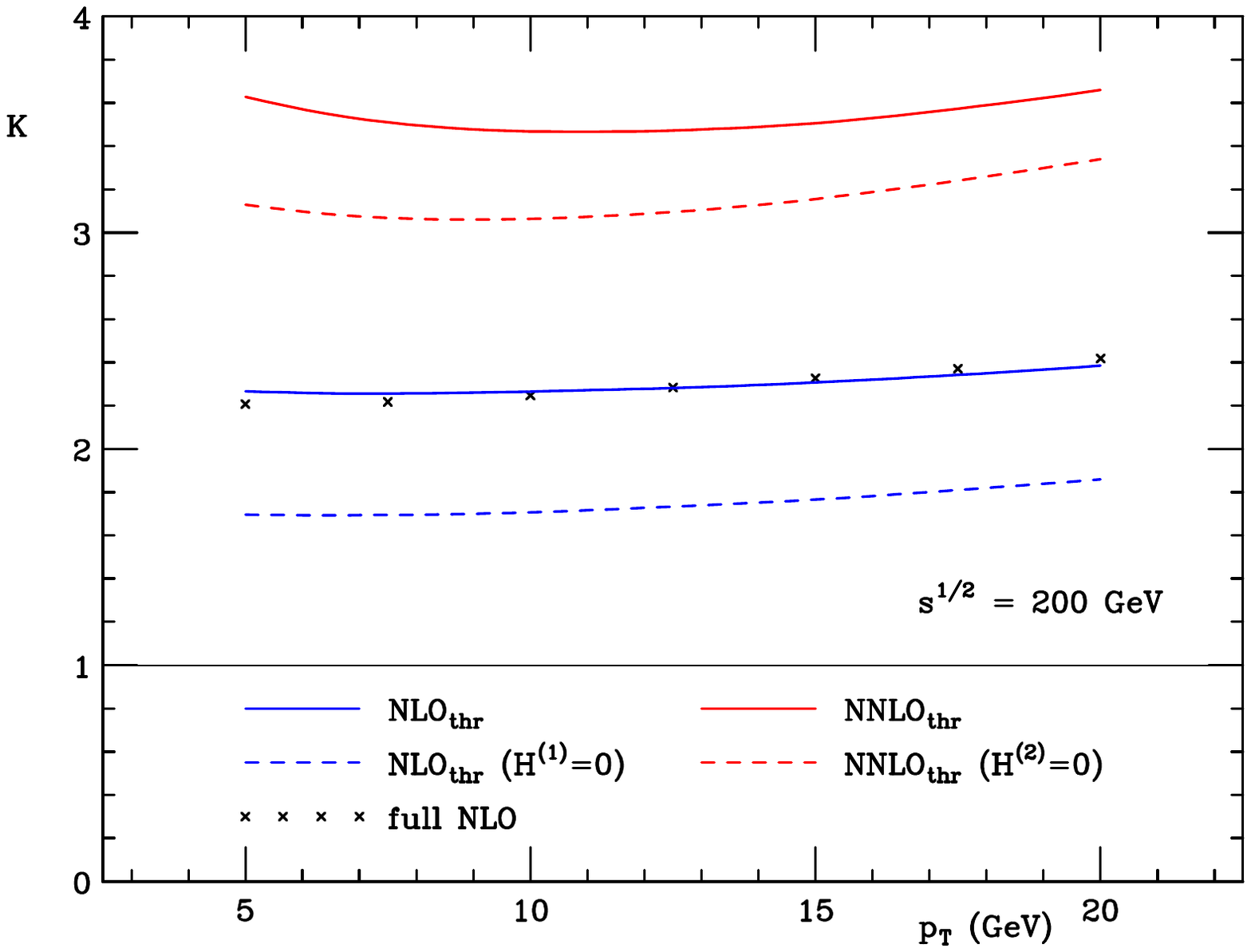,width=1\textwidth}

\vspace*{-6.4cm}
\caption{{\it Same as Fig.~\ref{fig:collider}, but for scales $\mu_F=\mu_R=2p_T$.}}
\label{fig2a}
\end{figure}

Figures~\ref{fig1a} and~\ref{fig2a} show the same calculations as Figs.~\ref{fig:fixed-order} and~\ref{fig:collider}, 
respectively, but now choosing scales $\mu_F=\mu_R=2p_T$. We observe that for this scale choice the $K$-factors
turn out to be even larger than the ones we found in the previous figures. The hard functions $H^{(1)}_{ab\to cr}$ 
and $H^{(2)}_{ab\to cr}$ turn out to be relatively less dominant, although still important, for this scale choice.

\begin{figure}[t!]
\vspace*{-3.4cm}

\hspace*{-2.1cm}
\epsfig{figure=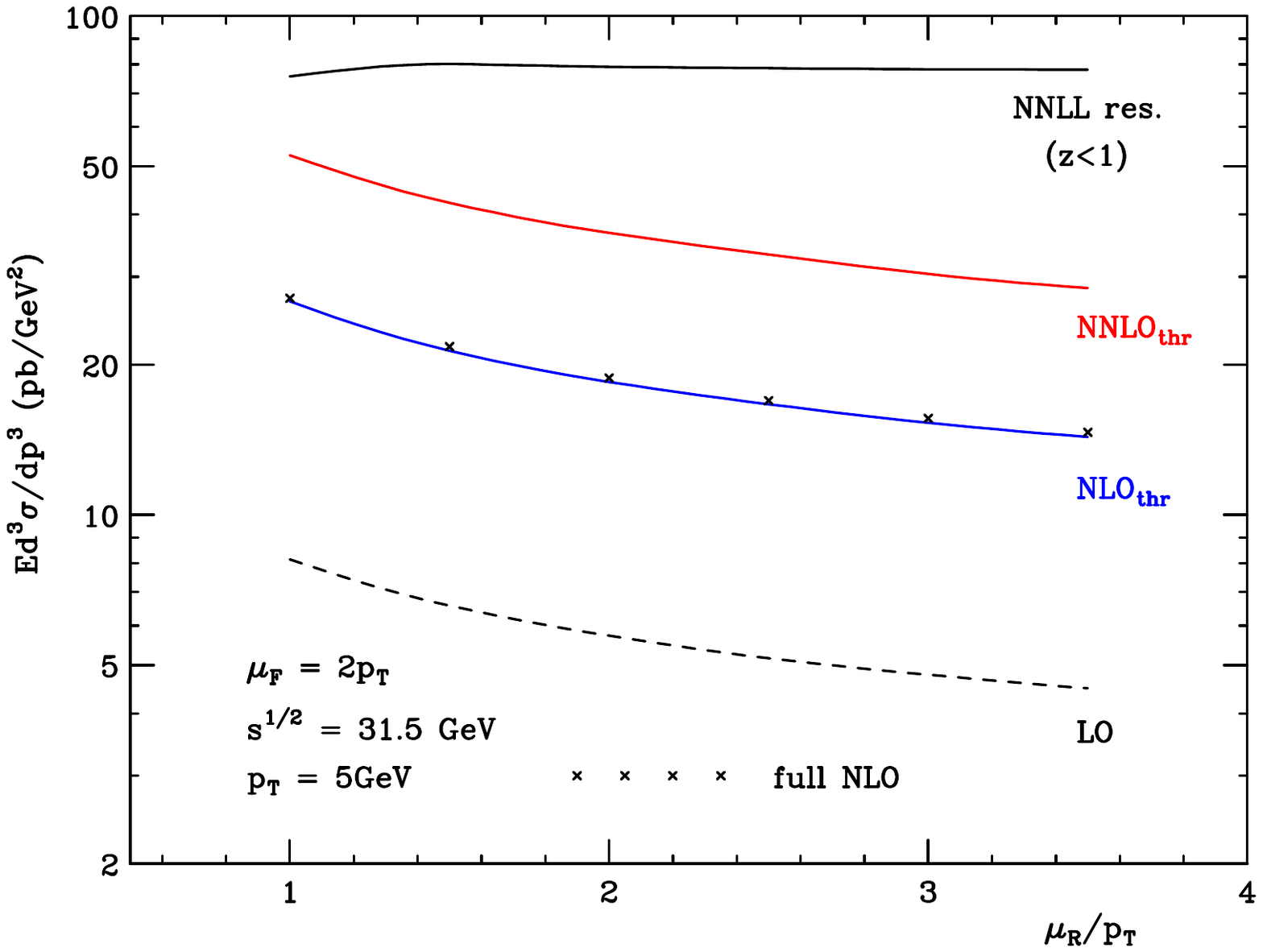,width=0.68\textwidth}

\vspace*{-15.4cm}
\hspace*{7cm}
\epsfig{figure=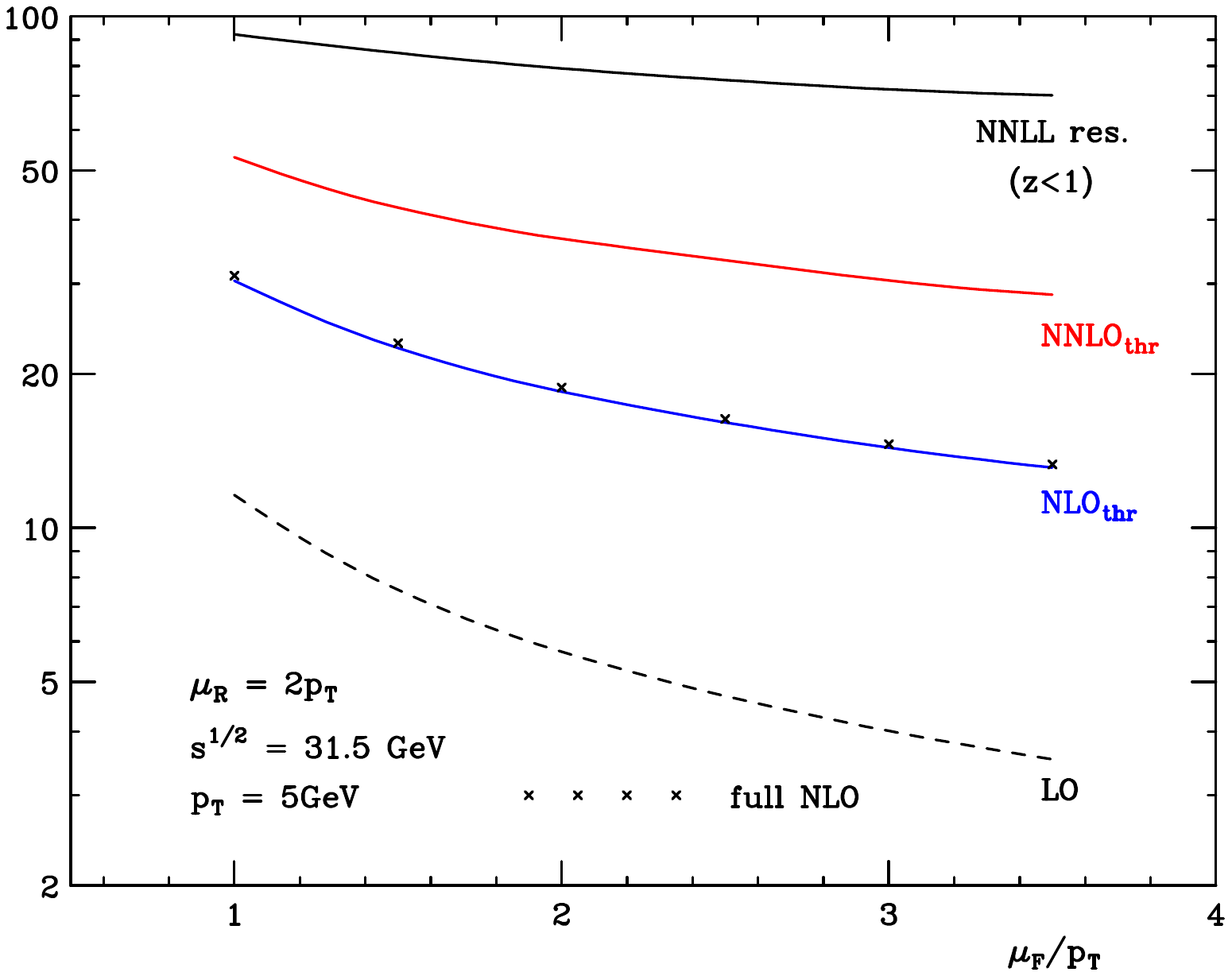,width=0.68\textwidth}

\vspace*{-7.9cm}
\hspace*{-2.1cm}
\epsfig{figure=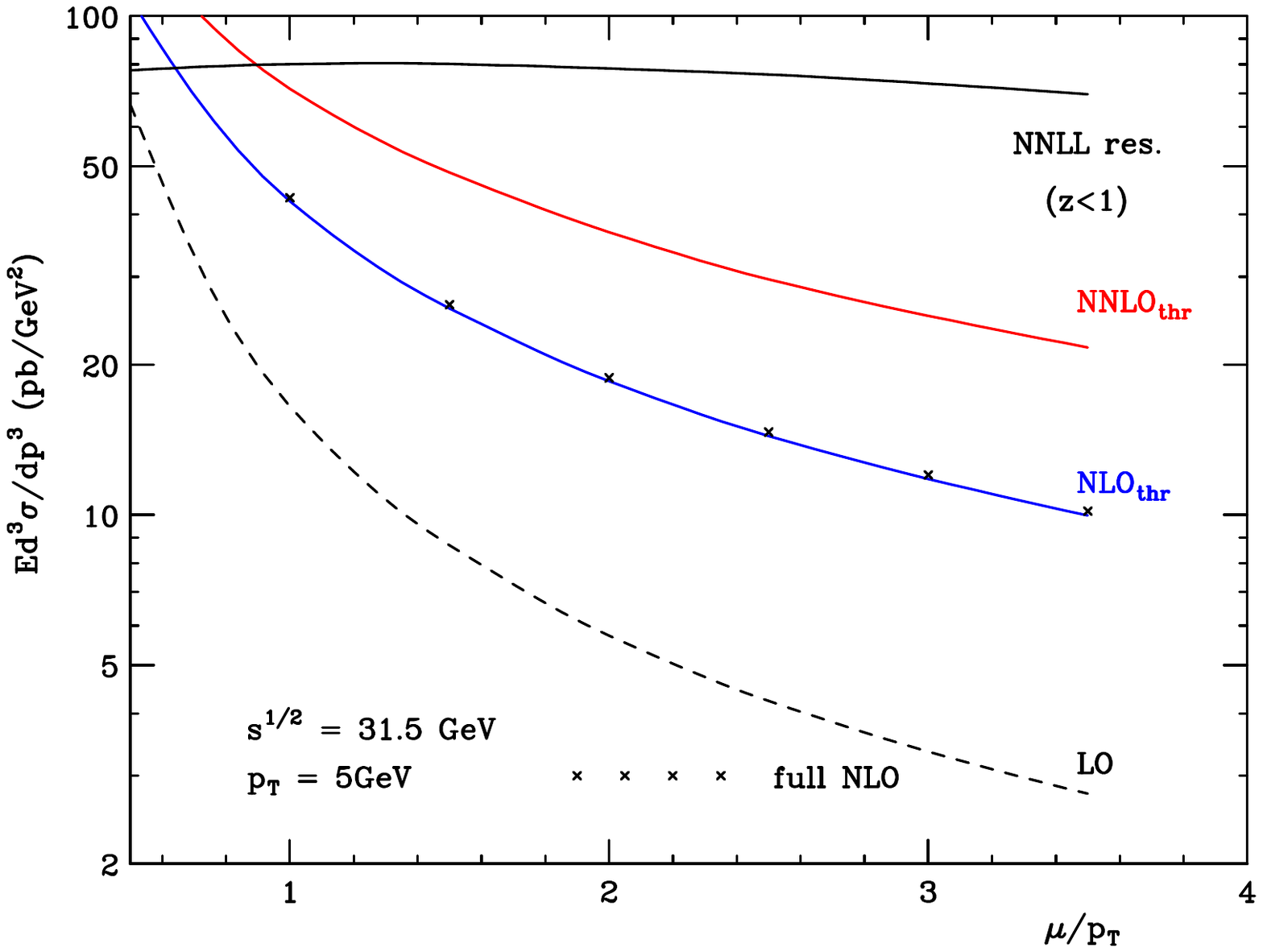,width=0.68\textwidth}

\vspace*{-15.4cm}
\hspace*{7cm}
\epsfig{figure=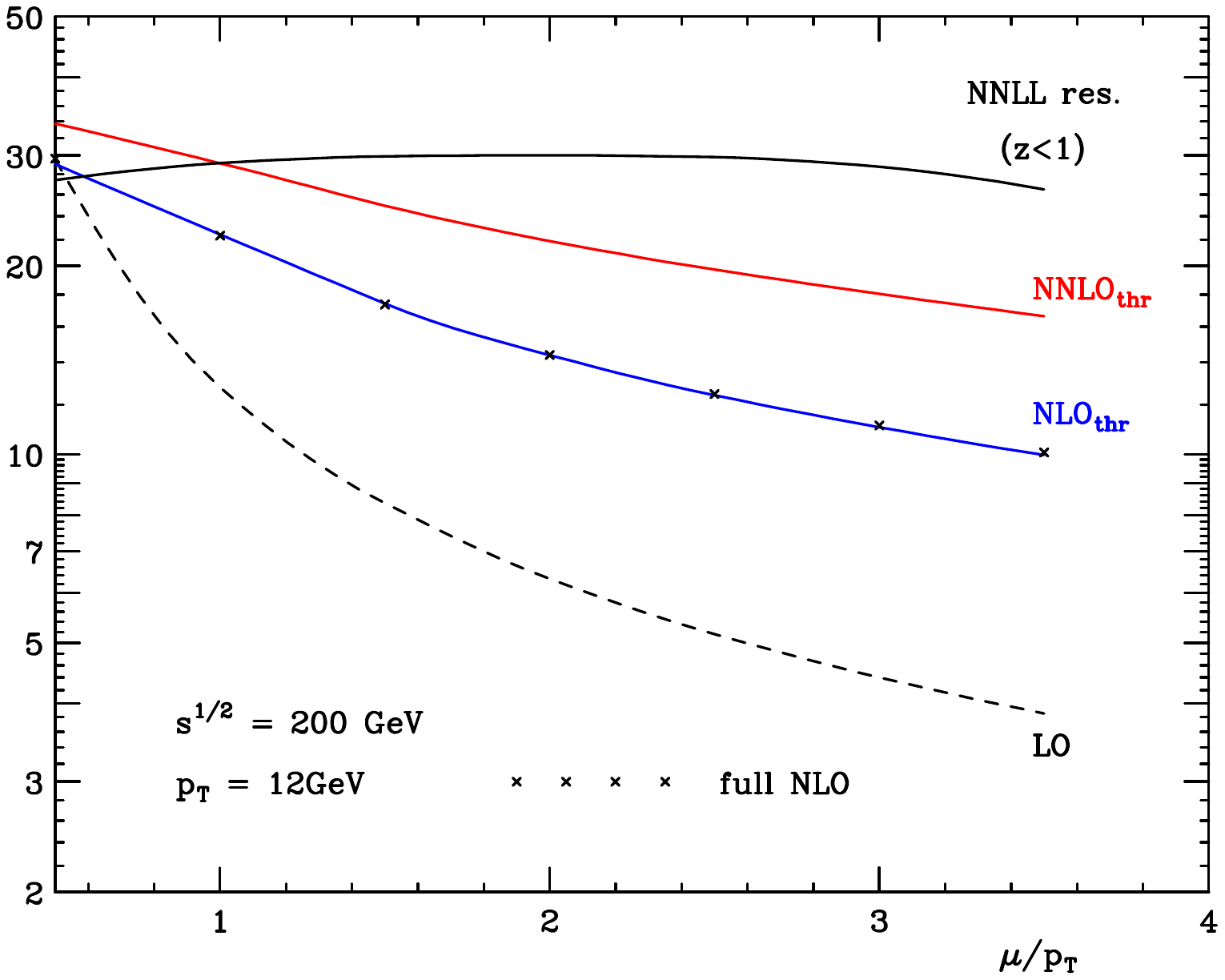,width=0.68\textwidth}

\vspace*{-4.2cm}
\caption{{\it Upper left: $\mu_R$-dependence of the invariant cross section for E706 energy at $p_T=5$~GeV for
fixed $\mu_F=2p_T$. Upper right: $\mu_F$-dependence of the invariant cross section for E706 energy at $p_T=5$~GeV for
fixed $\mu_R=2p_T$. Lower left: scale dependence of the invariant cross section for E706 energy at $p_T=5$~GeV, setting
$\mu_R=\mu_F\equiv \mu$. Lower right: scale dependence of the invariant cross section for RHIC energy at $p_T=12$~GeV, setting
$\mu_R=\mu_F\equiv \mu$.}}
\label{fig:fixed-order_scdep}
\end{figure}

The results we have shown so far suggest a relatively strong scale dependence of the fixed-order
expansions. To investigate this further, the upper left part of
Fig.~\ref{fig:fixed-order_scdep} shows the $\mu_R$ dependence of the invariant cross section 
for E706 kinematics at $p_T=5$~GeV, keeping a fixed value $\mu_F=2p_T$.
We vary $\mu_R$ no further down than $p_T$, in order to avoid having $\mu_F$ and $\mu_R$ 
too different.
The lower dashed line shows the LO result, which exhibits a very large
scale dependence. The two solid lines above show the NLO and NNLO expansions, respectively,
which show only a slightly weaker dependence on $\mu_R$.
This feature was also seen in expansions of the NLL resummed cross section of Ref.\ \cite{deFlorian:2005yj},
and should not be surprising, 
 given that both
the ${\cal O}(\alpha_s)$ and ${\cal O}(\alpha_s^2)$ corrections have such a large size.  
Although the NNLO contribution includes logarithms of $p_T/\mu_R$ that help decrease
the scale-dependence of NLO,  the full NNLO contains additional terms whose scale dependence
is compensated only at N${}^3$LO and beyond.    
The
crosses again show the full NLO result, which is again in remarkable agreement with the 
NLO expansion of the resummed result. The upper right plot of Fig.~\ref{fig:fixed-order_scdep} shows
the dependence of the invariant cross section on $\mu_F$ for fixed $\mu_R=2p_T$.
Similar features are found in this case.

In view of the large scale dependence still found at NNLO, one may wonder whether
resummed perturbation theory may ultimately help to stabilize the predicted cross
section with respect to scale changes. As described above (see discussion after Eq.~(\ref{eq:Omega-ff})), 
the ``minimal'' Mellin inversion of the resummed cross section introduces unphysical contributions
at $z>1$ in $\Omega^{ab\to c,{\mathrm{resum}}}(\hat\eta,z)$ in~(\ref{eq:Omega-inverse}), 
due to the presence of the Landau pole. 
In the present case, these contributions even turn out not to be controllable numerically. As remarked
earlier, 
we leave for future work the implementation of a practical resummation formalism that includes full Mellin moment dependence, without unphysical support in $z$.
Nonetheless, we can obtain finite and well-defined results by restricting $z$ to the physical 
regime $z<1$ in $\Omega^{ab\to c,{\mathrm{resum}}}(\hat\eta,z)$ and then in the convolution 
in Eq.~(\ref{eq:Omega-ff}). The upper lines in each of the two plots in the first row of Fig.~\ref{fig:fixed-order_scdep} 
show the all-order result obtained in this way. As one can see, both of them are much flatter, 
showing very little residual renormalization scale dependence, and much reduced factorization scale dependence. This is precisely as would be expected from
a full resummation formalism. We note that we find the same feature for RHIC energy at $p_T=
10$~GeV. Our findings provide confidence that, once the unphysical regime
$z>1$ is adequately treated, resummation will yield valuable physical results. We caution that for
the reasons just described the two results shown in the figures are not to be regarded as truly
meaningful predictions of the resummation formalism we have developed here.

We finally examine the scale dependence for equal renormalization and factorization scales,
$\mu_R=\mu_F\equiv \mu$, as is often done in phenomenological studies. 
The lower left part of Fig.~\ref{fig:fixed-order_scdep} 
shows again the invariant cross section for E706 conditions, varying $\mu$. The main 
patterns are as in the previous two figures. This time, we
follow the results all the way down to $\mu=p_T/2$, although we do not necessarily
favor such a small
scale for the inclusive-hadron cross section: Given that the hadron takes only a fraction $z\sim 0.5$
or so of the momentum of its parton progenitor, for a given hadron $p_T$ the hard-scattering typically 
will reside at a hard scale twice that value or so. The scale $p_T/2$ may thus not reflect the
hardness of the partonic interaction very realistically. Nonetheless, it is interesting to see
that the various results edge closer together when the scale is chosen to have a small
value. This becomes especially evident when going to RHIC energy in the lower right
plot in Fig.~\ref{fig:fixed-order_scdep}. There, LO and NLO even meet at scale $p_T/2$ for the
value of $p_T=12$~GeV we consider, NNLO is only moderately higher, and the resummed
result with $z<1$ is also very close. It is interesting to note that such a tendency for the scale variation to narrow
at low scales was observed in the literature also in various different contexts, for example 
early on in studies of prompt-photon production~\cite{Aurenche:1987fs}, but also recently
for $t\bar{t}$ production at the LHC~\cite{Czakon:2016dgf}. We stress again 
that we do not assign much significance to the precise location of the solid curve for 
the $z<1$ resummed result. The fact that at low scales already the NNLO expansion is 
higher just shows once more that the result with $z<1$ should be regarded as only a 
part of a fully resummed phenomenological cross section.

\section{Conclusions}  

We have developed a threshold resummation for single-particle inclusive cross sections in hadron-hadron scattering at next-to-next-to-leading logarithm, up to the ideal matching with exact next-to-next-to-leading order hard scattering functions.   As in previous work on this subject, threshold resummation organizes leading-power plus distributions in the variable $\hat s_4$, the invariant mass of all  radiation recoiling against the fragmenting parton.   

New results include the definition and one-loop calculation of the matrix that organizes coherent soft radiation for all 
single-particle inclusive
two-to-two partonic processes.   This enables us for the first time to separate exact one-loop corrections to these processes  between short-distance and long-distance factors, which are expanded in terms of the running coupling at hard and soft scales, respectively.  The one-loop expansions of the factorized jet and soft functions that we derive reproduce all leading-power singular terms in the exact NLO calculations for partonic subprocesses.   For completeness, the NLO singular terms are provided for each subprocess in appendices.  In a test at phenomenologically-relevant kinematics, the ${\cal O}(\as)$ expansions reproduce full NLO results to the accuracy of a few percent.  
The resummation analysis of this paper is given in Mellin moment space, and we compared with the closely-related NNLL resummed prompt-photon cross section of Ref.\ \cite{Becher:2009th}, developed using soft-collinear effective theory.   

We have seen that the resummed single-particle inclusive cross section has a number of unique features that distinguish it from dihadron and single-photon inclusive cross sections.  In particular, an inverse transform using the minimal prescription 
leads to unphysical contributions from $z>1$ that 
we find are not numerically stable for single-particle inclusive cross sections, due to an enhanced unphysical range 
of the partonic fractional momenta for this process, which can cause partonic rapidities to become very large.
   We have provided an analytic Mellin inverse for $N$-independent infrared scales, but given the apparent importance of threshold logarithms including $N$-dependence in the examples we studied, it seems natural to investigate the conventional ``direct-QCD" approach further.
Phenomenological applications thus will require further development, especially regarding the inverse transforms.   This will be the subject of forthcoming work.   
We anticipate that the analysis given in this paper will be relevant to  existing data at fixed-target energies, and to data from present and future hadronic colliders, in addition to the analysis of resolved photons at electron-hadron colliders.
We expect that the formalism presented here will also have valuable applications in resummation studies of single-inclusive
jet cross sections at hadron colliders. 

\section*{Acknowledgements}

We are grateful to Andreas Vogt for helpful communications and to 
Marco Stratmann for information on the DSS14 fragmentation functions, and to Andrea Ferroglia for a useful discussion.
GS thanks the Pauli Center and the Institute for Theoretical Physics, ETH Zurich, as well as the Department of Physics, 
University of Vienna for their hospitality. The work of GS was supported in part by the National Science Foundation,  grants 
PHY-1316617 and 1620628. FR is supported by the Department of Energy under 
Contract No. DE-AC0205CH11231, and the LDRD Program of Lawrence Berkeley National Laboratory.
The numerical calculations presented here were in part carried out on the HPC resource bwUniCluster
funded by the state of Baden-W\"{u}rttemberg.

\begin{appendix}

\section{Anomalous dimensions
\label{AppB}}

We present here the explicit low-order expansions of the various anomalous dimensions used
for the resummed jet functions to the extent that they have not yet been given in the text.
The functions $A_i$, $\hat{B}_i$ and $\hat{D}_i$ appearing 
in Eqs.~(\ref{eq:J-in-soln2}), (\ref{eq:out-jet-def}) and (\ref{eq:J-rec-MV}) are expanded 
as series in $\as$,
\bea
A_i(\as) & = & \frac{\as}{\pi} A_i^{(1)} +\left( \frac{\as}{\pi}\right)^2 A_i^{(2)} + 
\left(\frac{\as}{\pi}\right)^3 A_i^{(3)}+{\mathcal O}(\as^4) \,,\nn \\[2mm]
\hat{B}_i(\as) & = & \frac{\as}{\pi} \hat{B}_i^{(1)} +\left( \frac{\as}{\pi}\right)^2 \hat{B}_i^{(2)} + 
{\mathcal O}(\as^3) \,,\nn \\[2mm]
\hat{D}_i(\as) & = & \left(\f{\as}{\pi}\right)^2 \hat{D}_i^{(2)}+{\mathcal O}(\as^3) \,.
\eea
The coefficients of $A_i$ are familiar. To NNLL, we use~\cite{Vogt:2000ci,KT,Moch:2004pa,Harlander:2001is,eric}
\beeq 
\label{A12coef} 
A_i^{(1)} &=&C_i
\;,\;\;\;\;\quad A_i^{(2)}\;=\;\frac{1}{2} \; C_i  \left[ 
C_A \left( \frac{67}{18} - \frac{\pi^2}{6} \right)  
- \frac{5}{9} N_f \right] \; , \nn \\[2mm] 
A_i^{(3)}&=&\f{1}{4}C_i\left[C_A^2 \left(\f{245}{24}-\f{67}{9}\zeta(2)+
\f{11}{6}\zeta(3)+\f{11}{5}\zeta(2)^2 \right)+C_F N_f\left(-\f{55}{24}+2\zeta(3) \right)\right. \nn \\[2mm]
&& \qquad\;\, \left. +C_A N_f\left(-\f{209}{108}+\f{10}{9}\zeta(2)-\f{7}{3}\zeta(3) \right) -\f{1}{27}N_f^2 \right] \; , 
\eeeq
with $N_f$ is the number of flavors and 
\beq\label{cqcg} 
C_q\,=\,C_F\,=\,\frac{N_c^2-1}{2N_c}\,=\,\frac{4}{3}  
\;, \;\;\;C_g\,=\,C_A\,=\,N_c=3 \; .
\eeq
The coefficient $\hat{D}_i^{(2)}$ was already given in Eq.~(\ref{eq:Dhat2}). As shown in~(\ref{eq:Drel}), 
it is directly related to the also widely used coefficient $D_i^{(2)}$. For completeness,
we recall its explicit value~\cite{Vogt:2000ci,Catani:2001ic}:
\beq
D_i^{(2)}\,=\,
C_i\left[C_A\left(-\f{101}{27}+\f{11}{3}\zeta(2)+\f{7}{2}\zeta(3)\right) + N_f\left(\f{14}{27}-\f{2}{3}\zeta(2)  \right) \right] \,.
\eeq
In addition, for the recoiling jet in Eq.~(\ref{eq:J-rec-MV}) we also need
\beq
\hat{B}_q^{(1)}\,=\, -\frac{3}{4}\,C_F\,,\quad\quad \hat{B}_g^{(1)}\,=\,-\pi b_0\,,
\eeq
where as before $b_0\,=\,(11C_A-2N_f)/(12\pi)$. 
The two-loop coefficients $\hat{B}_i^{(2)}$ were already given in~(\ref{eq:B-Bhat}). 
As discussed in Eq.~(\ref{eq:Br-def}), they are obtained from the customary coefficients
$B_i^{(2)}$ which have been computed in~\cite{Moch:2009my}:
\beeq
B_q^{(2)}&=&\frac{C_F^2}{2} \left( -\frac{3}{16}+\frac{3}{2}\,\zeta(2) -3 \zeta(3)\right)+\frac{C_FC_A}{2}
\left( -\frac{3155}{432} + \frac{11}{6}\,\zeta(2)+5\zeta(3)\right)\nn\\[2mm]
&+&\frac{C_FN_f}{2} \left( \frac{247}{216}-\frac{1}{3}\,\zeta(2)\right) \,,\nn\\[2mm]
B_g^{(2)}&=&\frac{C_A^2}{2} \left( -\frac{611}{72}+\frac{11}{3}\,\zeta(2)+2 \zeta(3)\right)+
\frac{C_AN_f}{2}\left(\frac{107}{54}-\frac{2}{3}\,\zeta(2)\right)+\frac{C_F N_f}{8}-
\frac{5N_f^2}{108}\,.
\eeeq 
Finally, we also present the expansion of the $\delta$ function contributions
to the diagonal DGLAP splitting functions. These appear in Eq.\ (\ref{eq:BDPrel})
and also determine the factorization scale dependence of the hard function $H$ in Eq.~(\ref{HmuF}).
Writing
\be
P_{i,\delta}(\as) \,=\, \frac{\as}{\pi} P_{i,\delta}^{(1)} +\left( \frac{\as}{\pi}\right)^2 P_{i,\delta}^{(2)} + 
{\mathcal O}(\as^3)\,, 
\ee
we have 
\be
P_{i,\delta}^{(1)}\,=\,-\hat{B}_i^{(1)}\,,
\ee
and \cite{cfp}
\beeq
P_{q,\delta}^{(2)}&=&\frac{1}{4}\left[ C_F^2 \left( \frac{3}{8}-3\zeta(2) +6 \zeta(3)\right) + C_F C_A 
\left(\frac{17}{24}+\frac{11}{3}\zeta(2) -3 \zeta(3)\right) -
\frac{C_F N_f}{2} \left(\frac{1}{6}+\frac{4}{3} \zeta(2) \right)\right]\,,\nn\\[2mm]
P_{g,\delta}^{(2)}&=&\frac{1}{4}\left[ C_A^2 \left(\frac{8}{3} + 3 \zeta(3)\right) - 
\frac{C_F N_f}{2} - \frac{2}{3} C_A N_f\right]\,.
\eeeq

\section{Explicit NNLL forms of the exponents \label{AppC}}

Evaluating the exponent in Eq.~(\ref{eq:J-in-soln2}) and choosing $\mu_{rf}=\sqrt{\hat{s}}$, 
one obtains an explicit expression for the NNLL expansion of the function
$\tilde{J}_{\mathrm{in}}^{(i)}$:
\beeq\label{DfctNLL}
\tilde{J}_{\mathrm{in}}^{(i)}\left ( \frac {1}{\bar N^2},\frac{\mu_F}{\sqrt{\hat{s}}},\as(\sqrt{\hat{s}}) \right) 
& = &\widehat R_i\big(\as(\sqrt{\hat{s}})\big)\, \exp\left\{{\cal E}_i \left(\lambda, \frac{\mu_R^2}{\hat{s}},
\frac{\mu_F^2}{\hat{s}} \right)\right\}\; ,
\eeeq
where
\beq
\lambda = b_0 \as (\mu_R) \ln(N{\mathrm{e}}^{\gamma_E})\,,
\eeq
and
\beeq
{\cal E}_i \left(\lambda, \frac{\mu_R^2}{\hat{s}},
\frac{\mu_F^2}{\hat{s}} \right)&=&\f{A_i^{(1)}}{ 2 \pi \b0^2 \,\alpha_s(\mu_R)}
\left( 2 \lambda + (1-2\lambda)\ln ( 1- 2 \lambda) \right)\nn\\[2mm]
&-&\frac{A_i^{(2)}}{2 \pi^2 b_0^2} \left[2\lambda + \ln (1-2\lambda)\right] \nn\\[2mm]
&+&\frac{A_i^{(1)}b_1}{2\pi b_0^3}\left[2\lambda +
\ln(1-2\lambda)+ \frac{1}{2} \ln^2(1-2\lambda)\right] \nonumber  \\[2mm]
& -& \frac{A_i^{(1)}}{2\pi b_0} \left[2 \lambda +
\ln(1-2\lambda)\right] \ln\frac{\mu_R^2}{\hat s}  + \frac{A_i^{(1)}}{\pi b_0} \lambda
 \ln \frac{\mu_F^2}{\hat s} \nn\\[2mm]
 &+ &\alpha_s(\mu_R)\Bigg\{
- \f{A_i^{(2)} \bone }{2\pi^2 \b0^3} \f{1}{1-2\la}\left[2\la+\ln(1-2\la) +2 \la^2 \right]\nn \\[2mm]
& +& \f{A_i^{(1)}  \bone^2}{2\pi \b0^4 (1-2\la)}
\left[2\la^2+ 2\la \ln(1-2\la)+\f{1}{2} \ln^2(1-2\la) \right] \nn \\[2mm]
& +& \f{A_i^{(1)}  \btwo}{2\pi\b0^3}
\left[2\la+\ln(1-2\la) +\f{2\la^2}{1-2\la} \right]
+ \f{A_i^{(3)}}{\pi^3 \b0^2} \f{\la^2}{1-2\la} \nn \\[2mm]
&+& \f{A_i^{(2)}}{\pi^2 \b0} \,\la \ln\f{\mu^2_F}{\hat s}
- \f{A_i^{(1)}}{2\pi} \, \la  \ln^2\f{\mu^2_F}{\hat s}
+ \f{A_i^{(1)}}{\pi} \la \ln\f{\mu_R^2}{\hat s} \ln\f{\mu^2_F}{\hat s} \nn \\[2mm]
&-& \f{1}{1-2\la} \Big(
 \f{ A_i^{(1)} \bone}{2\pi \b0^2} \left[ 2\la+\ln(1-2\la) \right]
 - \f{2 A_i^{(2)}}{\pi^2 \b0}  \la^2 \Big)
\ln\f{\mu_R^2}{\hat s} \nn \\[2mm]
&+ &\f{A_i^{(1)}}{\pi} \f{\la^2}{1-2\la}\ln^2\f{\mu_R^2}{\hat s}
- \f{ \hat{D}_i^{(2)}}{2\pi^2 \b0} \f{\la}{1-2\la}\Bigg\} \; .
\label{EE}
\eeeq
Here $b_0,\,b_1,\,b_2$ are the first three coefficients of the QCD beta function which are given by~\cite{Tarasov:1980au,Larin:1993tp}
\bea\label{thebs}
b_0 & = & \f{1}{12\pi} \left(11C_A-2 N_f\right)\; , \qquad b_1 = \f{1}{24\pi^2}\left(17C_A^2-5C_AN_f-3C_F N_f\right) \; , \nn \\[2mm]
b_2 & = & \f{1}{64\pi^3}\left(\f{2857}{54} C_A^3- \f{1415}{54} C_A^2 N_f-\f{205}{18} C_A C_F N_f+C_F^2 N_f+
\f{79}{54} C_A N_f^2+\f{11}{9} C_F N_f^2\right) .\;\;
\eea
Note that we have obtained Eq.~(\ref{EE}) by expanding the running coupling in the integrand of 
Eq.~(\ref{eq:J-in-soln2}) as~\cite{Vogt:2000ci}
\be
\alpha_s(\mu)\,=\,\frac{\alpha_s(\mu_R)}{X}\left[1-
\frac{\alpha_s(\mu_R)}{X}\,\frac{b_1}{b_0}\,\ln X+\left(\frac{\alpha_s(\mu_R)}{X}\right)^2
\left(\,\frac{b_1^2}{b_0^2}\,\Big(\ln^2 X-\ln X+X-1\Big)-\frac{b_2}{b_0}\,\big(X-1\big)\right)\right],
\ee
where 
\be
X\,\equiv\,1+b_0 \,\alpha_s(\mu_R) \,\ln \big(\mu^2/\mu_R^2\big)\,.
\ee
In this way, our perturbative expansion of the resummed exponents, which 
necessarily truncates the perturbative series, introduces dependence on a
renormalization scale $\mu_R$; see discussion after Eq.~(\ref{Hformnew}).

The corresponding result for the outgoing recoil jet may be written in compact form as
\beeq\label{DfctNLL-2}
\tilde{J}_{\mathrm{rec}}^{(r)}\left ( \frac {1}{\bar N^2},\as(\sqrt{\hat{s}}) \right) 
& = &\frac{\tilde{\Sigma}_r\big(1,1,\as(\sqrt{\hat{s}})\big)}{\widehat R_{\bar{r}}
\big(\as(\sqrt{\hat{s}})\big)}\, \exp\left\{{\cal F}_r \left(\lambda, \frac{\mu_R^2}{\hat{s}}\right)\right\}\; ,
\eeeq
where
\beeq\label{JJ}
{\cal F}_r\left(\lambda, \frac{\mu_R^2}{\hat{s}}\right)&=&2\, {\cal E}_r \left(\frac{\lambda}{2}, \frac{\mu_R^2}{\hat{s}},1\right)-
{\cal E}_r \left(\lambda, \frac{\mu_R^2}{\hat{s}},1\right)\,+\,\frac{\hat{B}_r^{(1)}}{\pi b_0}\,\log(1-\lambda)\nn\\[2mm]
&-&\alpha_s(\mu_R)\,\frac{\hat{B}_r^{(1)}}{\pi b_0^2 (1-\lambda)}\left(
b_0^2\, \lambda \,\ln\f{\mu_R^2}{\hat s}-b_1 (\lambda+\log(1-\lambda))\right)\nn\\[2mm]
&-&\alpha_s(\mu_R)\,\left(2\hat{B}_r^{(2)}-\hat{D}_r^{(2)}\right)\frac{\lambda}{2\pi^2 b_0(1-\lambda)}\,.
\eeeq

\section{Ingredients for soft matrices}
\label{sec:soft-matrices-explicit}

Every one-loop soft function for a given partonic channel is given by (see Eq.~(\ref{Sid}))
\ba\label{Sid1}
\tilde{S}^{(1)}_0&=&\tilde{S}_0^{(0)}\Bigg[\frac{1}{4}(C_a+C_b+C_c-C_r) \, \left( \ln^2\left(\frac{1-v}{v}\right)+2\zeta(2)
\right)-\,C_a\ln^2(v)-C_b\ln^2(1-v)\Bigg]\nn\\[2mm]
&&\hspace*{9mm}\pm\, 2 \ln(1-v)\ln(v) \,{\cal R}_{12}\,,
\ea
where in the last term the positive sign applies to all processes with a $q\bar{q}^{\,(}{}'{}^{)}$ initial or final state, and 
the negative sign to all others. Below we present the matrices $S_0^{(0)}$ and ${\cal R}_{12}$ for
all partonic channels. We also recall the one-loop soft anomalous dimension matrices
$\Gamma^{ab\to cr,(1)}(\hat\eta)$, which may be found in this form in 
Ref.~\cite{Kidonakis:2000gi}.

For $qq'\to qq'$ and $qq\to qq$ scattering we have
\be
S_0^{(0)}\,=\,\left( \begin{array}{cc} \frac{C_A^2-1}{4} & 0 \\[2mm] 
0 & C_A^2 \end{array} \right)\,,\quad\quad
{\cal R}_{12}\,=\,-\frac{C_F}{2}\left( \begin{array}{cc} 1 & -C_A \\[2mm] 
-C_A &0 \end{array} \right)\,.
\ee
The soft anomalous dimension matrix has already been given in Eq.~(\ref{eq:Gqqp}).
For $q\bar{q}{\,'}\to q\bar{q}{\,'}$, $q\bar{q}\to q'\bar{q}{\,'}$, and $q\bar{q}\to q\bar{q}$ scattering we have
\bea
&&S_0^{(0)}\,=\,\left( \begin{array}{cc} C_A^2  & 0 \\[2mm] 
0 & \frac{C_A^2-1}{4} \end{array} \right)\,,\quad\quad
{\cal R}_{12}\,=\,\frac{C_F}{2}\left( \begin{array}{cc} 0 & C_A \\[2mm] 
C_A &\frac{1}{2}(C_A^2-2) \end{array} \right)\,,\nn\\[3mm]
&&\Gamma^{q {\bar q}\rightarrow q {\bar q},(1)}(\hat\eta)\,=\,\left(
                \begin{array}{cc}
                 2C_F T  &   -\frac{C_F}{C_A} U   \\[2mm]
                -2U    &-\frac{1}{C_A}\big(T-2 U\big)
                \end{array} \right)\, ,
\eea
where $T=\ln(1-v)+i\pi$, $U=\ln(v)+i\pi$.
For $q\bar{q}\to gg$ we have
\bea\label{S0qqb}
&&S_0^{(0)}\,=\,C_F \left( \begin{array}{ccc} 2 C_A^2 & 0 & 0 \\[2mm] 
0 &C_A^2-4 & 0\\[2mm]
0 & 0 & C_A^2 \end{array} \right)\,,\quad\quad
{\cal R}_{12}\,=\,\frac{C_F}{2C_A}\left( \begin{array}{ccc}  4 C_F C_A^3 & 0 & 0 \\[2mm] 
0 &C_A^2-4& 0 \\[2mm]
0 & 0 & -C_A^2
\end{array} \right)\,,\nn\\[3mm]
&&\Gamma^{q {\bar q}\rightarrow gg,(1)}(\hat\eta)\,=\,\left(
                \begin{array}{ccc}
                 0  &   0  & U-T  \vspace{2mm} \\[2mm]
                 0  &   \frac{C_A}{2}\left(T+U \right)    & \frac{C_A}{2}
\left(U-T\right) \vspace{2mm} \\[2mm]
                 2\left(U-T \right)  & \frac{C_A^2-4}{2C_A}
\left(U-T\right)  & \frac{C_A}{2}\left(T+U \right)
                \end{array} \right)\,.
\eea
For $gg\to q\bar{q}$ we have the same $S_0^{(0)}$ and $\Gamma^{(1)}$ as in~(\ref{S0qqb}),
but
\bea
{\cal R}_{12}&=&\frac{C_FC_A}{2}\left( \begin{array}{ccc}  4C_A^2 & 0 & 0 \\[2mm] 
0 &C_A^2-4& 0 \\[2mm]
0 & 0 & C_A^2
\end{array} \right)\,.
\eea
For $qg\to qg$ and $qg\to gq$ we have the same $S_0^{(0)}$ as in~(\ref{S0qqb}), 
but 
\bea
{\cal R}_{12}&=&\frac{C_FC_A}{4}\left( \begin{array}{ccc}  0 & 0 & 4C_A \\[2mm] 
0 &-(C_A^2-4)& C_A^2-4 \\[2mm]
4C_A & C_A^2-4 & -C_A^2
\end{array} \right)\,,\nn\\[2mm]
\Gamma^{qg\rightarrow qg,(1)}(\hat\eta)&=&\left(
                \begin{array}{ccc}
                 \left( C_F+C_A \right) T  &   0  & U  \vspace{2mm} \\[2mm]
                 0  &   C_F T+ \frac{C_A}{2} U     & \frac{C_A}{2} U
\vspace{2mm} \\[2mm]
                 2 U  & \frac{C_A^2-4}{2C_A} U  &  C_F T+ \frac{C_A}{2} U
                \end{array} \right)\,.
\eea
Finally, for $gg\to gg$ all three matrices have the block structure
\be
S_0^{(0)}\,=\,\left( \begin{array}{cc} S_{3\times 3} & 0_{3\times 5} \\[2mm]
0_{5\times 3} & S_{5\times 5}\end{array} \right)\,,\quad\quad
{\cal R}_{12}\,=\,\left( \begin{array}{cc} R_{3\times 3} & 0_{3\times 5} \\[2mm]
0_{5\times 3} & R_{5\times 5}\end{array} \right)\,,\quad\quad
\Gamma^{gg\rightarrow gg,(1)}(\hat\eta)\,=\,\left( \begin{array}{cc} \Gamma_{3\times 3} & 0_{3\times 5} \\[2mm]
0_{5\times 3} & \Gamma_{5\times 5}\end{array} \right)\,,
\ee
where, setting $C_A=3$ for simplicity,
\bea
&&S_{3\times 3}\,=\,\left( \begin{array}{ccc}  5 & 0 & 0 \\[2mm]
0 & 5 & 0 \\[2mm]
0 & 0 & 5
\end{array} \right)\,,\quad\quad 
S_{5\times 5}\,=\,\left( \begin{array}{ccccc}  1 & 0 & 0 & 0 & 0 \\[2mm]
0 & 8  & 0 & 0 & 0 \\[2mm]
0 & 0 & 8 & 0 & 0 \\[2mm]
0 & 0 & 0 & 20 & 0 \\[2mm]
0 & 0 & 0 & 0 & 27
\end{array} \right)\,,\nn\\[2mm]
&&R_{3\times 3}\,=\,-\left( \begin{array}{rrc}  \frac{15}{2} & 0 & 0 \\[2mm]
0 & \frac{15}{2}  & 0 \\[2mm]
0 & 0 & 0
\end{array} \right)\,,\quad\quad 
R_{5\times 5}\,=\,-\left( \begin{array}{ccccc}  0 & 0 & 3 & 0 & 0 \\[2mm]
0 & 6  & 6 & 12 & 0 \\[2mm]
3 & 6 & 6 & 0 & 9 \\[2mm]
0 & 12 & 0 & 30 & 18 \\[2mm]
0 & 0 & 9 & 18 & 54
\end{array} \right)\,,\nn\\[2mm]
&&\Gamma_{3\times 3}\,=\,
\left(  \begin{array}{ccc}
                  3 T  &   0  & 0  \\[2mm]
                  0  &  3 U & 0    \\[2mm]
                  0  &  0  &  3\left( T+ U \right)
                   \end{array} \right)\,,\quad
\Gamma_{5\times 5}\,=\,\left(\begin{array}{ccccc}
6 T & 0 & -6 U & 0 & 0 \vspace{2mm} \\[2mm]
0  & 3 T+\frac{3 U}{2} & -\frac{3 U}{2} & -3 U & 0 \vspace{2mm} \\[2mm]
-\frac{3 U}{4} & -\frac{3 U}{2} &3 T+\frac{3 U}{2} & 0 & -\frac{9 U}{4}
\vspace{2mm} \\[2mm]
0 & -\frac{6 U}{5} & 0 & 3 U & -\frac{9 U}{5} \vspace{2mm} \\[2mm]
0 & 0 &-\frac{2 U}{3} &-\frac{4 U}{3} & -2 T+4 U
\end{array} \right)\,.\nn\\[2mm]
\eea

\section{NLO expansions for other partonic channels \label{otherC}}

As before, we define
\ba
L&=&\ln(v)\,,\nn \\[2mm]
\bar{L}&=&\ln(1-v)\,.
\ea
For an arbitrary process $ab\to cr$ the first-order term in the product of the jet functions
may be written as
\bea
J^{(a),(1)}_{{\mathrm{in}}}(y)+
J^{(b),(1)}_{{\mathrm{in}}}(y)+J^{(c),(1)}_{{\mathrm{fr}}}(y)+J^{(r),(1)}_{{\mathrm{rec}}}(y)
&=&2 \left(C_a+C_b+C_c-\frac{1}{2}C_d\right)\, \left(\frac{\ln(\bar{y})}{\bar{y}}\right)_+\nn\\[2mm]
&-&\left( 2 C_a\,L + 2 C_b\,\bar{L}+\frac{1}{2}\,\gamma_c\right)\left(\frac{1}{\bar{y}}\right)_+\nn\\[2mm]
&+&\left(C_a\,L^2+C_b\,\bar{L}^2-\frac{3}{4}(C_a+C_b)\zeta(2)+\frac{1}{2}K_c\right)\delta(\bar{y})\,,\nn\\
\eea
where $C_q=C_F$, $C_g=C_A$, and 
\ba
&&\gamma_q \,=\, \frac{3}{2}\,C_F\,,\quad\quad \gamma_g \,=\, 2\pi b_0\,=\,\frac{1}{6}(11C_A-2N_f)\,,\nn\\[2mm]
&&K_q \,=\, \left(\frac{7}{2}-3\zeta(2)\right)C_F\,,\quad\quad K_g\,=\,
\left(\frac{67}{18}-3 \zeta(2)\right)C_A-\frac{5}{9}\,N_f\,.
\ea
The one-loop hard functions $H_{ab\to cr}^{(1)}$ used below may be found in 
Refs. \cite{Hinderer:2014qta,Kelley:2010fn,Broggio:2014hoa}.

\subsection{$q\bar{q}{\,'}\to q\bar{q}{\,'}$}

The term ${\mathrm{Tr}}\left\{ H^{(0)} \, S_0^{(0)}  \right\}$ is identical to that for the process $qq'\to qq'$ 
given in Eq.~(\ref{qqpexp}). Furthermore,
\bea
{\mathrm{Tr}}\left\{ H^{(0)} \,\left(\Gamma^{(1)}\right)^\dagger\, S_0^{(0)}+
H^{(0)} \, S_0^{(0)}\,\Gamma^{(1)}  \right\}_{q\bar{q}{\,'}\to q\bar{q}{\,'}}
&=&{\mathrm{Tr}}\left\{ H^{(0)} \,\left(\Gamma^{(1)}\right)^\dagger\, S_0^{(0)}+
H^{(0)} \, S_0^{(0)}\,\Gamma^{(1)}  \right\}_{qq'\to qq'}\nn\\[2mm]
&-&
\frac{2C_F(C_A^2-4)}{C_A^2}\,\frac{1+v^2}{(1-v)^2}\,\,L\,,
\eea
and
\bea
{\mathrm{Tr}}\left\{ H^{(1)} \, S_0^{(0)} +H^{(0)} \, S_0^{(1)}  \right\}_{q\bar{q}{\,'}\to q\bar{q}{\,'}} &=&
{\mathrm{Tr}}\left\{ H^{(1)} \, S_0^{(0)} +H^{(0)} \, S_0^{(1)}  \right\}_{qq'\to qq'}\nn\\[2mm]
&-&\frac{C_F(C_A^2-4)}{4C_A^2(1-v)^2}\,
\bigg(-(1-v^2) \big(L^2 + 2 \bar{L}^2 + 2 \bar{L} + \pi^2\big)\nn\\[2mm]
&+&2 L \big(1-v-\bar{L} \,(3 +5 v^2)\big)\bigg)\,,
\eea
with the trace terms for $qq'\to qq'$ also given in~(\ref{qqpexp}).

\subsection{$qq\to qq$}

We have
\bea
{\mathrm{Tr}}\left\{ H^{(0)} \,S_0^{(0)}\right\}&=&\frac{2 C_F}{C_A\,v^2 (1-v)^2}\,
\Big(C_A\big(1-3v+4v^2-2v^3+v^4\big)-v (1-v)\Big)\,, \nn\\[2mm]
{\mathrm{Tr}}\left\{ H^{(0)} \,\left(\Gamma^{(1)}\right)^\dagger\, S_0^{(0)}+
H^{(0)} \, S_0^{(0)}\,\Gamma^{(1)}  \right\}&=&  \frac{2 C_F}{C_A\,v^2 (1-v)^2}\,\nn\\[2mm]
&\times&\bigg[2 C_F \Big(L\, v^2(1+v)^2+\bar{L}(1-v)^2\big(1+(1-v)^2\big)\Big)
\nn\\[2mm]
&&-\frac{2}{C_A^2}(L+\bar{L})\Big(C_A\big(1-3v+4v^2-2v^3+v^4\big)-v (1-v)\Big)\bigg]\,.\nn\\
\eea
and
\bea
{\mathrm{Tr}}\left\{ H^{(1)} \, S_0^{(0)} +H^{(0)} \, S_0^{(1)}  \right\}&=&\frac{1}{
243\,v^2(1-v)^2}\bigg[3 L ^2 \left(46 v^4-200 v^3+285 v^2-251 v+72\right)+\nn\\[2mm]
&+&3 \bar{L}^2 \left(46 v^4+16 v^3-39 v^2+97 v-48\right)\nn\\[2mm]
&-&6 \bar{L} L  \left(148 v^4-296v^3+487 v^2-339 v+108\right)\nn\\[2mm]
&+&12 N_f\, \bar{L} \,v \left(3 v^3+4 v-1\right)
+12N_f\, L \, (v-1) \left(3
    v^3-9 v^2+13 v-6\right)
\nn\\[2mm]
&-&6 \bar{L} v \left(33 v^3-46
v^2+96 v-29\right)-6 L  (v-1) \left(33 v^3-53 v^2+103 v-54\right)\nn\\[2mm]
&-&40\, N_f \left(v^2-v+3\right)
    \left(3 v^2-3 v+1\right)+252 \left(v^2-v+3\right) \left(3 v^2-3 v+1\right)\nn\\[2mm]
&+&\pi ^2 \left(540 v^4-1080 v^3+2069
    v^2-1529 v+420\right)\bigg]\,.
\eea
In the last expression we have set $C_F=4/3$ and $C_A=3$ for simplicity.

\subsection{$q\bar{q}\to q'\bar{q}{\,'}$}

We have
\bea
{\mathrm{Tr}}\left\{ H^{(0)} \, S_0^{(0)}  \right\}&=&\frac{C_F}{C_A}\,\big(
v^2+(1-v)^2\big)\,,\nn\\[2mm]
{\mathrm{Tr}}\left\{ H^{(0)} \,\left(\Gamma^{(1)}\right)^\dagger\, S_0^{(0)}+
H^{(0)} \, S_0^{(0)}\,\Gamma^{(1)}  \right\}&=&  \frac{2 C_F}{C_A^2}\,
\big(v^2+(1-v)^2\big)\,\Big(2 L +\big(C_A^2-2\big)\bar{L}\Big)\,,
\eea
and
\bea
{\mathrm{Tr}}\left\{ H^{(1)} \, S_0^{(0)} +H^{(0)} \, S_0^{(1)}  \right\}&=&\frac{1}{81}\,
\bigg[-6 \,L^2\, \left(8 v^2-10 v+5\right)-3\,\bar{L}^2 \left(16 v^2-2 v+1\right)\nn\\[2mm]
&-&60 \,L\,\bar{L}\, \left( v^2+(1-v)^2 \right)+42\,v\,\bar{L}+12 \, (1-v)\,L\nn\\[2mm]
&+&2 \Big(5\, \pi^2  +63 -10\,N_f\Big)\,\left(v^2+(1-v)^2\right)\bigg]\,.
\eea
In the last expression we have set $C_F=4/3$ and $C_A=3$ for simplicity.

\subsection{$q\bar{q}\to q\bar{q}$}

We have
\bea
{\mathrm{Tr}}\left\{ H^{(0)} \,S_0^{(0)}\right\}&=&\frac{2 C_F}{C_A^2\, (1-v)^2}\,
\Big(C_A\big(1-2v+4v^2-3v^3+v^4\big)+v^2 (1-v)\Big)\,, \nn\\[2mm]
{\mathrm{Tr}}\left\{ H^{(0)} \,\left(\Gamma^{(1)}\right)^\dagger\, S_0^{(0)}+
H^{(0)} \, S_0^{(0)}\,\Gamma^{(1)}  \right\}&=&  \frac{2 C_F}{C_A\,(1-v)^2}\,\nn\\[2mm]
&\times&\bigg[2 C_F (1-v)^2\, \left(v^2+(1-v)^2\right)\,\bar{L}+
\frac{4C_F}{C_A}\,L\,v^2 (1-v)\nn\\[2mm]
&+&\frac{2}{C_A^2}\big(2L-\bar{L}\big)
\Big(C_A\big(1-2v+4v^2-3v^3+v^4\big)+v^2 (1-v)\Big)
\bigg]\,.\nn\\
\eea
and
\bea
{\mathrm{Tr}}\left\{ H^{(1)} \, S_0^{(0)} +H^{(0)} \, S_0^{(1)}  \right\}&=&\frac{1}{
243\,(1-v)^2}\bigg[-12\,L^2\, \left(12 v^4-43 v^3+58 v^2-30 v+15\right)\nn\\[2mm]
&-&3 \,\bar{L}^2 \left(48 v^4-97 v^3+39 v^2-16 v-46\right)\nn\\[2mm]
&-&12 \,L\,\bar{L}\,\left(30 v^4-110v^3+125 v^2-60 v+9\right)\nn\\[2mm]
&-&12\,\bar{L}\,N_f \,  \left(v^3-4 v^2-3\right)+6 \,\bar{L}\, \left(29 v^3-96 v^2+46v-33
\right)\nn\\[2mm]
&+&36 \,L\, (1-v) \left(v^2-2 v+2\right)\nn\\[2mm]
&+&4\,\big(63-10N_f\big)\left(v^2-3 v+3\right) \left(3 v^2-v+1\right)\nn\\[2mm]
&+&2 \pi ^2\left(30 v^4-130 v^3+259 v^2-60 v+111\right) \bigg]\,.
\eea
In the last expression we have set $C_F=4/3$ and $C_A=3$ for simplicity.

\subsection{$q\bar{q}\to gg$}

We have
\bea
{\mathrm{Tr}}\left\{ H^{(0)} \, S_0^{(0)}  \right\}&=&C_F\,
\frac{v^2+(1-v)^2}{v(1-v)}\left(v^2+(1-v)^2-\frac{1}{C_A^2}\right)\,,\nn\\[2mm]
{\mathrm{Tr}}\left\{ H^{(0)} \,\left(\Gamma^{(1)}\right)^\dagger\, S_0^{(0)}+
H^{(0)} \, S_0^{(0)}\,\Gamma^{(1)}  \right\}&=&\frac{2C_F}{C_A}\,\frac{v^2+(1-v)^2}{v(1-v)}\,
\Big(L\, \big(C_A^2 (1-v)^2-1\big)+\bar{L}\,\big(C_A^2 v^2-1\big)\Big)\,,\nn\\
\eea
and
\bea
{\mathrm{Tr}}\left\{ H^{(1)} \, S_0^{(0)} +H^{(0)} \, S_0^{(1)}  \right\} &=&\frac{1}{
81\,v(1-v)}\bigg[\,L^2\,\left(-288 v^4+909 v^3-1126
    v^2+605 v-144\right)\nn\\[2mm]
&+&\bar{L}^2 \left(-288 v^4+243 v^3-127 v^2+72 v-44\right)\nn\\[2mm]
&-&8 \,L\,\bar{L}\,  \left(v^2+(1-v)^2\right) 
\left(11-45 v(1-v)\right)+4 \,\bar{L}\, v (9 v-1) (11 v-5)\nn\\[2mm]
&+&4 \,\bar{L}\, (1-v) (9 v-8) (11 v-6)+30 \pi ^2 \left(v^2+(1-v)^2\right) \left(4-9v(1-v)
\right)\nn\\[2mm]
&-&2 \left(828 v^4-1656 v^3+1735 v^2-907 v+224\right) \bigg]\,.
\eea
In the last expression we have set $C_F=4/3$ and $C_A=3$ for simplicity.

\subsection{$qg\to qg$}

We have
\bea
{\mathrm{Tr}}\left\{ H^{(0)} \, S_0^{(0)}  \right\}&=&\frac{1}{C_A^2}\,
\frac{1+v^2}{v(1-v)^2}\left(C_FC_A(1+v^2)+v\right)\,,\nn\\[2mm]
{\mathrm{Tr}}\left\{ H^{(0)} \,\left(\Gamma^{(1)}\right)^\dagger\, S_0^{(0)}+
H^{(0)} \, S_0^{(0)}\,\Gamma^{(1)}  \right\}&=&\frac{1}{C_A^3}\,\frac{1+v^2}{v(1-v)^2}\,
\Big(2 C_F^2 C_A^2 \big(2L+\bar{L}(1+v^2)\big)\nn\\[2mm]
&+&2C_FC_A\big( (1+2v-v^2)\,L+(1-v+v^2)\,\bar{L} \big)\nn\\[2mm]
&+&v(2-v)\,L+ (1-v)^2\,\bar{L}\Big)\,,
\eea
and
\bea
{\mathrm{Tr}}\left\{ H^{(1)} \, S_0^{(0)} +H^{(0)} \, S_0^{(1)}  \right\} &=&\frac{1}{
216\,v(1-v)^2}\bigg[\,L^2\, \left(-44 v^4+104 v^3-175 v^2-29
    v-144\right)\nn\\[2mm]
&+&\bar{L}^2\, \left(-204 v^4+71 v^3-382 v^2+71 v-204\right)\nn\\[2mm]
&+&2 \,L\,\bar{L}\,\left(60 v^4-100v^3-117 v^2+33 v-164\right)\nn\\[2mm]
&+&4\,\bar{L}\, v \left(19 v^2+178 v+19\right)+4 \,L\,(1-v) (v+8) (5 v+6)\nn\\[2mm]
&-&2 \left(224 v^4+11 v^3+358 v^2+11 v+224\right)\nn\\[2mm]
&+&\pi ^2 \left(240 v^4+175v^3+393 v^2+42 v+140\right)\bigg]\,.
\eea
In the last expression we have set $C_F=4/3$ and $C_A=3$ for simplicity.

\subsection{$qg\to gq$}

We have
\bea
{\mathrm{Tr}}\left\{ H^{(0)} \, S_0^{(0)}  \right\}_{qg\to gq}&=&\left[
{\mathrm{Tr}}\left\{ H^{(0)} \, S_0^{(0)}  \right\}_{qg\to qg}\right]_{v\leftrightarrow 1-v}\,,
\nn\\[2mm]
{\mathrm{Tr}}\left\{ H^{(0)} \,\left(\Gamma^{(1)}\right)^\dagger\, S_0^{(0)}+
H^{(0)} \, S_0^{(0)}\,\Gamma^{(1)}  \right\}_{qg\to gq}&=&\left[
{\mathrm{Tr}}\left\{ H^{(0)} \,\left(\Gamma^{(1)}\right)^\dagger\, S_0^{(0)}+
H^{(0)} \, S_0^{(0)}\,\Gamma^{(1)}  \right\}_{qg\to qg}\right]_{v\leftrightarrow 1-v}\,,\nn\\
\eea
and 
\bea
{\mathrm{Tr}}\left\{ H^{(1)} \, S_0^{(0)} +H^{(0)} \, S_0^{(1)}  \right\}_{qg\to gq} &=&\left[
{\mathrm{Tr}}\left\{ H^{(1)} \, S_0^{(0)} +H^{(0)} \, S_0^{(1)}  \right\}_{qg\to qg}
\right]_{v\leftrightarrow 1-v}\nn\\[2mm]
&+&\frac{C_A^2+1}{8 C_A\,v^2 (1-v)}\,\big(1+(1-v)^2\big)\,\left(1+(1-v)^2-\frac{v^2}{C_A^2}
\right)\nn\\[2mm]
&\times&\left(3\,L^2-\bar{L}^2-2\,L\,\bar{L}-\zeta(2)\right)\,.
\eea

\subsection{$gg\to q\bar{q}$}

We have
\bea
{\mathrm{Tr}}\left\{ H^{(0)} \, S_0^{(0)}  \right\}_{gg\to q\bar{q}}&=&\frac{C_A^2}{C_A^2-1}\,
{\mathrm{Tr}}\left\{ H^{(0)} \, S_0^{(0)}  \right\}_{q\bar{q}\to gg}\,,
\nn\\[2mm]
{\mathrm{Tr}}\left\{ H^{(0)} \,\left(\Gamma^{(1)}\right)^\dagger\, S_0^{(0)}+
H^{(0)} \, S_0^{(0)}\,\Gamma^{(1)}  \right\}_{gg\to q\bar{q}}&=&\frac{C_A^2}{C_A^2-1}\,
{\mathrm{Tr}}\left\{ H^{(0)} \,\left(\Gamma^{(1)}\right)^\dagger\, S_0^{(0)}+
H^{(0)} \, S_0^{(0)}\,\Gamma^{(1)}  \right\}_{q\bar{q}\to gg}\,,\nn\\
\eea
and 
\bea
{\mathrm{Tr}}\left\{ H^{(1)} \, S_0^{(0)} +H^{(0)} \, S_0^{(1)}  \right\}_{gg\to q\bar{q}} &=&
\frac{C_A^2}{C_A^2-1}\,
{\mathrm{Tr}}\left\{ H^{(1)} \, S_0^{(0)} +H^{(0)} \, S_0^{(1)}  \right\}_{q\bar{q}\to gg}\nn\\[2mm]
&-&\frac{C_A^2+1}{8\,v (1-v)}\,\big(v^2+(1-v)^2\big)\,\left(v^2+(1-v)^2-\frac{1}{C_A^2}
\right)\nn\\[2mm]
&\times&\left( (L-\bar{L})^2-2\zeta(2)\right)\,.
\eea

\subsection{$gg\to gg$}
 
We have
\bea
{\mathrm{Tr}}\left\{ H^{(0)} \, S_0^{(0)}  \right\}&=&
\frac{C_A^2}{2}\,\frac{(1-v+v^2)^3}{v^2(1-v)^2}\,,\nn\\[2mm]
{\mathrm{Tr}}\left\{ H^{(0)} \,\left(\Gamma^{(1)}\right)^\dagger\, S_0^{(0)}+
H^{(0)} \, S_0^{(0)}\,\Gamma^{(1)}  \right\}&=&\frac{27}{2}\,
\frac{(1-v+v^2)^2}{v^2(1-v)^2}\,\left((1+(1-v)^2)\,L +(1+v^2)\,\bar{L} \,\right)\,,\nn\\
\eea
and
\bea
{\mathrm{Tr}}\left\{ H^{(1)} \, S_0^{(0)} +H^{(0)} \, S_0^{(1)}  \right\} &=&\frac{1}{32\,v^2(1-v)^2}\,
\bigg[-108\,L^2\,\left(2 v^6-7 v^5+15 v^4-18 v^3+14 v^2-6 v+2\right)\nn\\[2mm]
&-&108\,\bar{L}^2\,\left(2 v^6-5 v^5+10 v^4-12v^3+10 v^2-5 v+2\right)\nn\\[2mm]
&+&9\,L^2\, N_f\, (1-v) v^2 \left(v^2+v-1\right)+
9\,\bar{L}^2\,N_f (1-v)^2 v \left(v^2-3 v+1\right)\nn\\[2mm]
&-&18 \,L\,\bar{L}\,\Big(6\left(5 v^4-10v^3+15 v^2-10 v+4\right)+
N_f v(1-v)\left( v^2+(1-v)^2\right)\Big)\nn\\[2mm]
&+&36 \,L\,(1-v) \left(v^2-v+1\right) \left(7 v^2-22 v+22\right)\nn\\[2mm]
&+&36 \,\bar{L}\,v \left(v^2-v+1\right) \left(7 v^2+8 v+7\right)\nn\\[2mm]
&-&6 \,L\,N_f\, (1-v) \left(v^2-v+1\right)\left(5 v^2-8 v+8\right)\nn\\[2mm]
&-&6 \,\bar{L}\,N_f\, v \left(v^2-v+1\right) \left(5 v^2-2 v+5\right)\nn\\[2mm]
&+&9N_f\, \pi ^2 (1-v) v \left(v^2+(1-v)^2\right)\nn\\[2mm]
&+&2 N_f\,\left(40 v^6-120 v^5+267 v^4-334 v^3+267 v^2-120 v+40\right)\nn\\[2mm]
&+&18 \pi^2 \left(20 v^6-60 v^5+123 v^4-146 v^3+129 v^2-66 v+20\right)\nn\\[2mm]
&-&6 \left(268 v^6-804 v^5+1635 v^4-1930 v^3+1635 v^2-804v+268\right)
\bigg]\,.
\eea
We have set $C_F=4/3$ and $C_A=3$ for simplicity.

\end{appendix}



\end{document}